# KINETIC PHENOMENA IN METALLIC MULTILAYERS

L. V. Dekhtyaruk,[1] Yu. A. Kolesnichenko,[2] V. G. Peschansky[2]

[1] Kharkov State University of Civil Engineering and Architecture, 40 Sumskaya str., Kharkov, 61002, Ukraine.

[2] B.Verkin Institute for Low Temperature Physics and Engineering, National Academy of Sciences of the Ukraine, 47 Lenin Ave, Kharkov, 61164, Ukraine.

A series of kinetic phenomena in metallic multilayers has been considered. The kinetic properties of multilayers differ essentially from the properties of both massive metals and thin films. One of the main reasons of that is the influence of electron interaction of electrons with interfaces between layers. From one hand, this interaction leads to the additional electron scattering and conductivity of multilayer may be noticeably less than specific conductivities of composing metals. From the other hand, the electron reflection from interfaces in a strong magnetic field may results in considerable increasing of conducting properties in consequence of the static skin effect. Due to changing of electron trajectories after collisions with interfaces new types of periodic motion in the magnetic field and therefore new size and resonance phenomena in high frequency fields appear. In thin normal layers on the superconducting substrate, changing of trajectories is due to Andreev reflection and resonance effects, which do not exist in normal multilayers, films, and bulk monocrystals, take place. In multilayers consisting of ferromagnetic and nonmagnetic metals, the internal magnetic field in ferromagnetic layers must be taken into account, if the Larmor radius in this field is comparable with the layer thickness. Because of the mutual diffusion of metals, the kinetic coefficients of multilayers are changed in time. The investigation of this changing may be used for determination of diffusion coefficient for the bulk and grain-boundary diffusion. The effects, which have been analyzed theoretically in this review, can be used for the obtaining information on the electron interaction with interfaces in conducting multilayers.

## CONTENTS

# 1. INTRODUCTION

The significant progress in microelectronics in many respects is obliged to making use of layered structures, such as heterostructures, semiconducting superlattices, and different types of multilayers, which have unique physical properties. For this reason investigations of electron phenomena in various layered structures have arisen an extensive interest.  Multilayers, which are periodic systems formed by alternating layers of different metals or semiconductors, are widely used as elements of modern microelectronic devices. Their physical properties essentially depend on electron interaction with interfaces between layers.

Magnetic multilayers (MML) whose periodic structure contains a ferromagnetic layer as an element, display especially interesting properties (see, for example, the review [1]). Among them, the giant magnetoresistance (GMR) [2] is one of the most striking and significant for applied tasks effects displayed by MML. In the absence of magnetic field, the magnetic moments $\mathbf{M}$ in adjacent ferromagnetic $(F)$ layers separated by a nonmagnetic $(N)$ layer are antiparallel because of indirect RKKY interaction between the $F$-layers. The application of a magnetic field $\mathbf{H}$ applied parallel to the interfaces of a MML leads to an alignment of magnetic moments $\mathbf{M}$ in $F$-layers along the vector $\mathbf{H}.$ At the same time, the sample resistance decreases. The magnetoresistance decreasing even in a quite weak magnetic field may be very significant (sometimes exceeding 100%!) and so it was called giant magnetoresistance. The assumption about the dominating role of electron scattering at the internal boundaries, which depends on the spin of electrons, seems to be, the most well founded hypothesis about the origin of GMR [1 - 3]. It is based on the well-known fact that the cross-sections of scattering of charge carriers with different spin by magnetic impurities are different in view of the dependence of the density of electron states at the Fermi surface on a spin direction [4]. GMR was observed [2] for the first time in the MML Fe/Cr, and investigated subsequently for quite different combinations of ferromagnetic and nonmagnetic metals.

In last years, the interest to a study of the mutual influence of ferromagnetism and superconductivity in artificial heterostructures superconductor-ferromagnetic $(S - F)$ has grown too much [5-8]. It was found that a critical temperature of such systems is a nonmonotonic function of the $F$-layer thickness and very sensitive to the transparency of $S - F$ interface. The possibility of tunneling of Cooper pairs through the $S - F$ interface make it possible to realize the unconventional superconductivity in the $F$-layer.

Great number of theoretical and experimental papers deals with investigations of electron phenomena in semiconducting superlattices. A period of these artificial lattices is imposed by the technology of production and it essentially exceeds interatomic distances. A week conductivity of superlattice across the layers in comparison with conductivity along the layers causes a noticeable dissipation of an electron current, which flows across the layers. As a result, in semiconducting superlattice a current-voltage characteristic is nonlinear at relatively small current densities (see, for example, [9-12] and references in it).

The experimental investigations of conductivity of multilayers of nontransient metals were started in 1965 by Lukas [12], who measured the specific resistance of double-layer gold film (DLF). He has found that the dependence of the DLF resistance on its thickness was differing essentially from analogous dependence of a single layer film. Subsequent experiments had shown that the conductivity of double-layer system depends nonmonotonically on the sample thickness. This effect may result from periodically changes of reflection conditions for electrons at the interface or by the quantum size effect.

In a magnetic field, the dependence of kinetic coefficients of multilayers of nontransition metals on the electron interaction with interfaces is more essential, because at low temperatures in pure conductors the electron mean path is about few millimeters, and ballistic effects became most pronounced. In this connection, we conceive that multilayers of nontransition metals will be used for modern microelectronics.

The aim of our review is to show a variety of physical phenomena in multilayers consisting of metal layers, which possess high electrical conductivity.

## 2. FORMULATION OF THE PROBLEM AND SOLUTION OF THE KINETIC EQUATION.

Let us consider a multilayer consisting from two types of metal layers with interfaces $x = x_{\text{int}}$, which are parallel to surfaces $x = x_s$ of a sample (Fig.1). The boundaries between layers will be simulated by ideal plane (in real systems, the width of the transition layer does not exceed $10\,\overset{\circ}{\mathrm{A}}$ as a rule). A normal to the boundaries is parallel to $x$ axis. For simplicity the size of the specimen in $yz$ plane we assume to be infinite. Because of homogeneity of the problem in the plane of interfaces all characteristics, such as a current density, electrical field and so on, will depend only on $x$ coordinate. The thicknesses $d_n$ $(n = 1, 2)$ of the layers are assumed much larger than the de Broglie wavelength of electrons. Combined with the fact that magnetic quantization of the energy spectrum of electrons in a quite week magnetic field can be disregarded, this allows us to confine the analysis to the quasiclassical approximation.

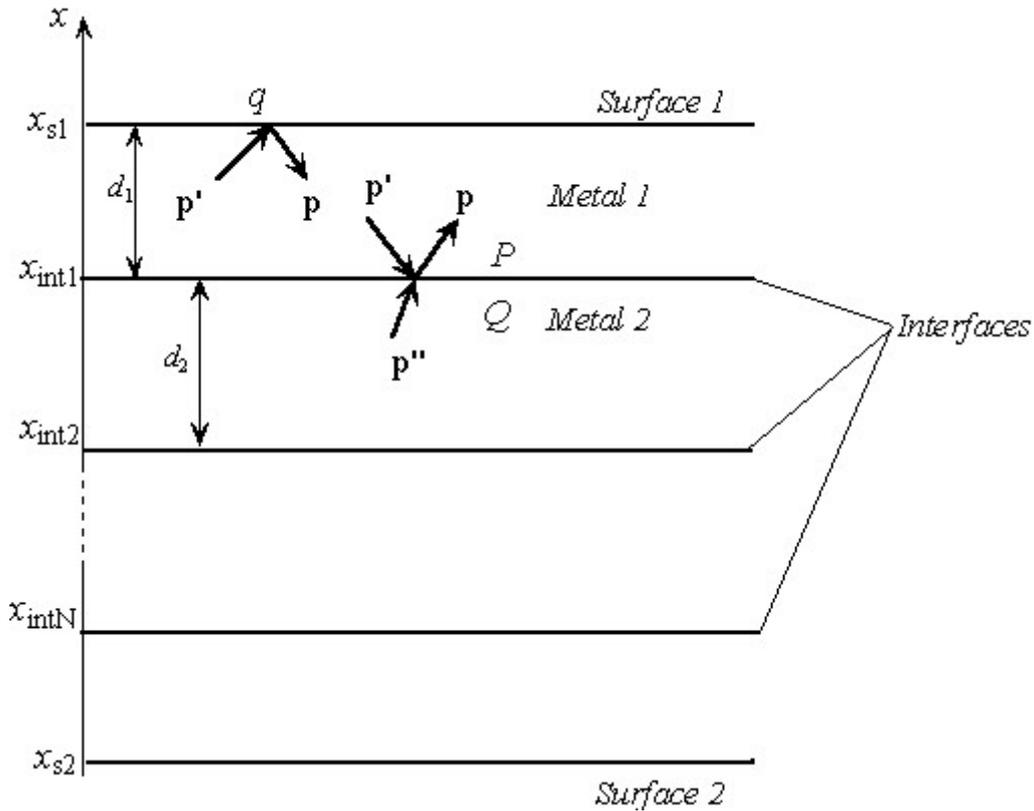

**Figure 1.** Model of a metallic multilayer. In the absence of diffuse electron scattering momenta $\mathbf{p}$, $\mathbf{p}'$ and $\mathbf{p}''$ are related by Eqs. (2.5), (2.6).

The microscopic analysis of an interaction of an electron with a real interface is extremely complicated. For this reason, the phenomenological approach, which was proposed by Kaganov and Fiks in Ref. [13], is applied frequently. In the quasiclassical approximation, the reflection and the transmission of charge carriers are described by using the boundary conditions for the nonequiltbrium electron distribution function. A subsequent comparison of the theoretical results with experimental data will permit determining the transmission and scattering probabilities for charge carriers colliding with the boundary separating the layers.

In this review, we shall consider only linear response of the metal to an external perturbation. The kinetic equation for the nonequiltbrium addition $\left(\frac{\partial f_0}{\partial \varepsilon_n}\right)\Psi_n(x,\mathbf{p},t_0)$ to the Fermi distribution function $f_0(\varepsilon_n)$ in the $n$-th layer, linearized with respect to the perturbation (for example, the electrical field $\mathbf{E}$), can be presented in the form

$$\frac{\partial \Psi_n}{\partial t_0} + v_{xn}\frac{\partial \Psi_n}{\partial x} + \frac{e}{c}\left[\mathbf{v}_n\mathbf{H}\right]\frac{\partial \Psi_n}{\partial \mathbf{p}} = g\left(x,\mathbf{p},t_0\right) + I\{\Psi_n\}, \tag{2.1}$$

where $e$, $\mathbf{r}$, and $\mathbf{p}$ are the charge, coordinate, and momentum of an electron, respectively; $\varepsilon_n(\mathbf{p})$ and $\mathbf{v}_n = \partial \varepsilon_n / \partial \mathbf{p}$ are an electron energy and velocity in $n$-th layer; $\mathbf{H}$ is a strength of a magnetic field; $t_0$ is a "time" (in stationary fields the derivative of $\Psi_n$ over $t_0$ is equal to zero); $I\{\Psi_n\}$ is collision integral, which describes a bulk scattering of electrons. Below we use the $\tau$-approximation for $I\{\Psi_n\}$

$$I\{\Psi_n\} = -\frac{1}{\tau_n}\Psi_n, \tag{2.2}$$

where $\tau_n$ is effective mean free time between bulk collisions in the $n-$th layer. The function $g\left(x,\mathbf{p},t_0\right)$ depends on the type of nonequilibrium perturbation of electrons, and for electrons in an electrical field it is equal to $e\,\mathbf{v}_n\mathbf{E}$. The index $n$ indicates the layers with which the corresponding quantity is related. Describing an interaction of electrons with interface, we use the boundary conditions obtained by Ustinov [14], which automatically ensures that the normal component of the current is conserved. The boundary conditions connect the distribution function $\Psi_n(\mathbf{p})$ of electrons flying into the *n-th* layer on the interface $x = x_{int}$ with the distribution functions of charge carriers, which incident on the boundary from the same layer $\Psi_n(\mathbf{p}')$ or the adjacent layer $\Psi_m(\mathbf{p}'')$:

$$\Psi_n(x_{int}, \mathbf{p}) = P\Psi_n(x_{int}, \mathbf{p}') + Q\Psi_m(x_{int}, \mathbf{p}'') + W_1 \frac{\langle v_{xn}\Psi_n \rangle_{-s_n}}{\langle v_{xn} \rangle_{-s_n}} + W_2 \frac{\langle v_{xm}\Psi_m \rangle_{s_m}}{\langle v_{xm} \rangle_{s_m}}. \qquad (2.3)$$

Here

$$\langle ... \rangle_{s_n} = \int\limits_{\varepsilon_n(\mathbf{p}) = \varepsilon_F} \frac{dS_\mathbf{p} \Theta(s_n)}{v_n} \{...\}; \qquad (2.4)$$

$s_n(\mathbf{p}) = \mathrm{sgn}(\mathbf{N} \ \mathbf{v}_n)$, $\Theta(s_n)$ is the unit step function, $\mathbf{N}$ is an inner normal to the interface in $n-th$ layer; $dS_\mathbf{p}$ is an area element at the Fermi surface $\varepsilon_n = \varepsilon_F$, and $v_n = |\mathbf{v_n}|$; $\varepsilon_F$ is the Fermi energy (we assume that Fermi energies in the layers are equal); $P + Q + W_1 + W_2 = 1$; $P$ and $Q$ are the probabilities for charge carriers to be reflected by and to tunnel through the boundary without a scattering; $W_1$ and $W_2$ are the probabilities for diffusive scattering without and with penetration into the neighboring layer. In order to avoid writing out cumbersome equations, we will assume that the probabilities $P$, $Q$, $W_1$ and $W_2$ do not depend on the momentum of the charge carriers. The momentum $\mathbf{p}$ of electrons reflected from or passing through the interface in Eq. (2.3) is related to the momenta $\mathbf{p}'$ and $\mathbf{p}''$ of the incident charges by the conservation of energy $\varepsilon_n(\mathbf{p})$ and the tangential component $\mathbf{p}_t$ of the quasimomentum $\mathbf{p}$:

$$\varepsilon_n(\mathbf{p}) = \varepsilon_n(\mathbf{p}'), \qquad \mathbf{p}_t = \mathbf{p}'_t, \quad s_n(\mathbf{p}) = -s_n(\mathbf{p}'), \qquad (2.5)$$

$$\varepsilon_n(\mathbf{p}) = \varepsilon_m(\mathbf{p}''), \qquad \mathbf{p}_t = \mathbf{p}''_t, \quad s_n(\mathbf{p}) = -s_n(\mathbf{p}''). \qquad (2.6)$$

We describe the scattering of electrons by the surface $x = x_s$ of the specimen with the help of the specularity parameter $q \ (\mathbf{p})$ [15]

$$\Psi_n(x_{si}, \mathbf{p}) = q_i \Psi_n(x_{si}, \mathbf{p}') + \frac{\langle (1 - q_i)v_{xn}\Psi_n \rangle_{-s_n}}{\langle v_{xn} \rangle_{-s_n}}; \quad i = 1, 2; \quad n = 1, N+1. \qquad (2.7)$$

In Eq. (2.7) the momenta $\mathbf{p}$ and $\mathbf{p}'$ are related by the condition of specular reflection (2.5).

The terms in the boundary conditions (2.3), (2.7), containing integrals of the distribution functions $\Psi_n$ with the velocity component $v_{xn}$, describe the change in the chemical potential of electrons reflected from and tunneling through the boundary as a result of the diffuseness of the scattering.

In the case of monochromatic perturbation (with frequency $\omega$)

$$g(x,\mathbf{p},t_0)=g(x,\mathbf{p})\exp(i\omega t_0);\quad \Psi(x,\mathbf{p},t_0)=\Psi(x,\mathbf{p})\exp(i\omega t_0); \qquad (2.8)$$

it is easy to solve Eq. (2.1) using the method of characteristics and to represent it in the form

$$\Psi_n(x,\mathbf{p})=F(x-x^{(n)}(t))\alpha(t,\lambda)+\int_{\lambda}^{t}dt'\alpha(t,t')g(x+x^{(n)}(t')-x^{(n)}(t),\mathbf{p})\,; \qquad (2.9)$$

$$\alpha(t,t')=\exp\left(i\omega^*\left(t-t'\right)\right); \qquad (2.10)$$

$$x-x^{(n)}(t)=x_{s,\mathrm{int}}-x^{(n)}\left(\lambda\right). \qquad (2.11)$$

Here $\omega^*=\omega+i\!\!\not\!\tau$, $\lambda<t$ is the "instant for the last collision" of an electron with the surface of the specimen or with the interface $x=x_{s,\mathrm{int}}$ ($\lambda(t)$ is a root of Eq.(2.11)); $t$ is the "time" that the electron moves along a ballistic trajectory;

$$x^{(n)}(t)=\int^{t}dt'\ v_{xn}(t')\,. \qquad (2.12)$$

The function $F(x-x^{(n)}(t))$, remaining constant along the ballistic trajectory of an electron, must be obtained with the help of the boundary conditions (2.3), (2.7).

Non-homogeneous electrical and magnetic fields in a specimen must be found from Maxwell equations. In metals placed in stationary fields the Poison equation for electrical field $\mathbf{E}=-\nabla\varphi$ can be replaced by the electroneutrality condition [16]

$$\left\langle \Psi_n(x,\mathbf{p})\right\rangle =0\,. \qquad (2.13)$$

It is obvious that in view of the homogeneity of the problem in the boundary plane, the potential $\varphi(\mathbf{r})=-yE_y-zE_z+\varphi(x)$ and Eq. (2.13) can be used for determining the field component $E_x(x)$ at right angle to the boundaries.

If the diffusive scattering of electrons at the surfaces and interfaces is negligible the boundary conditions (2.3) and (2.7) can be reduced to the form:

$$\Psi_n(x_{\mathrm{int}},\mathbf{p})=P\Psi_n(x_{\mathrm{int}},\mathbf{p}')+Q\Psi_m(x_{\mathrm{int}},\mathbf{p}'')\,; \qquad (2.14)$$

$$\Psi_n(x_{si},\mathbf{p})=q_i\Psi_n(x_{si},\mathbf{p}')\,, \qquad (2.15)$$

where $1-P-Q\ll 1;\;\;1-q_i\ll 1$. Note that the boundary conditions (2.14), (2.15) are valid for any surface scattering of electrons, if integral terms in Eqs. (2.3) and (2.7) are small. It is so if, for example, the function $\Psi_n(x,\mathbf{p})$ is a sharp function of the momentum on the Fermi surface. As a matter of fact, the using of Eqs. (2.14) and (2.15) leads to qualitatively right results for kinetic coefficients, which describe the electron transport along boundaries in a general case.

## 3. ELECTRICAL CONDUCTIVITY OF DOUBLE LAYER FILMS.

The existence of an interface between monocrystalline layers of a metal leads to a considerable disparity between the dependence of kinetic coefficients of double-layer films (DLF`s) on their thickness and the analogous dependence for a single-layer sample. This fact was firstly observed in Ref.12 in which the resistance of gold DLF was measured. Subsequent experimental investigations showed [12, 17-20] that the specific conductivity of DLF depends nonmonotonically on the sample thickness. The problem of electrical conductivity of DLF was firstly theoretically analyzed by Lukas [21], who considered a simple situation, in which the existence of the interface does not cause an additional scattering of the charge carriers. An attempt to take into account the tunneling of conduction electrons through the interface between layers was also made by the authors of Ref.22 who described the interaction of electrons with the internal boundary by considering the Fuchs generalized boundary conditions [15]. Later, the electrical conductivity of DLF, consisting of a base layer with a thin metal film deposited on it, was studied in Refs.23 while a general analytical expression was obtained in Refs. 24 and 25 for the electrical conductivity of a DLF having an arbitrary ratio of the plate thickness.

In this chapter, we analyze in detail the size dependence of the electrical conductivity of a DLF. It is shown that conductivity changes nonmonotonically with increasing thickness $d_2$ of a metal layer deposited on a base layer of thickness $d_1$. For $d_2 << d_1$ this change is determined by the interaction of electrons with the boundary between the layers. The electrical conductivity is analyzed numerically for a wide range of layer thickness for different values of parameters characterizing the interaction of change carriers with the interface between layers and the surfaces of a DLF [26, 27].

### 3.1 Films with monocrystalline structure

Let us consider a DLF of thickness $d = d_1 + d_2$ formed by monocrystalline layers of a metal (Fig.2), in which external electric field $\mathbf{E}$ is applied. By substituting $\Psi_i(x, \mathbf{p})$ in the form (2.9) (for $\omega = 0$) into the boundary condition (2.14) and (2.15), we obtain a set of four linear algebraic equations that allows the calculation of the functions $F_i$:

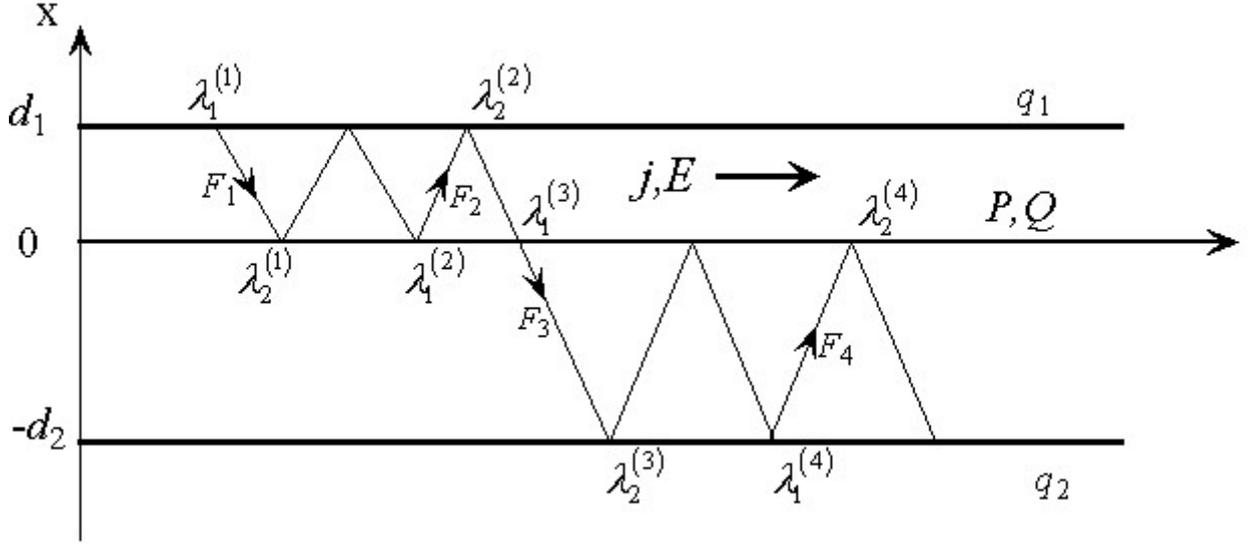

**Figure 2** Model of a double-layer metal film. The broken line shows schematically the possible trajectory of a conduction electron.

$$F_{1,4} = \frac{q_{1,2}}{D}\left\{\left(\varphi_k + P\alpha_k\varphi_i\right)\left(1 - q_k P\alpha_j\alpha_m\right) + Q\alpha_k\left(\varphi_j + q_k\left(\alpha_j\varphi_m + Q\alpha_j\alpha_m\varphi_i\right)\right)\right\}; \qquad (3.1)$$

$$F_{2,3} = \frac{1}{D}\left\{\left(\varphi_i + q_i\alpha_i\varphi_k\right)\left(P + q_k\left(Q^2 - P^2\right)\alpha_j\alpha_m\right) + Q\left(\varphi_j + q_k\alpha_j\varphi_m\right)\right\}. \qquad (3.2)$$

Here

$$\alpha_i = \exp\left\{\frac{\lambda_1^{(i)} - \lambda_2^{(i)}}{\tau_i}\right\}; \qquad (3.3)$$

$$\varphi_i = \int\limits_{\lambda_1^{(i)}}^{\lambda_2^{(i)}} dt' \mathbf{v}(t')\mathbf{E}\exp\left\{\frac{t' - \lambda_2^{(i)}}{\tau_i}\right\}, \qquad (3.4)$$

$$D = \left(1 - Pq_1\alpha_1\alpha_2\right)\left(1 - Pq_2\alpha_3\alpha_4\right) - Q^2 q_1 q_2\alpha_1\alpha_2\alpha_3\alpha_4;$$

$\lambda_1^{(i)}$ and $\lambda_2^{(i)}$ are two successive moments $(\lambda_1^{(i)} < \lambda_2^{(i)})$ of collision of an electron with the boundaries (see, Fig.2). The quantities $\alpha_i$ and $\varphi_i$ have the meaning of the probability of an electron moving without scattering in the bulk over a segment of the trajectory between two collisions and the energy acquired by this electron in an electric field over this segment.

Knowing the electron distribution function in each layer of the DLF, we can calculate the mean current density $\mathbf{j}$:

$$\mathbf{j} = \frac{2e}{dh^3}\sum_{i=1}^{2}\int\limits_0^{d_i} dx \int \frac{dS_{\mathbf{p}}}{v_i}\mathbf{v}_i\Psi_i(|x|, \mathbf{p}). \qquad (3.5)$$

The conductivity $\sigma = j_y / E$ of the DLF can be written in the form:

$$\sigma = \frac{1}{d} \sum_{i=1}^{2} d_i \sigma_{0i} \Phi_i \ , \qquad (3.6)$$

where $\sigma_{0i}$ is the specific conductivity of a bulk metal. The functions $\Phi_i$, defining the effect of the DLF size on $\sigma$, can be presented in the form

$$\Phi_i = 1 - \langle G_i \rangle \ , \qquad (3.7)$$

where

$$G_i = \frac{1}{\Delta} \left\{ \left( 2 - q_i - P + \left( q_i + P - 2q_i P \right) \varepsilon_i \right) \ \left( 1 - q_j P \varepsilon_j^2 \right) - q_j Q^2 \varepsilon_j^2 \left( 1 - \varepsilon_i + 2 q_i \varepsilon_i \right) - \right.$$
$$\left. Q \tau_{j,i} \left( 1 - \varepsilon_j \right) \left( 1 + q_i \varepsilon_i \right) \left( 1 + q_j \varepsilon_j \right) \right\} \ . \qquad (3.8)$$

Here

$$\Delta = 1 - P \left( q_i \varepsilon_i^2 + q_j \varepsilon_j^2 \right) - q_i q_j (Q^2 - P^2) \varepsilon_i^2 \varepsilon_j^2 . \qquad (3.9)$$

$$\varepsilon_i = \exp\left( -\frac{k_i}{\xi} \right); \qquad <...> = \frac{3}{4 k_i} \int_0^1 d\xi \left( \xi - \xi^3 \right) (1 - \varepsilon_i) \left\{ ... \right\} . \qquad (3.10)$$

For simplicity we assumed that the Fermi surface is a sphere of radius $p_F$.

One of the main parameters of the problem is the layer thickness $d_i$ normalized to the electron mean free path $k_i = d_i / l_i$. For limiting cases of thick ($k_i \gg 1$) and thin ($k_i \ll 1$) layers, the general formula (3.7) leads to the following asymptotic expressions for the functions $\Phi_i$:

1. Thick layers ($k_i \gg 1$):

$$\Phi_i = 1 - \frac{3}{16 k_i} \left\{ 2 - q_i - P - Q \tau_{j,i} \right\} . \qquad (3.11)$$

2. Thin layers ($k_i \ll 1$):

$$\Phi_i = \frac{3}{4} \frac{1 + q_i}{(1 - q_i P)(1 - q_j P) - q_i q_j Q^2} \left( (1 + P)(1 - q_j P) + q_j Q^2 + (1 + q_j) Q d_{j,i} \right) k_i \ell n \frac{1}{k_i} . \qquad (3.12)$$

3. Thin layer ($k_j \ll 1$), on a thick base layer ($k_i \gg 1$):

$$\Phi_i = 1 - \frac{3}{16 k_i} \left\{ 2 - q_i - P \right\} + \frac{1 + q_j}{2 \left( 1 - q_j P \right)} Q d_{j,i} ; \qquad (3.13)$$

$$\Phi_j = \frac{3}{4} \frac{(1 + q_j)(1 + P)}{1 - q_j P} \left\{ 1 - \frac{Q \tau_{i,j} (1 - q_j^2 + q_j (1 + P)(3 + q_j))}{2(1 + q_j)(1 + P)(1 - q_j P)} \right\} k_j \ell n \frac{1}{k_j} + \frac{1 + q_j}{2(1 - q_j P)} Q \tau_{i,j} ; \qquad (3.14)$$

$$q_i, \ (P+Q) << 1. \tag{3.15}$$

Here $\tau_{j,i} = \tau_j / \tau_i$, $d_{j,i} = d_j / d_i$.

The curves presented in Figs.3 *a-c* were obtained numerical computations using the general formula (3.6) and illustrate the dependence of the normalized conductivity of DLF on the ratio $d_{2,1} = d_2 / d_1$ of the layer thicknesses for different values of parameters characterizing the DLF ($k_2 = k_1 d_{2,1} l_{1,2}$; $l_{1,2} = l_1 / l_2$). The obtained dependences $\sigma(d_{2,1})$ (Fig.3) show that the effect of coating on the total electrical conductivity of the specimen is manifested significantly starting from a small thickness $d_2$. In the region of small $d_{2,1} << 1$, the change in $d_2$ associated with the effect of electron scattering at the interface while the dependence $\sigma(d_{2,1})$ for $d_{2,1} >> 1$ is determined by the ratio $l_{2,1}$ of electron mean free paths in each layer of the DLF.

Curve 1 in Fig.3*a* corresponds to the classical size effect in a single-layer film ($P = 0$, $Q = 1$), describing the increase in its conductivity with sample thickness. Curve 2 describes the dependence $\sigma \ (d_{2,1})$ for a system of two films separated by an insulating interlayer $Q = 0$:

$$\sigma(Q = 0) = \sigma_1 + \frac{d_2}{d_1 + d_2}\{\sigma_2 - \sigma_1\}, \tag{3.16}$$

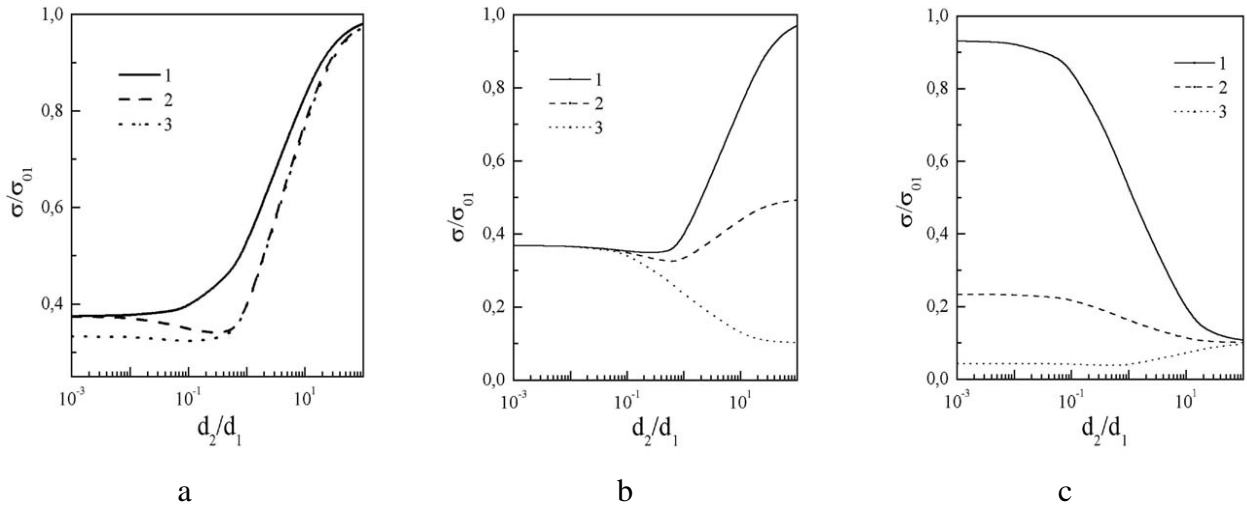

a b c

**Figure 3** Dependence of the electrical conductivity of a double-layer film on the ratio $d_2 / d_1$ of layer thickness for the, following of a) $q_i = 0.5, k_1 = 10^{-1}, l_1 / l_2 = 1$; 1) $P = 0$, $Q = 1$; 2) $P = 0.5, Q = 0$; 3) $P = 0.3, Q = 0.2$ b) $q_1 = 0.7, q_2 = 0.3, P = 0.3$, $Q = 0.2, k_1 = 10^{-1}$; $l_1 / l_2 = 1$ (1), $l_1 / l_2 = 2$ (2) $l_1 / l_2 = 10$ (3); c) $q_1 = 0.1$, $q_2 = 0.3$, $P = 0.1, Q = 0.2, l_1 / l_2 = 10$; $k_1 = 5$ (1), $k_1 = 10^{-1}$ (2), $k_1 = 10^{-2}$ (3).

where $\sigma_i$ is the electrical conductivity of a single-layer film of thickness $d_i$. It can be seen from this formula that if $d_2 \to 0$, for small values of $d_2/d_1$, the increment to the conductivity $\sigma_1$ is negative and reverses the sign at the point $d_2 \approx d_1$. The situation changes radically for a semitransparent (for a fixed value of $Q \neq 0$, $P \leq 1-Q$) boundary between the layers (curve 3 in Fig,3a), since the electrons tunneling into the coated layer with a nonzero probability of diffusive scattering may be scattered at the outer surface of the sample, and the quantity $\sigma$ differs from the conductivity $\sigma_1$ of the initial film even for $d_2 << d_1$. Note that the value of changes only insignificantly in the region $d_2 << d_1$ with increasing thickness of the coating and contains a minimum for $d_2 \approx d_1$, as in the case $Q = 0$.

Figure 3b illustrates the dependence $\sigma(d_{2,1})$ for different purities of the layer being deposited. Calculations show that for $d_2 << d_1$, the mean free path $l_2$ in the second layer plays an insignificant role, and the above-mentioned minimum can be observed only if a high-purity film with the specific conductivity $\sigma_{02} > \sigma_{01}$ is deposited. This conclusion is confirmed by the series of curves showing the emergence of the $\sigma(d_2)$ minimum in Fig. 3c (curve 4) upon a decrease in the thickness $d_1$ of the base layer (cf. curve 4 in Fig. 3b does not have a minimum for the same value of the ratio $l_2/l_1$).

### 3.2 Polycrystalline films

A theoretical analysis of the conductivity of a system consisting of alternating thin polycrystalline layers of two metals is made in Ref. [28] for the special case where there is no electron tunnelling through the interface between layers. The classical size effect in such films can be described on the basis of the model of Mayaadas and Shatzkes [29].

In this chapter we consider a double layer polycrystalline film of width $d$, substantially less than the free path of electrons (Fig.4). The collision integral $I(\Psi_i)$ in the kinetic equation (2.1), describes an electron scattering inside grains and on the boundaries of crystallites in polycrystalline layers. Following [29] we write $I(\Psi_i)$ in the form

$$I(\Psi_i) = -\Psi_i \frac{1}{\tau_i^*}, \tag{3.17}$$

where

$$\frac{1}{\tau_i^*} = \frac{1}{\tau_i}\left\{1 + \beta_i \frac{p_F}{|p_y|}\right\}, \tag{3.18}$$

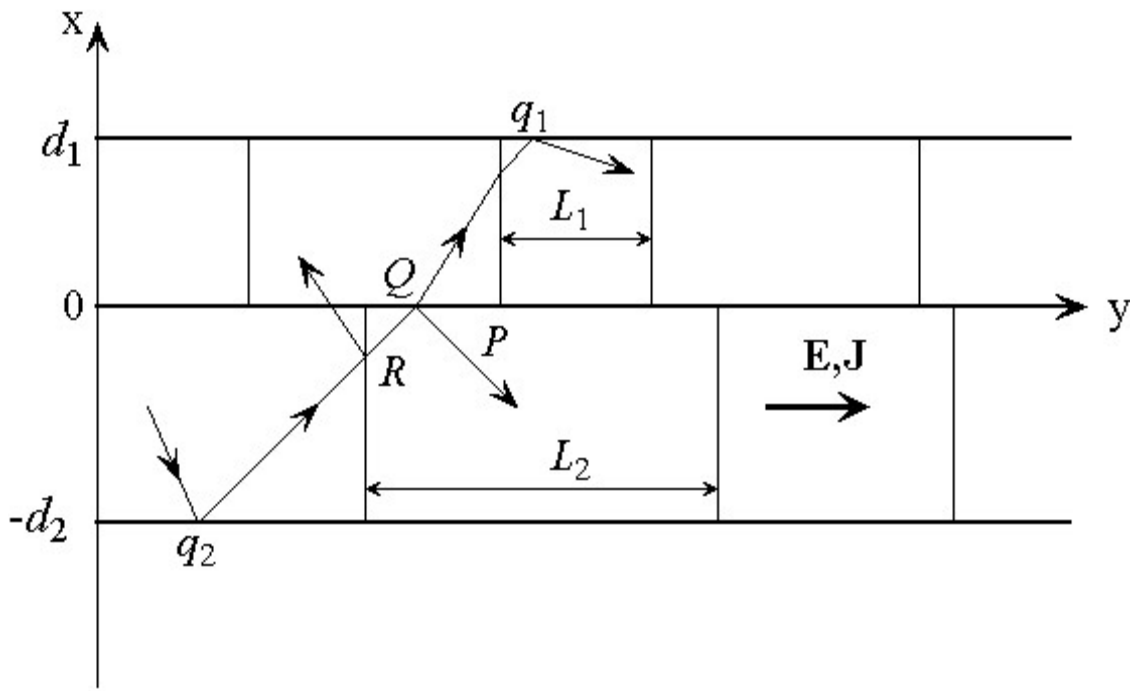

**Figure 4** Model of a double-layer polycrystalline film. The broken line represents a possible electron trajectory.

$p_F$ is the Fermi momentum; $\beta_i = \dfrac{l_i}{L_i}\dfrac{R_i}{1-R_i}$; $R_i$ is the coefficient of reflection of electrons by the interface; $L_i$ is the mean size of crystallites in the $i$-th layer.

Putting functions $\Psi_i$ in form (2.9) in boundary conditions (2.14), (2.15) we obtain a system of linear algebraic equations for the function $F$. The solutions of this system have the form of Eqs. (3.1), (3.2), but in functions $\alpha_i$ (3.3) and $\varphi_i$ (3.4) a mean free time $\tau_i$ must be replaced by effective value $\tau_i^*$ (3.18). Knowing the distribution function we can find the electrical current density (3.5) and conductivity of the specimen (3.6). In the case of a DLF with polycrystalline structure of layers the function $\Phi_i$ in Eq. (3.6) can be written in the form:

$$\Phi_i = T(\beta_i) - \langle\langle G_i \rangle\rangle\,; \tag{3.19}$$

$$T(\beta_i) = 1 - \frac{3}{2}\beta_i + 3\beta_i^2 - 3\beta_i^3 \ln\left(1 + \frac{1}{\beta_i}\right) \cong \begin{cases} 1 - \dfrac{3}{2}\beta_i + 3\beta_i^2, & \beta_i \ll 1; \\[2mm] \dfrac{3}{4\beta_i} - \dfrac{3}{5\beta_i^2}, & \beta_i \gg 1. \end{cases} \tag{3.20}$$

Functions $G_i$ are determined by Eq. (3.8), in which

$$\varepsilon_i = \exp\left\{-\frac{k_i H_i}{\xi}\right\}; \qquad H_i(\xi,\varphi) = 1 + \frac{\beta_i}{\cos\varphi\sqrt{1-\xi^2}}; \tag{3.21}$$

$$\langle\langle \ldots \rangle\rangle = \frac{3}{\pi k_i} \int\limits_0^{\pi/2} d\varphi \cos^2\varphi \int\limits_0^1 d\xi \frac{\left(\xi - \xi^3\right)\left(1 - \varepsilon_i\right)}{H_i^2(\xi,\varphi)} \{\ldots\}, \tag{3.22}$$

$\sigma_{0i}$ is the specific conductivity of a bulk single crystal.

If the thickness of the film layers is much larger than the mean free path of electrons, i.e., the inequality $d_i \gg l_i$ holds, the exponents in equation (3.19) are small and may be neglected. The angular integration gives the following expression for the conductivity of a DLF

$$\Phi_i = T(\beta_i) - \frac{3}{16k_i}\left\{(2 - q_i - P)\Gamma_{1,i} - Q\tau_{0j,i}\Gamma_{2,i}\right\}, \tag{3.23}$$

where [30]

$$\Gamma_{1,i} = 1 - \frac{32}{3\pi}\beta_i + 12\beta_i^2 + \frac{80}{\pi}\beta_i^3 - 40\beta_i^4 + \frac{16}{\pi}\beta_i^3(5\beta_i^2 - 4)I_i\,; \tag{3.24}$$

$$\Gamma_{2,i} = 1 - \frac{16}{3\pi\left(\beta_i - \beta_j\right)}\left\{\beta_i^2\left[1 - \frac{3\pi}{4}\beta_i - 3\beta_i^2 + \frac{3\pi}{2}\beta_i^3 - 3\beta_i^2(\beta_i^2 - 2)I_i\right] - \right.$$
$$\left. \beta_j^2\left[1 - \frac{3\pi}{4}\beta_j - 3\beta_j^2 + \frac{3\pi}{2}\beta_j^3 - 3\beta_j^2(\beta_j^2 - 2)I_j\right]\right\}; \tag{3.25}$$

$$I_i = \begin{cases} \dfrac{1}{\sqrt{1-\beta_i^2}}\ell n\dfrac{1 + \sqrt{1-\beta_i^2}}{\beta_i}, & \beta_i \le 1; \\[4mm] \dfrac{\arccos\left(\dfrac{1}{\beta_i}\right)}{\sqrt{\beta_i^2 - 1}}, & \beta_i > 1. \end{cases} \tag{3.26}$$

For small $(\beta_i \ll 1)$ and large $(\beta_i \gg 1)$ crystallites we obtain:

$$\Phi_i = 1 - \frac{3}{2}\beta_i - \frac{3}{16k_i}\left\{(2 - q_i - P)\left(1 - \frac{32}{3\pi}\beta_i\right) - Q\tau_{0j,i}\left(1 - \frac{16}{3\pi}\left(\beta_i + \beta_j\right)\right)\right\}, \quad \beta_i \ll 1; \tag{3.27}$$

$$\Phi_i = \frac{3}{4\beta_i}\left\{1 - \frac{1}{8k_i\beta_i}\left[\left(2 - q_i - P\right)\left(1 - \frac{512}{105\pi\beta_i}\right) - Q\tau_{0j,i}\frac{\beta_i}{\beta_j}\left(1 - \frac{256\left(\beta_i + \beta_j\right)}{105\,\beta_i\,\beta_j}\right)\right]\right\}, \quad \beta_i \gg 1. \tag{3.28}$$

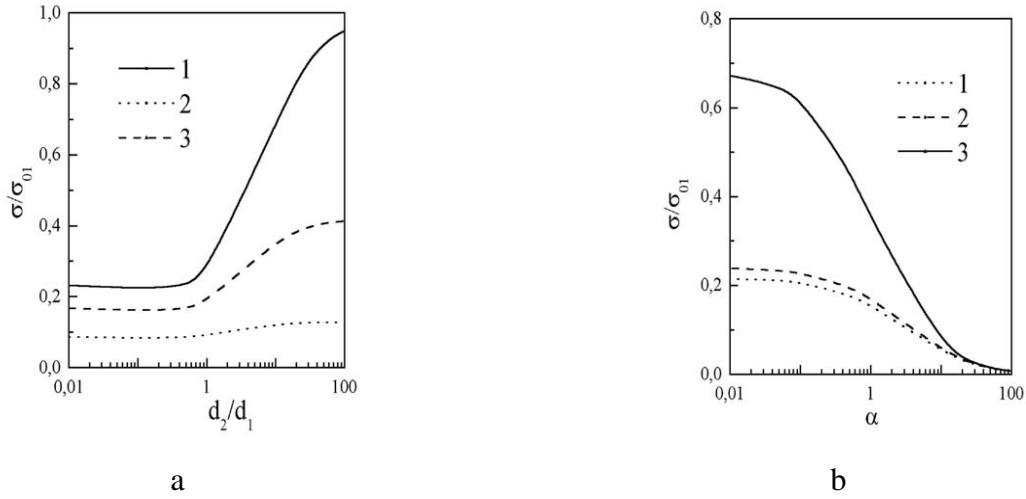

a                                          b

**Figure 5**  Dependence of electrical conductivity of double- layer polycrystalline film of the ratio $d_{2,1}$ for $\alpha_1 = \alpha_2 = \alpha$ and for parameter values: a) $q_1 = 0.1$, $q_2 = 0.2$, $P = 0.1$, $Q = 0.2$, $l_{2,1} = 1$, $k_1 = 0.1$, : $\alpha = 0.01$ (1), $\alpha = 5$ (2), $\alpha = 1$ (3);

b) $q_1 = 0.1$, $q_2 = 0.2$, $P = 0.1$, $Q = 0.1$, $l_1 / l_2 = 1$, $k_1 = 0.1$: $d_{2,1} = 10$ (1), $d_{2,1} = 1$ (2), $d_{2,1} = 0.1$ (3).

$$\Phi_i \approx \frac{3}{4}\frac{1 + q_i}{(1 - q_i P)(1 - q_j P) - q_i q_j Q^2}\cdot$$

$$\left((1 + P)(1 - q_j P) + q_j Q^2 + (1 + q_j)Q d_{j,i}\right)k_i\begin{cases}\ell n\dfrac{1}{k_i}, & \beta_i \le k_i \\[2mm] \ell n\dfrac{1}{k_i} - \dfrac{4}{\pi}\beta_i, & \beta_i > k_i \\[2mm] \ell n\dfrac{1}{k_i\beta_i}, & \beta_i \ll \dfrac{1}{k_i}\end{cases} \tag{3.29}$$

The curves shown in Figs. 5*a-b* were obtained by numerical calculation from formula (3.16), (3.19) and illustrate the dependence of the electrical conductivity of the DLF $\sigma$ on the size of one of the layers $d_2$ for different values of $P$, $Q$ and $\beta_i$ describing interaction of charge carriers with the interface and grain boundaries.

In the same way as we calculated the conductivity, other kinetic coefficients of DLF's and multilayers can be found [31,32].

# 4. GALVANOMAGNETIC EFFECTS IN DOUBLE LAYER FILMS

## 4.1 Sondheimer oscillations

The oscillatory dependence of the resistance of thin single-crystal conducting plates on the thickness $d$ and on the magnitude $H$ of a strong magnetic field was firstly predicted by Sondheimer [33]. These oscillations is due to electrons in the vicinity of the points of the Fermi surface, where the velocity is directed along a vector **H**, as well as to charge carriers with extreme displacement along the vector **H** during a period of rotation in the magnetic field [34]. For a long time it was considered that the oscillatory dependence takes place only in the presence of diffuse reflection of conduction electrons from the specimen boundaries. However, it was shown in Refs.[35, 36] that, generally speaking, the resistance of a plate with specular boundaries and a complex Fermi surface depends on the thickness of the conductor. In this case the resistance oscillations with the variation of $H$ or $d$ are due to the fact that the projection of the electron velocity on the direction of the electric current **j** is not conserved during specular reflection. The energy it acquires in the electric field $E$ is different from that of an electron, which does not interact with the specimen boundaries. This leads to the appearance of size effects.

In this chapter it is shown that the nonmonotonic dependence on the field $H$ of the resistance of DLS's is complex since along with the Sondheimer there may arise harmonics associated with the size of the layers $d_1$ and $d_2$ (Fig.6), and the investigation of the resistance provides information on the interaction of charge carriers with the interface [37,38].

Let us consider a DLF of thickness $d < l$ placed in a strong magnetic field ($r_N << d$, $r_N$ is the characteristic Larmor radius in the $N$-*th* layer, ($N = 1, 2$) lis the mean free path of electrons). At first, we shall assume that the magnetic field is perpendicular to the surface ($H = H_x$). Because the components $v_{Nz}, v_{Ny}$ of electron velocity in the plane perpendicular to the magnetic field are periodic functions of a time of electron motion along a ballistic trajectory, it can be expanded to a Fourier series, as well as nonequilibrium part of distribution function $\Psi$

$$v_{Nk}(t) = \sum_{n=-\infty}^{\infty} v_{Nk}^{(n)} e^{in\Omega_N t}; \quad \Psi_N = \sum_{n=-\infty}^{\infty} \Psi_N^{(n)} e^{in\Omega_N t} \quad ; \quad \left( N = 1,2; \ k = y, z \right) \qquad (4.1)$$

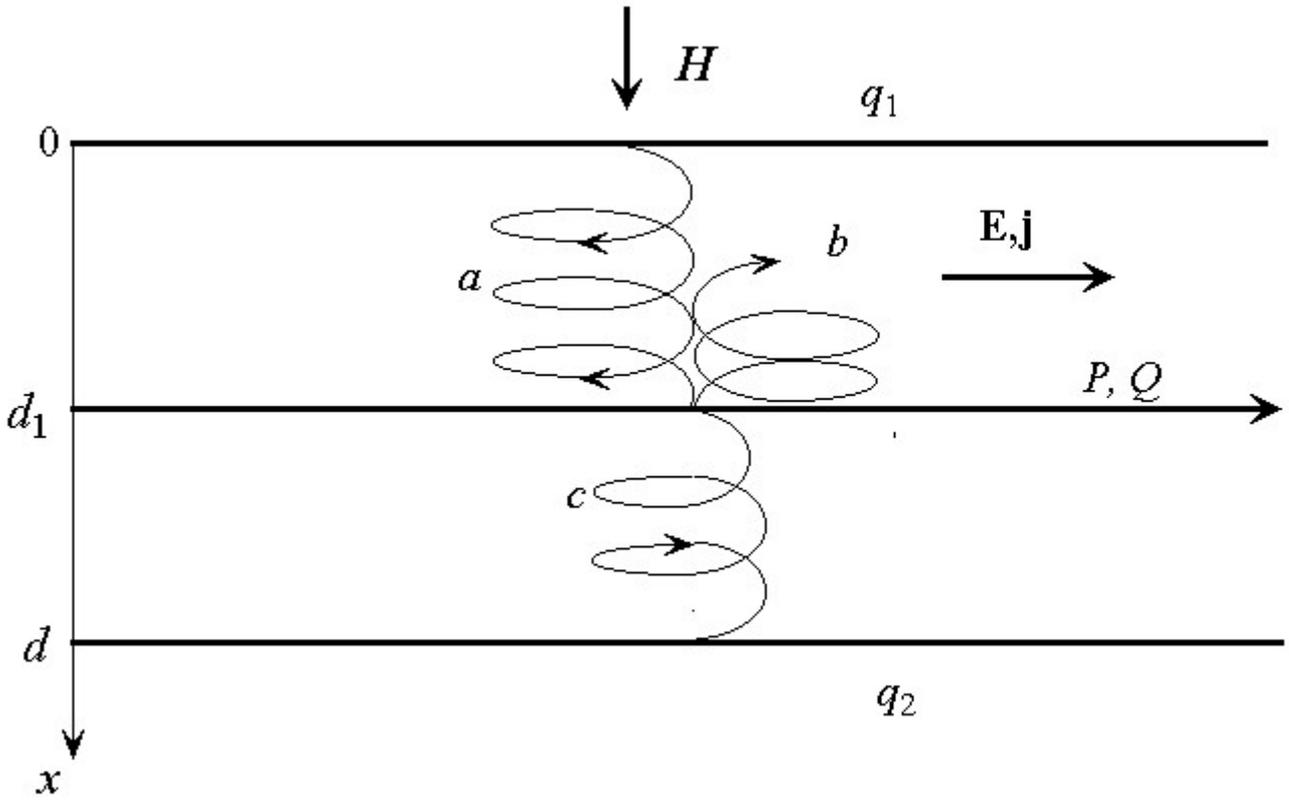

**Figure 6** Drawing of the observation of nonmonotonic dependence of the resistance of a DLF on the perpendicular field $H$. An electron incident on the interface (trajectory *a*) may be reflected (trajectory *b*) or may tunnel through it (trajectory *c*).

($\Omega_N$ is Larmor frequency in $N-th$ layer). The equation for $\Psi_N^{(n)}$ doesn't differ from analogous equation in zero magnetic field after formal replacement $\dfrac{1}{\tau} \rightarrow \dfrac{1}{\tau} + in\Omega$ and $\mathbf{v}_N \rightarrow \mathbf{v}_N^{(n)}$ (for simplicity we assume that a bulk relaxation in both layers can be described by the same mean free time $\tau$). In the case of spherical Fermi surface the projections of electron velocity are

$$v_{y,z}(t) = v_\perp \left( p_x \right) \begin{Bmatrix} \cos \Omega t \\ -\sin \Omega t \end{Bmatrix},$$

and we may search the solution of kinetic equation (2.1) in the form (4.1), where $n = \pm 1$ ($v_\perp = \sqrt{v_F^2 - v_x^2}$). The functions $\Psi_N^\pm$ can be found in a same way as it was done for DLF in the absence of magnetic field. Knowing the electron distribution function (see Eqs. (3.1), (3.2)), one may derive the total electric current (3.5) in the DLF.

We limit ourselves to the approximation of a uniform electric field in the DLF. After simple transformations, the conductivity tensor $\sigma_{ik}$ linking the current density with the electric field $E$ can be written in the form

$$\sigma_{yy} = \sigma_{zz} = \frac{\sigma_0}{1+\gamma^2}\left\{1 - \frac{l}{d}\left[\frac{1-\gamma^2}{1+\gamma^2}I_1 + \frac{2\gamma}{1+\gamma^2}I_2\right]\right\}; \tag{4.2}$$

$$\sigma_{yz} = -\sigma_{zy} = \frac{\sigma_0}{1+\gamma^2}\left\{\gamma - \frac{l}{d}\left[\frac{2\gamma}{1+\gamma^2}I_1 - \frac{1-\gamma^2}{1+\gamma^2}I_2\right]\right\}; \tag{4.3}$$

where

$$I_1 = \operatorname{Re} I; \quad I_2 = \operatorname{Im} I; \quad I = \frac{3}{4}\int_0^1 d\xi\frac{\xi\left(1-\xi^2\right)}{\Delta^2}\Phi\Delta^*;$$

$$\Phi = p_1 + p_2 + 2W\left(1 - p_1\beta_1 - p_2\beta_2\right) - 2p_1p_2Q\beta_1\beta_2 - W\left(1 - Q + P\right)\left(q_1\beta_1^2 + q_2\beta_2^2\right)$$

$$-P\left(1 - q_1q_2\right)\left(\beta_1^2 + \beta_2^2\right) - 2W(Q-P)\left(q_1p_2\beta_1^2\beta_2 + q_2p_1\beta_1\beta_2^2\right) -$$

$$2W(Q-P)q_1q_2\beta_1^2\beta_2^2 - \left(Q^2 - P^2\right)(q_1p_2 + q_2p_1)\beta_1^2\beta_2^2;$$

$$\Delta^* = 1 - P\left(q_1\beta_1^2 + q_2\beta_2^2\right) - q_1q_2\left(Q^2 - P^2\right)\beta_1^2\beta_2^2;$$

$$p_N = 1 - q_N; \quad W = 1 - P - Q;$$

$$\sigma_0 = \frac{8\pi}{3}\frac{e^2\tau p_F^3}{h^3m}; \quad l = \frac{p_F\tau}{m}; \quad \Omega = \frac{eH}{mc}; \quad \beta_N = \exp\left\{-\frac{d_N}{l\xi}\left(1 + i\gamma\right)\right\}; \quad \gamma = \Omega\tau.$$

The symbol "*" denotes the complex conjugate; $m$ is the electron effective mass.

The conductivity tensor can be presented in the form:

$$\sigma_{\alpha\beta} = \sum_{N,M=1}^{2}\sigma_{\alpha\beta}^{(N,M)}, \quad \alpha,\beta = y,z, \tag{4.4}$$

where $\sigma_{\alpha\beta}^{(N,N)}$ describes the contribution to the conductivity of electrons not leaving the $N$-th layer, and $\sigma_{\alpha\beta}^{(N,M)}$ is the contribution of the charge carriers tunneling across the interface.

Equations (4.2) - (4.3) solve, in principal, the problem completely, determining the dependence of the conductivity on the magnetic field and layer thickness. However, most informative is the nonmonotonic part $\sigma_{osc}$ of the total transverse conductivity. If $\exp\left(-\frac{d_N}{l}\right) \ll 1$ than in the oscillating part of the conductivity $\sigma_{osc}^{(N,M)}$ it is sufficient to retain only the first harmonic, and the components of the tensor $\sigma_{osc}^{(N,M)}$ will have the form:

$$\frac{\sigma_{osc}^{(N,M)}}{\sigma_{mon}} \approx \left(1-q_N\right)\left(1-P-Q\right)\frac{r^2}{dd_N}\exp\left(-\frac{d_N}{l}\right)\cos\frac{d_N}{r} \; ; \qquad (4.5)$$

$$\frac{\sigma_{osc}^{(N,M)}}{\sigma_{mon}} \approx Q\left(1-q_N\right)\left(1-q_M\right)\frac{r^2}{d^2}\exp\left(-\frac{d}{l}\right)\cos\frac{d}{r} \; , \qquad (4.6)$$

where $\sigma_{mon}$ is the monotonic part of the conductivity of the DLF.

It is not difficult to show that, in the absence of disorientation of the adjoining layers, the results remain qualitatively unchanged also for a nonisotropic Fermi surface, provided the film boundaries coincide with symmetry planes of the crystal. However, as already mentioned, in the monotonic part of the magnetoresistance there is a contribution due to electrons with extremal displacement along the magnetic field $H$ during a period.

As is known, the difference of the Fermi surfaces in the layers leads to a change in the velocity component normal to the interface of the tunneling electron. Therefore, the conductivity oscillations associated with the full thickness $d$ are mainly formed by the charge carriers for which the function

$$f(p_x) = \frac{d_1\Omega_1}{v_{x1}(p_x)} + \frac{d_2\Omega_2}{v_{x2}(p_{x2}(p_x))} , \qquad (4.7)$$

has an extremum $f = f^{extr}$ (the connection between $p_x$ and $p_{x2}$ is determined from the conservation of energy and of the tangential component of quasimomentum (2.6)). In the same way as the period, the amplitude of the oscillating term $\sigma_{osc}^{(N,M)}$ ($M \neq N$) depends on the disorientation of Fermi surfaces in the layers and is given in order of magnitude by

$$\frac{\sigma_{osc}^{(N,M)}}{\sigma_{mon}} \approx Q\left(\frac{r^*}{d}\right)^{3/2}\exp\left(-\frac{d_N}{l}\right)\cos\left(f^{extr}+\chi\right), \qquad (4.8)$$

where $r^* \approx r_1 \approx r_2$.

When the magnetic field is inclined to the normal of the specimen surface by an angle $\varphi$, the character of the oscillatory dependence of the conductivity on the magnetic field is not changed. In this case in expressions (4.5) and (4.6) the thickness $d_N$ must be replaced by $d_N^* = \frac{d_N}{\cos\varphi}$.

Thus, the analysis of the nonmonotonic dependence of the conductivity of DLF's on the magnetic field makes it possible to determine the degree of diffuseness of the interface and the probability of tunneling with conservation of the electron quasimomentum component tangent to it. Let us remark that the amplitude of the oscillations associated with the dimensions of the individual layer (4.5) does not depend on the ratio between the probabilities $P$ and $Q$ of reflection and transmission, respectively. There fore, the experimental investigation of $\sigma_{osc}(H)$ may prove to be the most convenient method of obtaining information on the diffuseness parameter for electron scattering from the interface.

*4.2 The static skin effect*

In a magnetic field $H$, which parallel to the surface (Fig.7), the electrical conductivity of the DLF's may be considerably higher than the electrical conductivity of monocrystalline plates [39]. This is due to the fact that under conditions of the static skin effect [40, 41], the electric current is concentrated not only near the surface of the sample but also near the interface. The static skin effect appears most clearly when the density of electrical current **j** and the vector **H** are perpendicular to each other and located in the plane of the interface. In this case, virtually the whole current flows along the interface, and the dependence of the resistance of the specimen on a strong magnetic field (the radius of curvature of the electron trajectory $r_i$ in *i-th* layer is much smaller than mean free path $l$ and the thickness of the layer $d_i$) is determined by the nature of the interaction of the charge carriers with the interface. An exception is a DLF with specular faces, whose electrical conductivity is contributed by electrons colliding with conductor surfaces $y = (-d_2, \ d_1)$ and with the interface $y = 0$.

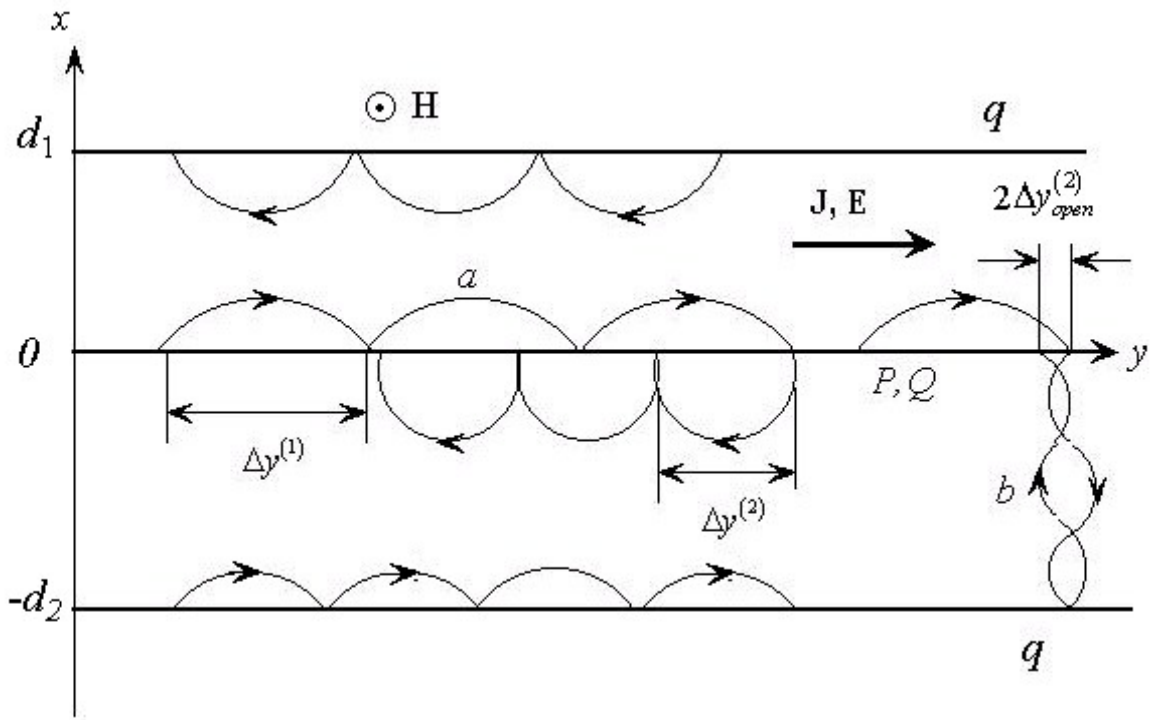

**Figure 7** Possible electron trajectories in a DLF in a parallel magnetic field.

To evaluate the electrical conductivity near the interface we use the solution (2.9) of the kinetic equation. From the calculations, we find for electrons that do not collide with the surface, but that do interact with the interface, the function of characteristics $F_k$ has the form

$$F_k\left(x - x^{(k)}(t)\right) = A_1\left\{\alpha_k \varphi_k\left[P\left(1 - P\alpha_i\right) + Q^2\alpha_i\right] + Q\alpha_i\varphi_i\right\}. \qquad i \neq k = 1, 2. \qquad (4.9)$$

Here

$$A_1^{-1} = (1 - P\alpha_1)(1 - P\alpha_2) - Q^2\alpha_1\alpha_2;$$

$$\varphi_i = \int_{\lambda_i}^{\lambda'_i} dt' \exp\left(-\frac{1}{\tau}(\lambda'_i - t')\right)\mathbf{v}(t')\mathbf{E}; \quad \alpha_i = \exp\left[-\frac{1}{\tau}(\lambda'_i - \lambda_i)\right],$$

$\lambda_i$ and $\lambda'_i$ are two successive instants of electron collision with interface.

In the basic approximation in the small parameter $r/l$ ($r = \max(r_1, r_2)$) the transverse electrical conductivity near the specular interface takes the form:

$$\sigma_{yy}^{(n)} = \left\langle \frac{v_{yn}\left\{\left[(1 - P\xi_m)P + Q^2\xi_m\right]\Delta y^{(m)} + Q\Delta y^{(n)}\right\}}{(1 - P\xi_1)(1 - P\xi_2) - Q^2\xi_1\xi_2}\right\rangle, \qquad (4.10)$$

where $1 - P - Q \ll r/l$; $v_{yn}$ is the electron velocity along the current direction, $\Delta y^{(n)} = y^{(n)}\left(\lambda'_n\right) - y^{(n)}\left(\lambda_n\right)$ is its displacement during the time between two collisions with the interface $T_n = \lambda'_n - \lambda_n$ and $\xi_n = \exp(-T_n/\tau)$. The indexes $m$ and $n$ indicate the number of the layer. The angular brackets denote integration over the Fermi surface (see Eq. (2.4)).

When $Q \gg r/l$, the electrons performing a periodic motion along one of the sides of the interface make an independent contribution to the total conductivity of a DLF and the transverse electrical conductivity of the layer near the boundary with a thickness of $4r_H = 2(r_1 + r_2)$ is of the order of $\sigma_0$ the conductivity of an bulk metal in the absence of a magnetic field. The specific electrical conductivity the skin layer near the interface

$$\sigma_\perp^{bound} = \frac{1}{d_1}\int_0^{2r_1}\sigma_{yy}^{(1)}(x)dx + \frac{1}{d_2}\int_{-2r_2}^0\sigma_{yy}^{(2)}(x)dx \approx \sigma_0\left\{\frac{r_1}{d_1} + \frac{r_2}{d_2}\right\}, \qquad (4.11)$$

considerably exceeds not only the conductivity of the conductor core but also the contribution to the total transverse electrical conductivity $\sigma_\perp$ of the electrons interacting with the diffuse surface of the sample. In a crystal with specular surfaces, the surface parts $\sigma_\perp^{bound} \approx \sigma_\perp^{surface}$ and $\sigma_\perp$ nevertheless exceeds the conductivity of a plate, which has no interface.

If $Q \gg r/l$ i.e., the electron is able during the mean free time $\tau$ to tunnel through the interface, then in the case of different Fermi surfaces in the layers the electron motion occurs along an open trajectory (trajectory a in Fig.7). The conductivity of the boundary layer, as follows from (4.10), does not depend on the probability of tunneling and, as before, remains high:

$$\sigma_\perp^{bound} \approx \sigma_0 \frac{\overline{(\Delta y^{(1)} + \Delta y^{(2)})^2}}{(r_1 + r_2) \; d}. \qquad (4.12)$$

Here, the averaging has been performed over all possible electron trajectories. It is easy to see that in the case of identical Fermi surfaces $\Delta y^{(1)} = \Delta y^{(2)}$ in Eq.(4.12) and it is necessary to take account of the following term in the expansion in the parameter $r/l$.

If for a specified orientation of the field $H$ closed and open cross-sections of the Fermi surface are possible, then during tunneling the electron can go over on to a closed orbit and move deep into the layer. In thick layers ($d_2 \ll l$) such an electron, which has undergone bulk collisions, does not again return to the interface. In this case, the electrical conductivity of the skin layer is described by Eq.(4.10), in which the second term in braces in the numerator should be dropped and we should take $\xi_2 = 0$. As a result, we find that transverse electrical conductivity is equal to

$$\sigma_\perp^{bound} \approx \sigma_0 \frac{r_1^2}{ld} \frac{P}{Q} \delta + \sigma_0 \frac{\overline{(\Delta y^{(1)} + \Delta y^{(2)})^2}}{(r_1 + r_2) \; d}(1 - \delta); \qquad (4.13)$$

and depends significantly on the probability $P$. In Eq.(4.13) $\delta$ is the width of the layer of open orbits on the Fermi surface divided by the Fermi momentum. In a thin layer ($d_2 \ll l$) an electron moving along an open orbit interacts with both faces of the layer. In the case of specular reflection of electrons from both faces, the electrons return to the interface (trajectory $b$ in Fig.7). If $Q \gg d/l$, $(1 - q) \ll 1$, electrons are again able to tunnel through it leading to an increase in the electrical conductivity of the boundary layer:

$$\sigma_\perp^{bound} \approx (1 - \delta)\sigma_0 \frac{\overline{(\Delta y^{(1)} + \Delta y^{(2)})^2}}{r_1 d} + \delta\sigma_0 \frac{\overline{(\Delta y^{(1)} + 2\Delta y_{open}^{(2)})^2}}{d_2 d}. \qquad (4.14)$$

Here $2\Delta y_{open}^{(2)}$ is the electron displacement along the $y$-axis during the time between two collisions with the interface. Thus, the presence of interface improves the conducting properties of metal sample in a strong magnetic field. Experimental study of conductivity of metallic DLF's as a function of the magnitude and direction of $H$ provides information on the nature of the scattering of conduction electrons at the interface and on the probability of penetration through it. This effect was observed experiment in Refs. [42,43].

# 5. HIGH FREQUENCY PHENOMENA IN METALLIC DOUBLE LAYER FILMS.

Under conditions of the anomalous skin effect, when the depth of the skin layer $\delta$ is much smaller than the characteristic radius of curvature of the trajectory $r \ll l$ in a magnetic field $\mathbf{H,}$ parallel to the surface of the specimen, the high-frequency (HF) field penetrates into the metal in the form of narrow spikes [44-46]. The position of the HF field spikes is determined by the extremal diameters of the electron orbit, which permits using this effect to study the dispersion law of conduction electrons. The nature of the penetration of the HF field into the metal also depends on the state of the conductor surface. In specimens with a perfect boundary, the field spikes have a much higher intensity over a wide range of frequencies, including the radio-frequency range. If the reflection of the charge carriers by the specimen surface is multichannel reflection, while the spikes are formed by all electrons for which umklapp processes are possible [47,48].

Observation of the radio-frequency size effect in bicrystalline aluminum plates [49], having a twin boundary parallel to the outer surface of the specimen, has shown that this effect is also useful for studying the nature of the interaction of charge carriers with intercrystallite boundaries and interfaces. In particular, the results of Ref. 49 indicate the high probability for conservation of the tangential component, relative to the intercrystallite boundary, of the quasimomentum of the electron interacting with it.

In this chapter, we examine the anomalous penetration of an electromagnetic wave into a metallic double-layer film along a chain in electronic trajectories [50] (Fig. 8). If the maximum size of an electron trajectory $D_1$, in the direction normal to the surface, is larger than the thickness of the layer $d$ ($d >> \delta$, $D_1 - d >> \delta$), then the amplitude and position of spikes are determined by the smoothness of the boundary and by Fermi surfaces of the metals. Already when the inequality $Q >> r/l$ is satisfied, the electrons over the time $\tau$ tunnel from the upper to lower layer. In this case, the electron efficiently absorbs energy from HF field in the skin layer, moving along arcs $a$, and creates a field spike while moving along the segments $b$ of its trajectory (Fig. 8). If the reflection at the separation boundary occurs with small diffuseness, then the average energy contributed by a single electron to the spike is of the same order of magnitude, as for $Q \approx 1$ (in the leading approximation with respect to the parameter $r/Q\ l$), while its amplitude is comparable to the amplitude of the electromagnetic field penetrating into the single crystal. On the other hand for low probabilities $Q << r/l$, the amplitude of the HF field in a spike is small and, generally speaking, proportional to $Q$.

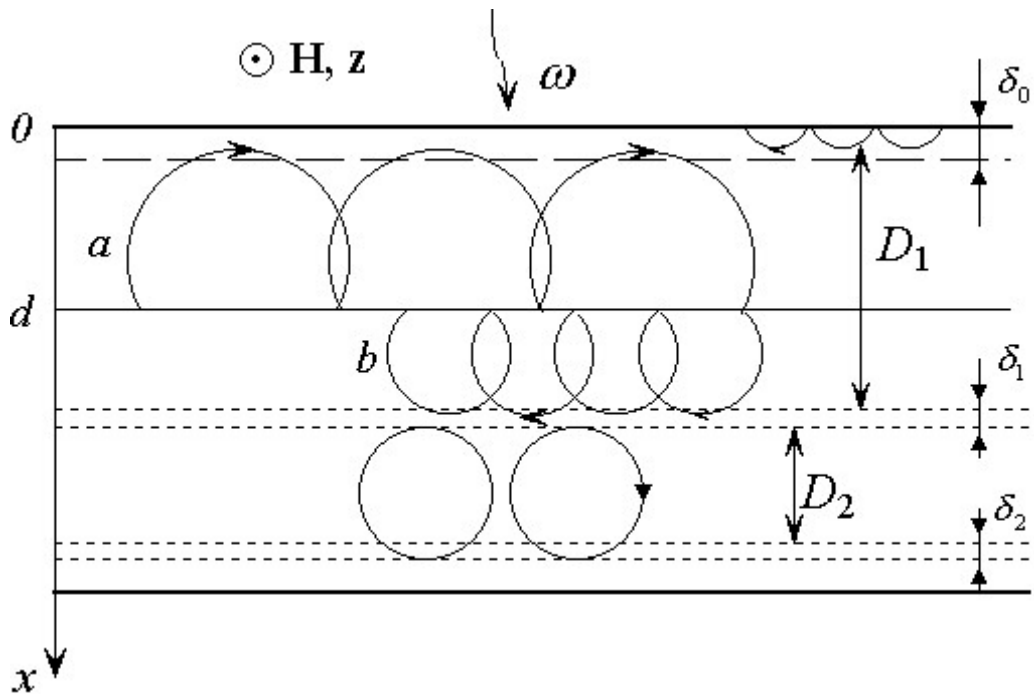

**Figure 8** Schematic drawing of experiment by observation of resonance and RF size effects in a DLF.

Resonant HF effects of the cyclotron resonance type are more sensitive to the magnitude of $Q$ [51]. If the boundary is almost transparent to conduction electrons, $(1 - Q \ll 1)$, then the resonant frequencies are related to the period of motion of the charge carriers along a trajectory intersecting the interface, while the quantity $1 - Q$ determines the broadening of the resonance lines. For an almost nontransparent separation boundary, $Q << 1$, reflecting charge carriers in an almost specular manner, a different type of resonance, related to the periodic motion of electrons returning into the skin layer due to collisions with the interface, is possible. This effect is recall of the cyclotron resonance in thin conductors [52,53] and the size of the "upper" layer plays the role of the specimen thickness. The width of the resonance lines permits judging the tunneling probabilities and the diffuse scattering on reflection.

Thus, the experimental study of HF effects of a DLF will make it possible to obtain detailed information on the interaction of conduction electrons with interfaces.

### 5.1 Spikes of high frequency electromagnetic field

Let us examine a situation, in which the interface is parallel to the surface of the specimen, while the magnetic field is parallel to it (Fig.9). If the maximum size of the electron trajectory along the $x$ axis is greater than the thickness of the upper layer $d$, then the function $F$ in a solution of kinetic equation (2.9) has six values, corresponding to motion along one of the trajectory segments illustrated in Fig.9. In this case, the boundary conditions (2.14), (2.15) lead to a system of linear equations, which can be solved exactly.

From the calculations, we find that for electrons that do not collide with the surface, but that do interact with the interface, the function of characteristics has the form

$$F_k\left(x - x^{(k)}(t)\right) = A_1\left\{\alpha_k\varphi_k\left[P(1 - P\alpha_i) + Q^2\alpha_i\right] + Q\alpha_i\varphi_i\right\}. \quad i \neq k. \tag{5.1}$$

If, on the other hand, the charge carrier also collides with the outer boundary of the crystal, then $F$ assumes three values:

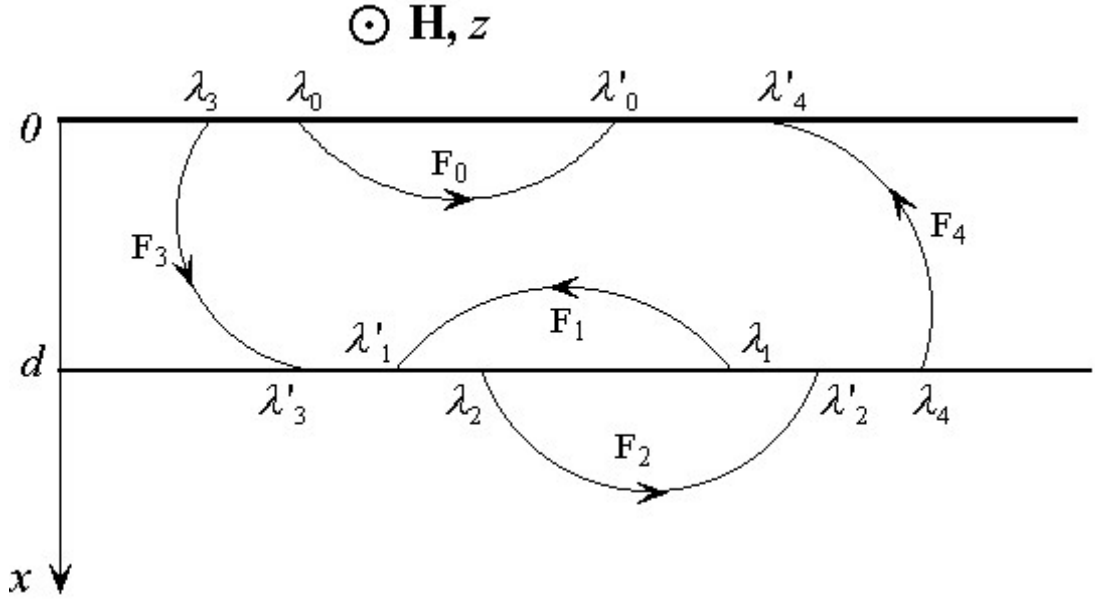

**Figure 9.** Possible segments of electron trajectories in a DLF in parallel magnetic field.

$$F_2\left(x - x^{(2)}(t)\right) = A_2\left\{Q\alpha_3\varphi_3 + P(1 - q_1P\alpha_3\alpha_4)\alpha_2\varphi_2 + q_1Q\alpha_3\alpha_4\varphi_4\right\};$$

$$F_3\left(x - x^{(1)}(t)\right) = q_1A_2\alpha_4\left\{\alpha_3\varphi_3\left[P(1 - P\alpha_2) + Q^2\alpha_2\right] + Q\alpha_2\varphi_2 + \varphi_4(1 - P\alpha_2)\right\};$$

$$F_4\left(x - x^{(1)}(t)\right) = A_2\left\{\alpha_3(\varphi_3 + q_1\alpha_4\varphi_4)\left[P(1 - P\alpha_2) + \alpha_2Q^2\right] + Q\alpha_2\varphi_2\right\}. \tag{5.2}$$

Here

$$A_1^{-1} = (1 - P\alpha_1)(1 - P\alpha_2) - Q^2\alpha_1\alpha_2;$$

$$A_2^{-1} = (1 - P\alpha_2)(1 - q_1P\alpha_3\alpha_4) - q_1Q^2\alpha_2\alpha_3\alpha_4;$$

$$\alpha_i = \exp\left[i\omega^*(\lambda'_i - \lambda_i)\right];$$

$$\varphi_i = \int_{\lambda_i}^{\lambda'_i} dt'\alpha(\lambda_i - t')g(x_s + x^{(1,2)}(t') - x^{(1,2)}(\lambda'_i)); \quad x_s = 0, d.$$

$$\alpha(t) = \exp\{i\omega^*t\}. \tag{5.3}$$

In Eqs. (5.1) and (5.2), we dropped terms related to the change in the chemical potential of the electrons as a result of collisions with the surface or the interface. This is related to the fact that under conditions of the anomalous skin effect, the normal components of the field and the current are small, and in the leading approximation with respect to the parameter $\delta/r \ll 1$, the current $j_x$ at any depth $x$ can be neglected.

The function $F_0$ [41], corresponds to the charge carriers interacting only with the surface of the conductor, is

$$F_0 = \frac{q_1 \varphi_1}{1 - q_1 \varphi_1}.$$  (5.4)

Maxwell equation for the Fourler component of the HF electric field and current

$$\mathrm{E}_\mu(k) = 2 \int_0^\infty E_\mu(x) \cos kx \ \ dx;$$  (5.5)

$$j_\mu(k) = 2 \int_0^\infty j_\mu(x) \cos kx \ \ dx;$$  (5.6)

has the following form:

$$k^2 \mathrm{E}_\mu(k) + 2 \frac{\partial E_\mu(0)}{\partial x} = \frac{4\pi i \omega}{c^2} j_\mu(k).$$  (5.7)

(We made use of the even continuation of the functions $E_\mu(x)$ and $j_\mu(x)$ into the region outside the metal). The HF electrical conductivity tensor, which is the kernel of the integral operator that couples $\mathrm{E}_\mu(k)$ and $j_\mu(k)$,

$$j_\mu(k) = \int_0^\infty K_{\mu\nu}(k, k') \mathrm{E}_\nu(k') dk', \qquad \mu, \nu = y, z,$$  (5.8)

can be found, if Eq.(2.1) is used for the nonequilibrium correction to the electron distribution function. Under conditions of the anomalous skin effect, when the depth of the skin layer $\delta$ is the smallest parameter with the dimension of length, large $k \approx \delta^{-1}$ are most important, and in order to find the surface impedance and HF field distribution in the specimen, it is sufficient to know the asymptotic expression for $K_{\mu\nu}$, when $kr \gg 1$, $k'r \gg 1$. In these case, we will assume that the flight time of the electron through a thin skin layer is negligibly small compared to its effective transit time $1/|\omega^*|$ i.e., the following inequality is satisfied:

$$\left|\omega^*/\Omega\right| \sqrt{\frac{\delta}{r}} \ll 1,$$  (5.9)

($\Omega$ is the characteristic frequency for a motion of the electron in the field $H$). Then, the HF electrical conductivity can be represented as a sum of several terms:

$$K_{\mu\nu}(k,k') = K_{11}^{(0)}(k,k') + K_{11}^{(1)}(k,k') + K_{22}^{(1)}(k,k') + K_{12}(k,k') + K_{22}(k,k'). \qquad (5.10)$$

Here and in what follows, we will assume that the electric field vector of the linearly polarized wave is directed along one of the axes in the coordinate system $y$ axes in which the kernel $K_{\mu\nu}(k,k')$ is diagonal, and in what follows, in order to simplify equations, we will not write the tensor indices. The kernel $K_{11}^{(0)}(k,k')$ is related to electrons skimming along the surface of the specimen and not leaving the skin layer over the mean free time $\tau$. It is these particular electrons that make the determining contribution to the formation of the HF screening current [54]. As it was shown by Falkovskii [55] the scattering of such charge carriers by a slightly rough surface is nearly specular and can be described with the help of the specularity parameter, which depends linearly on the angle $\varphi$ of incidence of the electron on the surface: $q_i = 1 - q_i'(0) \ \varphi$. When the operator $K_{11}^{(0)}$ operates on a smooth function of $k$, it can be represented in the following form:

$$K_{11}^{(0)}(k,k') = \frac{c^2}{4\pi\omega} \frac{k_0^{5/2}}{\sqrt{k \ k'}} \left[ \frac{1}{|k-k'|^{1/2}} - \frac{1}{|k+k'|^{1/2}} \right]. \qquad (5.11)$$

The dependence of $K_{11}^{(0)}$ on $k$ and $k'$ is the same as for an ideal surface, while its small roughness manifests itself in a change in the quantity $k_0$, into which the coefficient $q_i'(0)$ enters:

$$k_0^{5/2} = \frac{8\pi^{3/2}\omega \ e^3 H}{(ch)^3 \ \left(q_1'(0)\Omega_{11} - 2i\omega^*\right)} \int dp_z v_{y1}\left(s_2^{(1)}\right) \ v_{y1}\left(s_2^{(1)}\right) \left|v'_{x1}\left(s_2^{(1)}\right)\right|^{-1/2}; \qquad (5.12)$$

$\Omega_{11}$ is the Larmor frequency of an electron in the upper layer; $s_k^{(i)}$ are the points of the stationary phase on the trajectory of the electron in the $i$-$th$ layer $\left(v_{xi}\left(s_k^{(i)}\right) = 0, \ v'_{xi}\left(s_1^{(i)}\right) > 0, \ v'_{xi}\left(s_2^{(i)}\right) < 0\right); \ p_z$ is the projection of the momentum along the magnetic field direction.

The terms $K_{11}^{(1)}(k,k')$ and $K_{22}^{(1)}(k,k')$ in the electrical conductivity stem from the charge carriers, moving in the upper or lower layer and absorbing energy from the HF field along these same segments of the trajectory:

$$K_{11}^{(1)}(k,k') = \int dp_z \theta(D_1 - d) \ \rho_{11}\left(s_1^{(1)}, s_1^{(1)}\right) \left\{\frac{1}{k} \hat{S}_1\left(s_1^{(1)}, \tau_1; x^{(1)}\left(s_1^{(1)}\right)\right) f_{11}\left(\tau_1\right) -$$

$$\frac{2}{\pi^2} f_{11}(\tau_1) \frac{\ln(k/k')}{k^2 - k'^2} \bigg\};$$ (5.13)

$$K_{22}^{(1)}(k,k') = \int dp_z \theta(D_1 - d) \rho_{22}\left(s_2^{(2)}, s_2^{(2)}\right) \frac{1}{k'} \hat{S}_2\left(s_1^{(2)}, s_2^{(2)}, x^{(2)}\left(s_2^{(2)}\right)\right) f_{22},$$ (5.14)

where

$$S_i(a,b,x) f = \frac{1}{\pi} \int_a^b d\lambda \; \frac{f'(\lambda)}{k-k'} \sin\left[(k-k') \; \left(d - x^{(i)}(\lambda) + x\right)\right];$$ (5.15)

$$\rho_{ik}\left(s_\alpha^{(i)}; \; s_\beta^{(k)}\right) = \frac{2\pi e^3 H}{c h^3} \frac{v_{yi}\left(s_\alpha^{(i)}\right) \; v_{yk}\left(s_\beta^{(k)}\right)}{\left|v'_{xi}\left(s_\alpha^{(i)}\right) \; v'_{xk}\left(s_\beta^{(k)}\right)\right|^{1/2}};$$ (5.16)

$$f_{ii}(t_i') = A_1\left[(1+P\alpha_i)(1-P\alpha_k) + Q^2\alpha_1\alpha_2\right], \; if \; t_i \le \tau_i, \; (i \ne k);$$ (5.17)

$$f_{22}(t_2') = A_2\left[(1+P\alpha_2) \; (1-q_1 P\alpha_3\alpha_4) + q_1 Q^2\alpha_2\alpha_3\alpha_4\right], \; if \; t_2 > \tau_2;$$ (5.18)

$$D_1 = x^{(2)}\left(s_2^{(2)}\right) - x^{(2)}(\tau_2) + x^{(1)}(\tau_1) - x^{(1)}\left(s_1^{(1)}\right),$$ (5.19)

is the size of the electron orbit along the $x$ axis; $\tau_1$ is a root of the equation

$$x^{(1)}(\tau_1) - x^{(1)}\left(s_1^{(1)}\right) = d,$$ (5.20)

$\tau_2$ is related to $\tau_1$ by the conditions (2.6); $\theta(x)$ is the Heaviside function (the unitstep function).

The kernel $K_{12}(k,k')$ is formed by electrons crossing the interface and creating the first spike at a depth $D_1$ in the bulk of the conductor:

$$K_{12}(k,k') = 2\sqrt{\frac{2\pi}{\beta_1}} \rho_{12}\left(s_1^{(1)}, s_2^{(2)}\right)\left\{\frac{1}{\sqrt{kk'}} \hat{C}(D_1) f_{12} - \frac{f_{12}(\tau_1)}{k+k'} R(0, D_1)\right\}\bigg|_{p_z = p_i'};$$ (5.21)

$$R(a, D_i) = \frac{1}{\pi\sqrt{kD_i}} \frac{1}{k-k'}\left\{\cos\left[(k-k')a - k'D_i - \frac{\pi}{4}\gamma_i\right] - \right.$$

$$\sqrt{\frac{k}{k'}}\cos\left[\left(k-k'\right)a+kD_i+\frac{\pi}{4}\gamma_i\right]\right\};\qquad(5.22)$$

$$\hat{C}\left(D_1\right)f_{12}=\frac{1}{\pi}\int\limits_0^{\tau_1}d\lambda\, f_{12}''\quad R\left(d-x^{(1)}\left(\lambda\right)+x^{(1)}\left(s_1^{(1)}\right),D_1\right);\qquad(5.23)$$

$$f_{12}\left(t_1'\right)=A_1Q\left(e^{i\omega^*s_2^{(2)}}+\alpha_1\alpha_2e^{-i\omega^*s_2^{(2)}}\right);\qquad(5.24)$$

$$\beta_i=\left|\frac{1}{D_i}\frac{\partial^2 D_i}{\partial p_z^2}\right|;\quad\gamma_i=sign\frac{\partial^2 D_i}{\partial p_z^2}.\qquad(5.25)$$

All quantities in Eq. (5.21) are evaluated at $p_z=p_1^e$, for which $\partial D_1/\partial p_z\big|_{p_z=p_1^e}=0$.

In the high-frequency range $\left(\omega\geq\Omega\right)$, Eqs. (5.13), (5.14) and (5.21) are valid, while for small displacement of the electron orbit by an amount of the order of $\delta$, the relative change in the period of the motion is negligibly small compared to the broadening of the resonance line. In this case, the dependence of the characteristic frequency $\Omega$ on the time of the collision $\lambda'_i$ with the interface in the functions $f_{ik}\left(\lambda'_i\right)$ is practically unimportant, and they can be taken out from under the integral sign at the points of stationary phase. For electron trajectories that slightly touch the surface of the specimen, whose contribution is taken into account in expressions (5.13), (5.14), and (5.21), the inequality sought has the form

$$\left(r/\delta\right)^{1/2}>>\omega\tau.\qquad(5.26)$$

However, the qualitative results obtained below remain valid also for $r/\delta>>\omega\tau$ when the contribution of "tangent" electrons to the current can be neglected and the condition (5.26) indicated above is satisfied only for charge carriers interacting only with the interface. For $\omega\tau>>r/\delta$, averaging the functions $f_{ik}$ over the exit angles from the surface $x=d$ leads to appreciable attenuation of the resonance amplitude and spike intensity at the depth $D_1$, in contrast to the situation occurring in a single crystal [45].

The last term in Eq. (5.10) $K_{22}\left(k,k'\right)$ describes the contribution of electrons moving only in the lower layer, and further extending the electromagnetic field spike into the bulk of the specimen:

$$K_{22}\left(k,k'\right)=\frac{1}{\sqrt{kk'}}\int dp_z\left[\rho_{22}\left(s_1^{(2)},s_1^{(2)}\right)\, B(d)+\rho_{22}\left(s_2^{(2)},s_2^{(2)}\right)\, B\left(d+D_2\right)\right]+$$

$$+2\sqrt{\frac{2\pi}{\beta_2}}\frac{\rho_{22}\left(s_1^{(2)},s_2^{(2)}\right)}{\sqrt{kk'}\sinh\left(-i\omega^*T_{22}/2\right)}\left\{\frac{2}{\sqrt{kD_2}}\sin\left(kD_2+\frac{\pi}{4}\gamma_2\right)\delta\left(k-k'\right)-R\left(d,D_2\right)\right\}\bigg|_{p_z=p_2^e},\qquad(5.27)$$

where

$$B(a) = \cot\left(-i\omega^* T_{22}/2\right)\left[\delta(k-k') - \frac{1}{\pi}\frac{\sin(k-k')a}{k-k'}\right];\tag{5.28}$$

$D_2 = x^{(2)}\left(s_2^{(2)}\right) - x^{(2)}\left(s_1^{(2)}\right)$; $T_{22}$ is the period of electron motion in the second layer in the field $\mathbf{H}$. The functions of $p_z$ in the second term of Eq. (5.27) are evaluated on the sections of the Fermi surface for which $\partial D_2/\partial p_z\big|_{p_z=p_z^c}=0$.

Starting from the structure of the kernel $K(k,k')$, it is natural to look for a solution of Eq, (5.7) in the form of a sum:

$$\mathrm{E}(k) = \mathrm{E}_0(k) + \mathrm{E}_1(k) + \mathrm{E}_2(k); \qquad \mathrm{E}_0(k) = \mathrm{E}_0^*(k) + \Delta\mathrm{E}_0(k),\tag{5.29}$$

where $\mathrm{E}_0^*(k)$ is the Fourier component describing the field in the main skin layer, formed by skimming electrons; $\Delta\mathrm{E}_0(k)$ is a small correction to $\mathrm{E}_0^*$, related to charge carriers that can absorb the energy in the electromagnetic wave in a resonant manner. The function $\mathrm{E}_1(k)$ is responsible for the formation of the HF field spike at a depth $D_1$, while the Fourier component $\mathrm{E}_2(k)$ describes the penetration of the field to a depth larger than $D_1$. Substituting (5.29) into Maxwell equation (5.7) leads to the appearance of characteristic integrals on the right side of the equation that are easy to estimate asymptotically for $kr \gg 1$ and $k'r \gg 1$. Thus, for example,

$$\int_0^\infty dk' \hat{S}_1 f_{11} \mathrm{E}_i(k') \approx f_{11}(\tau_1)\mathrm{E}_i(k)\delta_{0i};\tag{5.30}$$

$$\int_0^\infty \frac{dk'}{k'}\hat{S}_2 f_{22}\mathrm{E}_i(k') \approx \frac{1}{k}f_{22}(\tau_2)\mathrm{E}_i(k)\delta_{1i};\tag{5.31}$$

$$\frac{4\pi\omega}{c^2}\int_0^\infty dk' K_{12}(k,k')\mathrm{E}_i(k') \approx \frac{\chi_1^3}{2k\sqrt{kD_1}}\Big\{\mathrm{E}_i(k)\sin\left(kD_1+\frac{\pi}{4}\gamma_1\right) -$$
$$\hat{G}\mathrm{E}_i(k)\cos\left(kD_1+\frac{\pi}{4}\gamma_1\right)\Big\}\delta_{0i},\tag{5.32}$$

where

$$\chi_1^3 = \frac{8\pi\omega}{c^2}\rho_{12}\left(s_1^{(1)}, s_2^{(2)}\right)\; f_{12}(\tau_1)\;\sqrt{\frac{2\pi}{\beta_1}}\Big|_{p_z=p_z^c};\tag{5.33}$$

$$\hat{G}\mathrm{E}_i(k) = \frac{2}{\pi}\int_0^\infty \frac{\mathrm{E}_i(xk)\,dx}{x^2-1}; \qquad \delta_{ik} = \begin{cases}1, & i=k,\\ 0, & i\neq k.\end{cases}\tag{5.34}$$

The operators $B(a)$ (5.28) and $R(a,D_i)$ (5.22) give a nonvanishing result only when operating on the term $\mathrm{E}_i(k)$, corresponding to the HF field at a depth larger than $a$.

We solve Maxwell equation (5.7) using perturbation theory, making use of the smallness of the "spike" terms compared to the amplitude of the field in the main skin layer $E_0^*(k)$ [56]:

$$E_0^*(k) = -\frac{\partial E(0)}{\partial x}\frac{1}{i\pi k_0^2}\int_{c-i\infty}^{c+i\infty}dz M(z)\left(\frac{k}{k_0}\right)^z; \quad -2 < c \leq 0; \quad (5.35)$$

where

$$M(z) = \Gamma^{-1}\left(\frac{7}{5}\right)\left(\frac{4}{25\sqrt{2\pi}}\right)^{\frac{2(z+2)}{5}}\exp\left(\frac{i\pi}{5}(z+2)\right)\cos\frac{\pi z}{2}\Gamma\left[\frac{1-2z}{3}\right]\Gamma\left[\frac{3-2z}{5}\right]\Gamma[z+1]. \quad (5.36)$$

The HF electric field distribution near the spike at the depth $D_1$ is:

$$E_1(x) = -\frac{\partial E(0)}{\partial x}\frac{i}{2\pi\sqrt{k_0 D_1}}\frac{\chi_1^3}{k_0^{3/2}}\int_0^\infty\frac{dk}{\sqrt{k(k^3-ik_1^3)}}\left\{F_0\left(\frac{k}{k_0}\right)\sin\left[k(D_1-x)+\frac{\pi}{4}\gamma_1\right]-\right.$$

$$\left.\hat{G}F_0\left(\frac{k}{k_0}\right)\cos\left[k(D_1-x)+\frac{\pi}{4}\gamma_1\right]\right\}, \quad (5.37)$$

where

$$F_0(k/k_0) = -k_0^2\left(\frac{2\partial E(0)}{\partial x}\right)^{-1}E_0^*(k); \quad (5.38)$$

$$\chi_1^3 = \frac{8\pi\omega}{c^2}\left|\frac{1}{2\pi D_1}\frac{\partial^2 D_1}{\partial p_z^2}\right|^{-1/2}\rho_{12}\left(s_1^{(1)}, s_2^{(2)}\right)\cdot$$

$$\frac{Q\left(1+e^{i\omega^* T_{12}}\right)}{\left(1-Pe^{i\omega^* T_1}\right)\left(1-Pe^{i\omega^* T_2}\right)-Q^2 e^{i\omega^* T_{12}}}; \quad (5.39)$$

$$k_1^3 = \frac{4\pi\omega}{c^2}\int dp_z\left[\left[\theta(D_1-d)\rho_{22}\left(s_2^{(2)}, s_2^{(2)}\right)\frac{\left(1+Pe^{i\omega^* T_2}\right)\left(1-Pe^{i\omega^* T_1}\right)+Q^2 e^{i\omega^* T_{12}}}{\left(1-Pe^{i\omega^* T_1}\right)\left(1-Pe^{i\omega^* T_2}\right)-Q^2 e^{i\omega^* T_{12}}}+\right.\right.$$

$$\left.\left.+\rho_{22}\left(s_1^{(2)}, s_1^{(2)}\right)\cot\left(-i\omega^*\frac{T_{22}}{2}\right)\right]\right]; \quad T_{12} = T_1 + T_2;$$

$$\rho_{MN}\left(s_i^{(M)}, s_k^{(N)}\right) = \frac{2\pi e^3 H}{ch^3}\frac{v_{M\mu}\left(s_i^{(M)}\right)v_{N\mu}\left(s_k^{(N)}\right)}{\left|v'_{Mx}\left(s_i^{(M)}\right)v'_{Nx}\left(s_k^{(N)}\right)\right|^{1/2}}. \quad (5.40)$$

Hear $T_1$ and $T_2$ are the times for electron motion along arc $a$ and $b$ (Fig.6). All quantities in Eq.(5.37) are evaluated at $p_z = p_{ze}$, for which $\left(\frac{\partial D_1}{\partial p_z}\Big|_{p_z=p_{zr}=0}\right)$.

Since the presence of the interface does not change in the form of the spike, depending on the nature of the extremum $D_1$ as a function of $p_z$, we considered in detail only its intensity when the quantity $D_1(p_z)$ has a maximum, i.e., $\gamma_1 = -1$. The electric field amplitude has its maximum value near the center of the spike $(|D_1 - x| \leq k_1^{-1})$, where it equals:

$$E_1(D_1) \approx -\frac{e^{i\pi/3}}{6\sqrt{6}} \frac{\partial E(0)}{\partial x} \frac{1}{k_0 \sqrt{k_0 D_1}} \frac{\chi_1^3}{k_1^2 k_0}. \qquad (5.41)$$

It follows from Eqs. (5.39) and (5.40) that in the radio-frequency range $\left|\omega^*/\Omega\right| \ll 1$ for scattering of charge carriers by the interface, whose degree of diffuseness is small $(1 - Q - P \ll 1)$, the probability $Q$ drops out of the final expressions, it in the leading approximation with respect to the parameter $(Q \; \Omega_1 \; \tau)^{-1}$ if

$$Q \gg \frac{1}{\Omega_1 \tau}; \quad \left( \frac{1}{\Omega_1 \tau} \ll 1; \quad T_1 = \frac{2\pi}{\Omega_1} = \tau_1 - \lambda_1(\tau_1) \right). \qquad (5.42)$$

The characteristic scale of the HF field in the spike $k_1^{-1}$ coincides in order of magnitude with the "width" of the spike in the single crystal. The amplitude (5.41) is comparable to the value of the electric field for the first spike in a monocrystal. Inequality (5.42), apparently, was satisfied in the experiment in Ref.9, in which the amplitude of the RF size effect lines in the DLF even turned out to be somewhat larger than the amplitude of the line in the single-crystal film. The latter circumstance could be related to the fact that the characteristics of the "effective" electron trajectory, refracted by the interface, differ from these quantities in the single crystal. (The change in the form of the orbit is indicated by the shift in the lines on the magnetic field scale.)

We note that for sufficiently small thickness of the upper layer $d < D_1/2$ (in the experiment in Ref. 9, $d/D_1 \approx 0.42$), when the time $T_1$ for motion along the arc $a$ is smaller than the time $T_2$ for motion along arc $b$ (Fig.8), it is possible for an intense spike to appear at depth $D_1$ even for $Q < D_1/l$, if inequality (5.42) is satisfied.

If the probability for the electron to penetrate through the interface is small:

$$Q \ll \frac{1}{\Omega_1 \tau},$$

when the amplitude of the HF field in a spike is small:

$$\frac{\chi_1^3}{k_1^2} \approx \begin{cases} Q\dfrac{l}{D_1}, & Q << \dfrac{1}{\Omega_2 \tau}, \quad T_2 = \dfrac{2\pi}{\Omega_2}; \\[2ex] \left(Q\dfrac{l}{D_1}\right)^{2/3}, & Q >> \dfrac{1}{\Omega_2 \tau}. \end{cases} \tag{5.43}$$

In thicker specimens $L >> D_1$, electrons that do not interact with the interface ensure that the electromagnetic wave extends into the bulk of the specimen. The distribution of HF field near the second spike ($x \approx D_1 + D_2$) is described by following integral:

$$E_2(x) = -\frac{\partial E(0)}{\partial x} \frac{1}{4\pi k_0^2} \frac{(\chi_1 \chi_2)^3}{\sqrt{D_1 D_2}} \cdot$$

$$\int_0^\infty \frac{dk}{k} \frac{F_0\left(k/k_0\right)\cos\left[k(D_1+D_2-x)+\dfrac{\pi}{4}\gamma\right] + \hat{G}F_0\left(k/k_0\right)\sin\left[k(D_1+D_2-x)+\dfrac{\pi}{4}\gamma\right]}{\left(k^3-ik_1^3\right)\left(k^3-ik_2^3\right)};$$

where

$$k_2^3 = \frac{4\pi\omega}{c^2}\int dp_z \left[\rho_{22}\left(s_1^{(2)},s_1^{(2)}\right)+\rho_{22}\left(s_2^{(2)},s_2^{(2)}\right)\right]\cot\left(-i\omega^*\frac{T_{22}}{2}\right);$$

$$\gamma_2^3 = \frac{4\pi\omega}{c^2}\rho_{22}\left(s_1^{(2)},s_1^{(2)}\right)sh^{-1}\left(-i\omega^*\frac{T_{22}}{2}\right)\Big|_{p_z=p_2^*}; \qquad \frac{\partial D_2}{\partial p_z}\Big|_{p_z=p_{22}}=0; \qquad \gamma=\gamma_1+\gamma_2.$$

$$D_2 = x^{(2)}\left(s_2^{(2)}\right)-x^{(2)}\left(s_1^{(2)}\right); \quad \gamma_2 = sign\frac{\partial^2 D_2}{\partial p_z^2}. \tag{5.44}$$

Its amplitude contains the same information as the spike at depth $D_1$, but is a factor of $(k_1 \quad r)^{1/2}$ smaller than the intensity of the first spike.

The anomalous penetration of an electromagnetic wave into a DLF can also be examined in a similar manner for stronger magnetic fields $D_1 < d$, when only the intensity of the second spike contains information on the interaction of charge carriers with the interface.

## 5.2. Resonance phenomena

In the ultra high frequency range, the absorption of energy in the electromagnetic wave has a resonance character and in the DLF two types of resonance are possible. One of them stems from the periodic motion (with period $T_{12} = 2\pi/\Omega_{12}$) of electrons tunneling through the interface and occurs at frequencies

$$\omega = n\Omega_{12}; \qquad n=1,2,3,..., \qquad \frac{\partial T_{12}}{\partial p_z}\Big|_{p_z=p_1}=0; \qquad T_{12}=T_1+T_2. \tag{5.45}$$

The second type of resonance, which is specific to DLF's, turns out to be possible due to the motion of charge carriers entering the skin layer with frequency $\Omega_1 = 2\pi / T_1$ after reflection from the interface:

$$\omega = m\Omega_1; \quad m = 1,2,3...; \quad \left.\frac{\partial T_1}{\partial p_z}\right|_{p_z = p_2} = 0. \tag{5.46}$$

Using the Fourier component of the HF electric field, we can find the surface impedance of the DLF $Z$

$$Z = \frac{4\pi i\omega}{c^2}\left(\frac{\partial E(0)}{\partial x}\right)^{-1}\int_0^\infty dk\, \mathrm{E}(k). \tag{5.47}$$

The largest term is connected with the electric field in the skin-layer (5.35) and equals

$$Z_0 = -\frac{8i\omega}{c^2 k_0}M(-1), \tag{5.48}$$

where $k_0$ and $M(-1)$ are defined by Eqs.(5.12) and (5.36).

The resonance correction to the high-frequency impedance $Z_0$ is easy to calculate using perturbation theory:

$$\Delta Z_{res} = 1.99\cdot 10^{-2}\frac{8\omega}{c^2}\left(\frac{\eta}{k_0}\right)^3\exp\left(\frac{4\pi i}{5}\right). \tag{5.49}$$

Near the resonance (5.45) the function $\eta$ has the following form $\left(\omega\tau << \left(\frac{r}{\delta}\right)^{\frac{1}{2}}\right)$:

$$\eta^3 = \frac{4\pi\omega}{c^2}\rho_{11}\left(s_1^{(1)},\ s_1^{(1)}\right)\ \Psi\left(\Delta, T_{12}, \nu\right)\ \left[2\left(1 - Pe^{-i\omega^* T_1}\right) - \nu\right]\ \big|_{p_z = p_1}; \tag{5.50}$$

where

$$\Psi(\Delta, T, \alpha) = \frac{1}{2\xi}\frac{1}{n\sqrt{2\chi}}\left\{\sqrt{\xi + s\Delta} + is\sqrt{\xi - s\Delta}\right\}; \tag{5.51}$$

$$\xi = \sqrt{\Delta^2 + \varsigma^2}; \qquad \chi = \frac{1}{2T}\frac{\partial^2 T}{\partial p_z^2}; \quad s = sign\chi; \qquad \varsigma = \frac{1}{\omega\tau} + \frac{\alpha}{2\pi n};$$

$$\nu = 2P\left(1 - \cos\omega^* T_1\right) + \left(1 - P\right)^2 - Q^2,$$

where $\Delta = \dfrac{|H - H_1|}{H_1} << 1$, is the detuning of the resonance, while the resonant values of the magnetic field $H_1$ are determined by Eq. (5.45).

In magnetic fields close to the values $H = H_2$ $\left(\Delta = \dfrac{|H_2 - H|}{H_2} << 1\right)$, which ensure that the resonance condition (5.46) is satisfied, we obtain

$$\eta^3 = \frac{4\pi\omega}{c^2}\rho_{11}\left(s_1^{(1)},s_1^{(1)}\right)\Psi(\Delta,T_1,\mu)\left[2\left(1-Pe^{i\omega^*T_{12}}\right)-\mu\right]\Big|_{p_z=p_{z2}}; \qquad (5.52)$$

$$\mu = (1-P)\left(1-e^{i\omega^*T_{12}}\right)+\left((1-P)^2-Q^2\right)e^{i\omega^*T_{12}}.$$

As is evident from Eq. (5.51), the tunneling probability $Q$ and the probability for diffuse scattering $1-P-Q$ enter into the width of the function $\Psi(\Delta)$. For this reason, their resonant character, generally speaking, is retained only if $|\nu|\ll 1$ and $|\mu|\ll 1$ i.e., when the collision occurs with a low amount of diffuseness $1-P-Q\ll 1$, while the probability for a conduction electron to pass through the interface is large ($P\ll 1$) or, vice versa, is small ($Q\ll 1$), and the motion of the charge carriers is almost periodic. In the opposite case, the resonance curve is strongly eroded. For $\frac{1}{\Omega_1\tau}\ll Q\ll 1-\frac{1}{\Omega_{12}\tau}$, the width of the resonance lines is determined by the transmission probability, as a result of which the impedance of the DLF in the ultra HF range is more sensitive to the magnitude of $Q$ than in the radio frequency range and similar experiments could give additional information concerning the properties of the interface.

The amplitude of the HF field in the spike at $x\approx D_1$ (5.41) is also of a resonant nature. In the case when the period $T_{12}$ is an extreme for electrons forming the spike, the characteristic scale of variation of the field in the spike in resonance at frequencies (5.45) is "compressed" $k_1\approx\left[\Psi(\Delta,T_{12},\nu)\right]^{-1/3}$, while its amplitude is proportional to $\left[\Psi(\Delta,T_{12},\nu)n\right]^{4/3}$. If, on the other hand, the frequency of electron motion $\Omega_{12}$ for the orbit intersecting the interface is not an extreme, then the function $f_{22}$ (5.18), which determines the width of the spike, is insensitive to the detuning of the resonance. The resonance character of the spike is related to the behavior of the function $f_{12}$ (5.24), which is given by $f_{12}\approx\left[\exp(-i\omega^*T_1)-1+\mu\right]^{-1}$. The main characteristics of the spike for resonance at frequencies (5.46) show an analogous behavior:

$$k_1\approx const(\Delta); \qquad E(D_1)\approx\left[\exp(-i\omega^*T_1)-1+\mu\right]^{-1}.$$

The resonance in the spike also appears when the frequency $\omega$ is a multiple of the frequency $\Omega_2$ of the electrons entering into it due to the almost specular reflection from the interface. Then $k_1\approx\left[\Psi(\Delta,T_2,\mu)\right]^{-1/3}$, while $E(D_1)\approx\left[\Psi(\Delta,T_2,\mu)\right]^{4/3}$, if the period $T_2$ is extremal. An exception is the case of a double resonance, when conditions (5.45) and (5.46) are satisfied simultaneously. In this case, the change in the type of the trajectory after specular reflection or transmission through the interface does not take the electron out of resonance, while the broadening of the lines related to the presence of the interface is small.

# 6. EFFECT OF SPONTANEOUS MAGNETIZATION ON THE ELECTRICAL CONDUCTIVITY OF FERROMAGNETIC- BASED METALLIC MULTILAYERS

Magnetic multilayers (MML) whose periodic structure contains a ferromagnetic as an element display especially interesting properties (see, for example, the review [1]). Giant magnetoresistance (GMR), involving a sharp decrease in the sample resistance (sometimes exceeding 100%!) in a quite weak magnetic field is without doubt among the most striking and significant effects displayed by MML. GMR was observed [2] for the first time in the MML Fe/Cr, and investigated subsequently for quite different combinations of ferromagnetic and nonmagnetic metals. As a rule, the MML having an antiferromagnetic structure (spontaneous magnetic moments $\mathbf{M}$, in adjacent ferromagnetic ($F$) layers separated by a nonmagnetic ($N$) layer are antiparallel) display GMR. This structure is determined by the thickness $d$ of the $N$-layer and is formed as a result of indirect RKKY interaction between the $F$-layers. The application of a magnetic field $\mathbf{H}$ parallel to the boundaries of a MML leads to an alignment of magnetic moments $\mathbf{M}$ in $F$-layers along the vector $\mathbf{H}$. At the same time, the sample resistance decreases significantly. The most well founded hypothesis about the origin of GMR can be the assumption about the dominating role of electron scattering at the internal boundaries, which depends on the spin of electrons [1,2]. It is based on the well-known fact that the cross-sections of scattering of charge carriers with different spin by magnetic impurities are different [3] in view of the dependence of the density of electron states at the Fermi surface on a spin direction [4]. A comparison of theoretical calculations [57-61] with the experimental results shows that GMR can be described quite accurately by using Boltzmann's kinetic equation for quasiclassical electron distribution functions with boundary conditions in which the probability of scattering of charge carriers at the layer boundaries depends on spin.

In this chapter we describe the results of theoretical investigations [62] of the effect of internal magnetic field $\mathbf{B}$ on electron trajectories, which is manifested in the conductivity of ferromagnetic-based magnetic multilayers (MML). This effect may be due to a significant variation of the dynamics of charge carriers colliding with interfaces in a field $\mathbf{B}$, just like the emergence of static skin effect in thin films caused by electrons "hopping" over the surface in a magnetic field [63-65]. These ballistic effects, which are extremely sensitive to the mutual orientation of the electric current $\mathbf{J}$ and magnetization $\mathbf{M}_s$, may be manifested as anisotropy of giant magnetoresistance. We show that the experimental investigation of the MML conductivity for various angles between the vectors $\mathbf{J}$ and $\mathbf{M}_s$ can provide information about the probability of tunneling of charge carriers through the interfaces. A consideration of such trajectory effects in a ferromagnetic with a domain structure [66, 67] made it possible to explain the negative magnetoresistance observed in them.

Let us consider an infinite periodic system consisting of alternating layers of ferromagnetic and nonmagnetic metals of the thicknesses $d$ and $d_f$. An electron having spin $\sigma$ colliding with the interface has a probability $Q_\sigma$ of tunneling into the adjoining metal without scattering, and a probability $P_\sigma$ of specular reflection. Analyzing the role of trajectory effects, let us consider the case when the width $d_f$ of the $F$-layer is larger than the characteristic Larmor radius $r$ of electron trajectories in the field $\mathbf{B}$ $(r < d_f \ll l)$. The induction $\mathbf{B}_0$ of the external magnetic field required for the change of the system from the antiferromagnetic to ferromagnetic state is assumed to be quite small. Consequently, $r_0 \gg r,l$ and its effect on the orbits of charge carriers can be disregarded $(r_0 = cp_F/eB_0)$. For example, $B \approx 2.5$ $T$ for Fe, a giant magnetoresistance is observed in the Fe/Cr system in fields $B_0 \approx 0.1$ $T$.

In order to calculate the current in MML for a given electrical field we must find the electron distribution function (2.9) taking into account dependence of probabilities $Q_\sigma$ and $P_\sigma$ on the spin $\sigma$. The values of the functions $F_{ik}^{(\sigma)}$ of characteristics in the solution (2.9) of the kinetic equation should be defined with the help of the boundary conditions [15, 58] connecting the distribution function $\Psi_{i,v_{ni}>0}^{(\sigma)}$ of electrons flying into the $i$-$th$ layer on the interface $x = x_{\text{int}}$ with the distribution functions of charge carriers incident on the boundary from the same layer $\Psi_{i,v_{ni}<0}^{(\sigma')}$ or the adjacent layer $\Psi_{k,v_{ni}>0}^{(\sigma'')}$:

$$\Psi_{i,v_{ni}>0}^{(\sigma)}(x_{\text{int}},\mathbf{p}) = P_{\sigma'}\Psi_{i,v_{ni}<0}^{(\sigma')}(x_{\text{int}},\mathbf{p}') + Q_{\sigma''}\Psi_{k,v_{ni}>0}^{(\sigma'')}(x_{\text{int}},\mathbf{p}''), \qquad (6.1)$$

where $v_{ni} = \pm v_{xi}$ is the projection of the velocity $\mathbf{v}_i$ on the inward normal to the boundary of the $i$-$th$ layer. The momenta $\mathbf{p}$, $\mathbf{p}'$, $\mathbf{p}''$ are connected through the conditions of conservation of energy and their tangential component relative to the plane $x = x_{\text{int}}$ (see, Eqs. (2.5), (2.6)). In the case of MML with antiferromagnetic order of spontaneous magnetic moments $\mathbf{M}_s$ the conditions $\sigma = \sigma' = -\sigma''$ and $\sigma = -\sigma' = \sigma''$ must be satisfied for electrons flying into the $F$-layer and for charge carriers falling into the $N$-layer, respectively. If the moments $\mathbf{M}_s$ are parallel (ferromagnetic order in MML), we should put $\sigma = \sigma' = \sigma''$. The boundary condition (6.1) is valid if there is no diffuse scattering of electrons at the boundaries, i.e., $P_\sigma + Q_\sigma = 1$. Consideration of the weak diffusivity of the boundary ($P_\sigma + Q_\sigma = 1 - \rho_\sigma$, $\rho_\sigma \ll 1$), which violates the ballistic motion of electrons, causes just a slight decrease in the trajectorial contribution to the conductivity of MML.

*6.1 Solution of the kinetic equation and general expression for electrical current in a MML*

For ferromagnetic and antiferromagnetic structure of MML the function $F_{ik}^{(\sigma)}$ $(i, k = 1, 2)$ assumes (for a given $\sigma$) four values each corresponding to the motion along a certain segment of the trajectory (see Fig.10 *a* and *b*). Substituting the solution in the form (2.9) into the boundary conditions (6.1), we arrive at a system of eight algebraic equations, whose solution can be written in the form

$$F_{ii}^{(\sigma)} = D^{-1} \left\{ \varphi_{ii}^{(\sigma)} \left[ P_\sigma G_k^{(\sigma')} + \left( Q_\sigma^2 - P_\sigma^2 \right) \alpha_{ik}^{(\sigma)} \alpha_{ki}^{(\sigma')} A_k^{(\sigma')} \right] + Q_\sigma \varphi_{ik}^{(\sigma)} \alpha_{ki}^{(\sigma')} A_k^{(\sigma')} + \right.$$
$$\left. Q_\sigma \varphi_{ki}^{(\sigma')} G_k^{(\sigma')} + Q_\sigma Q_{\sigma'} \varphi_{kk}^{(\sigma')} \alpha_{ki}^{(\sigma')} \right\}; \quad i \neq k, \tag{6.2}$$

for parts of trajectory lying in the $F$-layer, and

$$F_{ik}^{(\sigma)} = D^{-1} \left\{ Q_\sigma \varphi_{ii}^{(\sigma)} G_k^{(\sigma')} + \varphi_{ik}^{(\sigma)} \alpha_{ki}^{(\sigma')} A_k^{(\sigma')} A_i^{(\sigma)} + \varphi_{ki}^{(\sigma')} G_k^{(\sigma')} A_i^{(\sigma)} + Q_{\sigma'} \varphi_{kk}^{(\sigma')} \alpha_{ki}^{(\sigma')} A_i^{(\sigma)} \right\}; \quad i \neq k. \tag{6.3}$$

Here

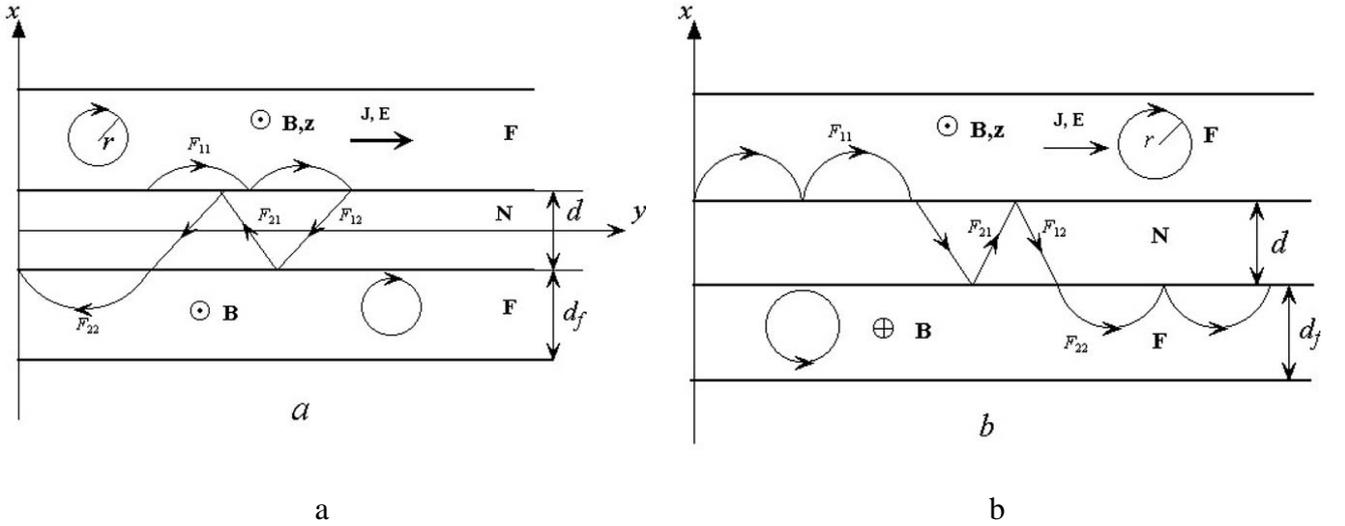

a

b

**Figure 10.** Electron trajectories for ferromagnetic (*a*) and antiferromagnetic (*b*) structure of multilayer.

$$D = G_k^{(\sigma')} G_i^{(\sigma)} - \alpha_{ik}^{(\sigma)} \alpha_{ki}^{(\sigma')} A_k^{(\sigma')} A_i^{(\sigma)}; \tag{6.4}$$

$$G_i^{(\sigma)} = 1 - P_\sigma \alpha_{ii}^{(\sigma)}; \tag{6.5}$$

$$A_i^{(\sigma)} = P_\sigma + (Q_\sigma^2 - P_\sigma^2) \alpha_{ii}^{(\sigma)}; \tag{6.6}$$

$$\alpha_{ik}^{(\sigma)} = \exp\left( \frac{\lambda_1^{(i)} - \lambda_2^{(k)}}{\tau_\sigma} \right); \tag{6.7}$$

$$\varphi_{ik}^{(\sigma)} = \int\limits_{\lambda_1^{(i)}}^{\lambda_2^{(k)}} dt' \mathbf{v}_i(t') \mathbf{E} \exp\left(\frac{t' - \lambda_2^{(k)}}{\tau_\sigma}\right), \qquad (6.8)$$

where $\lambda_1^{(i)}$ and $\lambda_2^{(k)}$ are two successive instants $(\lambda_1^{(i)} < \lambda_2^{(k)})$ of collision of an electron with the same $(i = k)$ or different $(i \neq k)$ boundaries. The quantities $\alpha_{ik}^{(\sigma)}$ and $\varphi_{ik}^{(\sigma)}$ have the meaning of the probability of an electron to move without scattering in the bulk over the corresponding segment of the trajectory and the energy acquired by this electron in an electric field over this segment. For charge carriers moving in closed Larmor trajectories in the bulk of the ferromagnetic, we should put $\lambda = -\infty$ in formula (2.9). Their contribution to the current can be calculated easily by using the well-known magnetic conductivity tensor for an infinite metal [16].

The current produced by electrons colliding with two adjacent MML boundaries can be presented in the form

$$I_\alpha = 2I_{11}^{(\alpha)} + I_{12}^{(\alpha)}; \qquad \alpha = y, z, \qquad (6.9)$$

where $I_{11}$ is the current passing in the boundary region of width $2r$ $(r = cp_F/eB)$ in the $F$-layer:

$$I_{11}^{(\alpha)} = -\frac{e^3 B}{ch^3} \sum_\sigma \int dp_z \int\limits_0^{T_H/2} d\lambda\, v_{xi}(\lambda) \int\limits_0^{T_H} dt\, \theta(t - \lambda) v_{\alpha i}(t) \cdot$$

$$\left\{ F_{11}^{(\sigma)} \exp\left(\frac{\lambda - t}{\tau_\sigma}\right) + \int\limits_\lambda^t dt' \mathbf{v}_i(t') \mathbf{E} \exp\left(\frac{t' - t}{\tau_\sigma}\right) \right\} \qquad (6.10)$$

and $I_{12}$ is the current in the $N$-layer:

$$I_{12}^{(\alpha)} = \frac{2e^2}{h^3} \sum_\sigma \int \frac{dS_{\mathbf{p}}}{v_i} \theta(-v_{xi}) v_{\alpha i}$$

$$\left\{ \mathbf{v}_i \mathbf{E} \tau_\sigma d + v_{xi} \tau_\sigma \left[ 1 - \exp\left(\frac{d}{v_{xi} \tau_\sigma}\right) \right] (\mathbf{v}_i \mathbf{E} \tau_\sigma - F_{12}) \right\}, \qquad (6.11)$$

where $p_z$ is the momentum projection on the direction of the magnetic field $\mathbf{B}$. While writing the expression (6.10), we used Eq. (2.11) to go over from integration over the coordinate $x$ to integration over the collision instant $\lambda$. Formula (6.11) describes the doubled electron current with negative velocity $v_{xi}$. It can easily be verified that, in view of the symmetry of the problem, the contributions from charge carriers with $v_{xi} > 0$ and $v_{xi} < 0$ to the conductivity of the $N$-layer are identical.

Formulas (6.10) and (6.11) are valid for any form of the Fermi surface of charge carriers. While calculating the current in the MML in the following analysis, we use the frequently employed model of a compensated metal with identical quadratic isotropic energy-momentum relations for electrons and "holes". Naturally, such a simple model does not take into consideration a whole range of very fine effects associated with the existence of a contact potential difference at the boundaries, peculiarities of electron tunneling due to a difference in the Fermi surfaces in magnetic and nonmagnetic metals, etc. However, this model makes it possible to find the explicit form of the dependence of current on the thicknesses $d$ and $d_f$ and the magnetic induction $\mathbf{B}$, and provides a qualitative description of the conductivity of an MML with an arbitrary closed Fermi surface just as in the case of a bulk conductor. The condition of compensation (equality of concentrations $n_1$ of electrons and $n_2$ of holes) allows us to neglect the Hall components of the MML conductivity and to disregard Eq. (2.13) for the field $E_x$, which can be solved only numerically. Note that the strict equality $n_1 = n_2$, which is quite normal for nonmagnetic metals, is not satisfied for pure ferromagnetics. However, it was shown by Kaganov and one of the authors [68] that so far as the galvanomagnetic properties are concerned, a metal for which the inequality $|n_1 - n_2|/(n_1 + n_2) << r/l$ is satisfied behaves like a compensated metal. Such a situation can be expected, for example, in $4f$-ferromagnetic metals.

### 6.2 MML conductivity parallel to the interfaces

In the model of spherical constant energy surfaces, we obtain the following expressions for the components of electron (hole) velocities:

$$v_x = -v_\perp \sin \Omega \; t; \quad v_y = v_\perp \cos \Omega \; t; \quad v_z = p_z/m, \tag{6.12}$$

where $\Omega = |e| B / mc$ is the cyclotron frequency, $v_\perp = \left(p_F^2 - p_x^2\right)^{1/2}/m$.. For the instant $t = 0$, we chose a point on the Fermi surface corresponding to charge carriers moving parallel to the boundaries (Fig. 11). In this reference system, the "instants" $\lambda_{1,2}^{(i)}$ of collisions are defined as

$$\lambda_1^{(1)} = -\lambda; \quad \lambda_1^{(2)} = \lambda_2^{(1)} = \lambda; \quad \lambda_2^{(2)} = T_H - \lambda; \quad T_H = \frac{2\pi}{\Omega}; \tag{6.13}$$

for the ferromagnetic structure of MML, and

$$\lambda_1^{(1)} = \lambda_2^{(2)} = -\lambda; \quad \lambda_2^{(1)} = \lambda_1^{(2)} = \lambda. \tag{6.14}$$

for the antiferromagnetic structure.

Let us first consider the current flowing at right angles to the vector $\mathbf{B}$ and present it in the

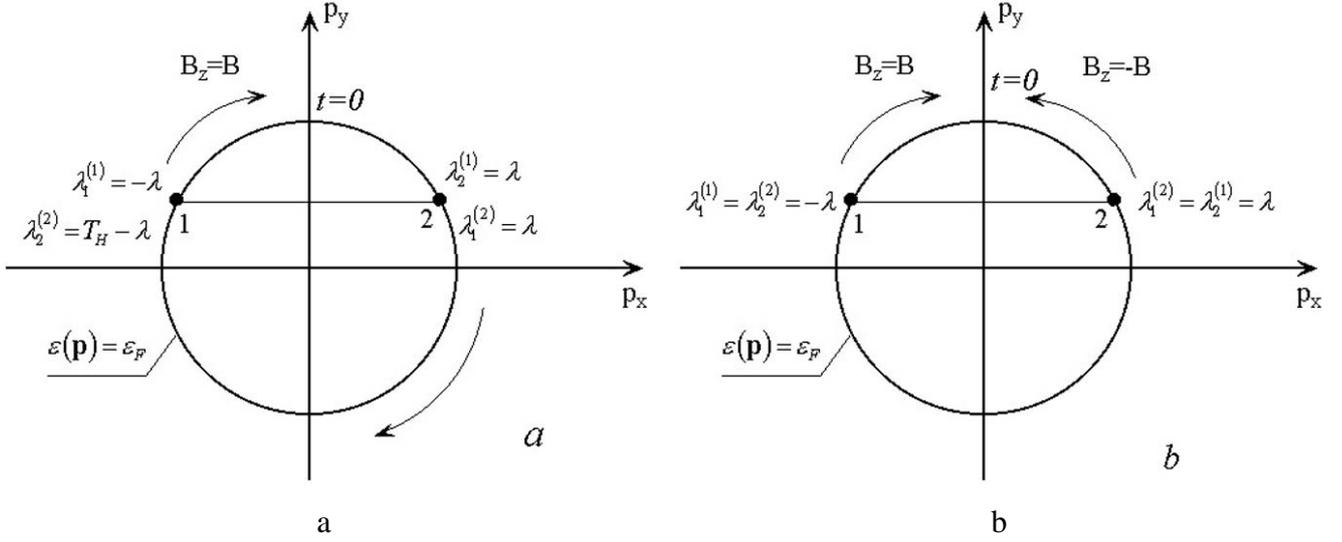

**Figure 11.** Electron trajectories in the momentum space for ferromagnetic (*a*) and antiferromagnetic (*b*) order of spontaneous magnetic moments. Points 1 and 2 correspond to the motion at a constant velocity in the nonmagnetic layer.

following form by using formula (6.9):

$$I_y = \left[ 4\sigma_{yy}^{(F)} r + \sigma_{yy}^{(N)} d \right] E_y, \tag{6.15}$$

where $\sigma_{yy}^{(F)}$ is the conductivity of the boundary region of the $F$-layer, and $\sigma_{yy}^{(N)}$ the conductivity of the $N$-layer. In order to pay special attention to the effect of the boundaries on the conductivity of MML, we can present the expression for $\sigma_{yy}^{(F,N)}$ for the case when the bulk relaxation frequencies in the layer are identical ($\tau_\sigma = \tau_{\sigma'} = \tau_{\sigma^*} = \tau$):

$$\sigma_{yy}^{(F)} = \frac{9\pi}{64} \frac{\sigma_0}{1+\gamma^2} + \frac{3\sigma_0}{16\pi} \frac{\gamma^2}{\left(1+\gamma^2\right)^2} \sum_\sigma \int_0^\pi d\varphi \sin\varphi \int_{-1}^1 du (1-u^2)^{3/2} D^{-1} \cdot$$

$$\left\{ a_{11} \cos\varphi + \gamma(2 - a_{22}) \sin\varphi \right\} \cdot$$

$$\left\{ \sin\varphi \left[ 2Q_\sigma A_2^{(\sigma')} + Q_{\sigma'}(2 - Q_\sigma a_{22}) + (2a_{12} - a_{12}^2)(2 - 3Q_\sigma) A_2^{(\sigma')} \mp Q_\sigma Q_{\sigma'}(1 - a_{12})(2 - a_{22}) \right] +$$

$$\cos\varphi \gamma a_{12} Q_\sigma \left[ 2A_2^{(\sigma')} + Q_{\sigma'} a_{22} - a_{12} A_2^{(\sigma')} \right] \right\}; \tag{6.16}$$

$$\sigma_{yy}^{(N)} = \sigma_0 + \frac{3}{4\pi} \frac{\sigma_0}{1+\gamma^2} \sum_\sigma \int_0^\pi d\varphi \sin\varphi \cos\varphi \int_{-1}^1 du (1-u^2)^{3/2} D^{-1} a_{12} \left\{ \gamma \sin\varphi \left[ Q_\sigma(2 - a_{11}) A_2^{(\sigma')} \mp \right. \right.$$

$$Q_{\sigma'}(1 - a_{12})(2 - a_{22}) A_1^{(\sigma)} \left] - \gamma^2 \cos\varphi \left[ Q_\sigma Q_{\sigma'}(a_{11} + a_{22}) + \right.$$

$$(Q_\sigma + Q_{\sigma'} - 3Q_\sigma Q_{\sigma'}) a_{11} a_{22} - Q_{\sigma'} a_{12} a_{22} A_1^{(\sigma)} \left] \right\}, \tag{6.17}$$

where

$$D = Q_\sigma a_{11} A_2^{(\sigma')} + Q_{\sigma'} a_{22} A_1^{(\sigma)} + Q_\sigma Q_{\sigma'} a_{11} a_{22} - (2a_{12} - a_{12}^2) A_1^{(\sigma)} A_2^{(\sigma')};$$

$$A_i^{(\sigma)} = Q_\sigma + (1 - 2Q_\sigma)\alpha_{ii}; \quad \gamma = \Omega\tau;$$

$$a_{12} = 1 - \exp(-(d/l)\sin\varphi\sqrt{1-u^2}); \quad l = \frac{p_F \tau}{m}; \quad \sigma_0 = \frac{(n_1 + n_2)e^2\tau}{m},$$

$\sigma_0$ is the conductivity of the bulk metal. For an MML with ferromagnetic structure, we should choose the upper sign in formulas (6.16) and (6.17) and put $\sigma = \sigma'$:

$$a_{11} = 1 - \exp(-2\varphi/\gamma); \quad a_{22} = 1 - \exp(-2(\pi - \varphi)/\gamma). \tag{6.18}$$

The antiferromagnetic structure of MML corresponds to the lower sign and $\sigma \neq \sigma'$:

$$a_{11} = a_{22} = 1 - \exp(-2\varphi/\gamma). \tag{6.19}$$

The condition of strong magnetic field $\gamma = l/r \gg 1$ allows us to expand $a_{ii}$, up to the first nonvanishing term, and we shall treat these quantities to be small and of the order of $\gamma^{-1}$.

In spite of a number of simplifying assumptions, the expressions (6.16) and (6.17) remain quite cumbersome. We shall present below their asymptotic forms for the most interesting cases.

1. $Q_\sigma \ll r/l$. For such a low tunneling probability, there are practically no electron trajectories passing through two $F$-layers, and the conductivities $\sigma_{yy}^{(F,N)}$ are independent of the mutual orientation of magnetic moments $\mathbf{M}_s$:

$$\sigma_{yy}^{(F)} \cong \sigma_0 \begin{cases} \dfrac{9}{32}g - \dfrac{3}{8}(Q_\sigma + Q_{\sigma'})\gamma \quad g_2, \ Q_\sigma \ll d/l, \ \sigma \neq \sigma'; \ (a) \\[2mm] \dfrac{9}{32}g - \dfrac{1}{4\pi}\dfrac{d}{2r}g_1, \ Q_\sigma \gg d/l; \ (b) \end{cases} \tag{6.20}$$

$$\sigma_{yy}^{(N)} \cong \sigma_0 \begin{cases} 1 - \dfrac{3}{16}(Q_\sigma + Q_{\sigma'})\dfrac{l}{d}\left(1 - \dfrac{g_2}{2}\right), \ Q_\sigma \ll d/l, \ \sigma \neq \sigma'; \ (a) \\[2mm] \dfrac{3}{4}\dfrac{Q_\sigma + Q_{\sigma'}}{Q_\sigma Q_{\sigma'}}\dfrac{d}{l}\ln\dfrac{l}{d}, \ Q_\sigma \gg d/l; \ (b) \end{cases} \tag{6.21}$$

where

$$g = \frac{1}{4}\big[3Si(\pi) - Si(3\pi)\big]; \quad g_1 = Si(2\pi); \quad g_2 = \frac{3}{4}\big[Ci(\pi) - Ci(3\pi) + \ln 3\big].$$

2. $r/l \ll Q_\sigma \ll d/l$. These inequalities mean that charge carriers freely penetrate into nonmagnetic metal from the ferromagnetic metal, but still the ballistic motion of electrons does not lead to a connection between $F$-layers because of a large thickness of the $N$-layer:

$$\sigma_{yy}^{(F)} = \sigma_0 \frac{r}{l}\left(\frac{1}{Q_\sigma} + \frac{1}{Q_{\sigma'}}\right), \ \sigma \neq \sigma', \ (a);$$

$$\sigma_{yy}^{(N)} = \sigma_0\left(1 - \frac{3r}{8d}\right), \ (b). \qquad (6.22)$$

3. $Q_\sigma \gg r/l, \ d/l$. In this case, the boundary is practically transparent to charge carriers, and the dominating role in the conductivity of MML is played by periodic trajectories intersecting two $F$-layers. The nature of these trajectories depends on the direction of the magnetic field $\mathbf{B}$ in $F$-layers because of which the asymptotic forms of the conductivities $\sigma_{yy}^{(F,N)}$ are found to be quite different for different types of magnetic order in MML.

The ferromagnetic structure of MML is characterized by the relations

$$\sigma_{yy}^{(F)} \cong \sigma_0 \sum_\sigma \begin{cases} \dfrac{5}{8}(1-Q_\sigma)\dfrac{r}{l} + \dfrac{3}{16}(3Q_\sigma-2)\dfrac{rd}{l^2}, \ d \gg r, \ (a); \\[2mm] \dfrac{3}{16}\dfrac{1-Q_\sigma}{Q_\sigma}\dfrac{r}{l}, \ d \ll r, \ (b); \end{cases} \qquad (6.23)$$

$$\sigma_{yy}^{(N)} \cong \sigma_0 \begin{cases} 1 - \dfrac{3r}{8d}, \ d \gg r, \ (a); \\[2mm] \dfrac{3}{\pi}\dfrac{d}{r}\ln\dfrac{l}{d} + \dfrac{(1-Q_\sigma)}{Q_\sigma}\dfrac{r}{l}, \ d \ll r, \ (b); \end{cases} \qquad (6.24)$$

while for the antiferromagnetic structure of MML we can write

$$\sigma_{yy}^{(F)} \cong \sigma_0 \begin{cases} \dfrac{6}{5}\dfrac{r}{d}, \ d \gg r, \ (a); \\[2mm] \dfrac{9}{32}g, \ d \ll r, \ (b); \end{cases} \qquad (6.25)$$

$$\sigma_{yy}^{(N)} \cong \sigma_0 \begin{cases} 1 - \dfrac{3r}{8d}, \ d \gg r, \ (a) \\[2mm] \dfrac{4}{\pi}\dfrac{d}{r}\ln\dfrac{l}{d} + \dfrac{1}{\pi}g_1, \ d \ll r, \ (b). \end{cases} \qquad (6.26)$$

It is well known that the bulk conductivity $\sigma_{\perp b}$ of a compensated metal in a direction perpendicular to the magnetic field $\mathbf{B}$ is given by [16]

$$\sigma_{\perp b} = \sigma_0\left(\frac{r}{l}\right)^2 \ \text{for} \ \ r \ll l. \qquad (6.27)$$

Hence, for thicknesses $d_f \ll l$ of the $F$-layer, the main contribution to the MML conductivity comes from the $N$-layers and the boundary region of the ferromagnetic.

Using formulas (6.10) and (6.11) to calculate the current flowing along the vector **B** in an MML, it can be easily verified that

$$\sigma_{zz}^{(F)} = \sigma_{zz}^{(N)} = \sigma_0 \qquad (6.28)$$

for any value of the tunneling probability $Q_\sigma$.

### 6.3 Qualitative description of the results

Let us now discuss the analytic results for conductivities $\sigma_{yy}^{(F,N)}$ (formulas (6.20)-(6.26)) obtained with the help of the exact formulas (6.16), (6.17) (within the limits of the model) using the concepts about the effective number of charge carriers $n_{eff}$ and their mean free path $l_{eff}$. Such an approach, which allows a qualitative description of transport phenomena in conductors, is used frequently in the physics of normal metals [16]. It is based on the extremely simple assumption that the conductivity of a complex system can be described roughly by the simple Drude-Lorentz relation containing the relative number $n_{eff}$ of electrons participating in the conduction, and the length $l_{eff}$ characterizing the energy $\Delta\varepsilon = eEl_{eff}$ acquired during the mean free time $\tau$.

$$\sigma = \frac{n_{eff} e^2 l_{eff}}{p_F} \equiv \sigma_0 \frac{n_{eff} l_{eff}}{nl}. \qquad (6.29)$$

In the limit $Q \to 0$ (for simplicity, we shall assume that the tunneling probability $Q$ does not depend on spin), the electrons interacting with the boundaries perform independent periodic motion in the $N$-layer (along a trajectory broken by specular reflections) and in the $F$-layer (along a "hopping" trajectory). The path traversed by them along the electric field is defined as $l_{eff} \cong l$ and the contribution to the conductivity is of the order of the conductivity $\sigma_0$ of the bulk sample. For $Q \neq 0$ however, $Q << r/l,\ d/l$, and most of the electrons fail to tunnel to the adjoining metal. However, as a result of each collision with the boundary, $Q\ n$ charge carriers are left out from the conduction process and go over to another layer. The total number of such carriers is $\delta n = n - n_{eff} \cong MQn$, where $M$ is the number of collisions with the boundary during a time $\tau$ ($M \approx l/r$ for the $F$-layer and $M \approx l/d$ for the $N$-layer). Accordingly, the conductivities $\sigma_\perp^{(F)}$ and $\sigma_\perp^{(N)}$ [see Eqs. (6.20a) and (6.21a)] are found to be of the order of

$$\sigma_\perp^{(F)} \approx \sigma_0 \left(1 - \frac{Ql}{r}\right), \qquad Q << r/l, d/l; \qquad (6.30)$$

$$\sigma_{\perp}^{(N)} \approx \sigma_0 \left(1 - \frac{Ql}{d}\right), \qquad Q << r/l; d/l. \tag{6.31}$$

For $Q >> d/l$ the electrons tunnel freely from $N$-layer to the $F$-layer, but their return during the mean free time is unlikely in view of the inequality $Q << r/l$. The situation arising in the $N$-layer is analogous to that in a thin plate with charge carriers scattered diffusely at the surface [15,69], when $l_{eff}$ is of the order of its thickness $d$, which leads to a decrease in the conductivity by a factor $l/d$. Before tunneling, the electrons manage to cover a distance $l_{eff} \approx d/Q$ in the $N$-layer, making a contribution

$$\sigma_{\perp}^{(N)} \approx \sigma_0 \frac{d}{lQ} \ln\frac{l}{d}, \qquad d/l << Q << r/l, \tag{6.32}$$

to the conductivity [see (6.21b)]. As usual, $\ln(l/d)$ reflects the contribution to $\sigma_{\perp}^{(N)}$ from electrons flying nearly parallel to the boundaries for which $l_{eff} \approx l$. Conversely, the conductivity of the $F$-layer even becomes slightly higher since almost all of the $\delta n \cong (l/r)Q \; n$ electrons tunneled through it return, and the difference between $\sigma_{\perp}^{(F)}$ and $\sigma_0$ is associated with a decrease in their effective mean free path by their path $d/Q$ in the $N$-layer [see (6.20b)]:

$$\sigma_{\perp}^{(F)} \approx \sigma_0 \left(1 - \frac{d}{r}\right), \quad d/l << Q << r/l. \tag{6.33}$$

Conversely, if $r/l << Q << d/l$ the charge carriers in the ferromagnetic go over to the $N$-layer after covering a path $l_{eff} = r/Q << l$ along the boundary, and the conductivity associated with them (see (6.22a)) is of the order of

$$\sigma_{\perp}^{(F)} \cong \sigma_0 \frac{r}{Ql}, \qquad r/l << Q << d/l. \tag{6.34}$$

The conductivity of the $N$-layer in the main approximation in the small parameter $d/(Ql)$ coincides with the conductivity of a thin plate with specular boundaries [15,69] (i.e., equal to $\sigma_0$). Its decrease is determined by the effective mean free path $l_{eff} = l - r/Q$ for $\delta n = (l/d)Qn$ electrons entering the $F$-layer (see (6.22b)):

$$\sigma_{\perp}^{(N)} = \sigma_0 \left(1 - \frac{r}{d}\right), \qquad r/l << Q << d/l. \tag{6.35}$$

For a highly transparent boundary $Q >> r/l,\ d/l$ a connection between $F$-layers is established due to trajectorial motion of electrons, and the conductivity of MML depends on the mutual orientation of magnetic field in them.

For the ferromagnetic structure of MML, electrons in different $F$-layers move in directions opposite to the electric field. In the limit $Q \to 1,\ d \to 0$, the magnetic field in the sample is uniform, and the conductivity $\sigma_\perp^{(F)}$ coincides with the corresponding value for an infinite conductor $\sigma_{\perp b}$ defined by Eq. (6.27). Proceeding from the concepts about the effective mean free path, the quantity $\sigma_{\perp b}$ can be estimated as follows. The displacement of an electron along the vector $\mathbf{E}$ over a period is equal to zero on account of its periodic motion, i.e., $l_{eff} = 0$ in the main approximation in $r/l$. However, taking into consideration the collisions and the fact that lengths $l_1$ and $l_2$ of trajectory segments along and against the field $\mathbf{E}$ are different $\ (|l_1 - l_2| \cong r)$, we obtain the following estimate for $l_{eff}$:

$$l_{eff}^{(1)} \cong r \left| \exp\left(-\frac{l_1}{l}\right) - \exp\left(-\frac{l_2}{l}\right) \right| \cong \frac{r^2}{l} \ \text{for} \ r << l, \tag{6.36}$$

which corresponds to formula (6.27). For $Q \neq 1,\ d \neq 0$ (but for $Q >> r/l, d/l$), the difference between the quantities $\sigma_\perp^{(F)}$ and $\sigma_{\perp b}$, is due to two reasons. First, as a result of reflection at the boundaries, $n_{eff} = (1 - Q)n$ electrons have displacements in opposite directions, differing by a quantity of the order of $l_{eff}^{(1)} \approx r$. Second, bulk scattering at trajectory segments lying in the $N$-layer leads, as in a bulk metal, to a nonzero effective mean free path

$$l_{eff}^{(2)} \cong r \left| \exp\left(-\frac{l_1}{l}\right) - \exp\left(-\frac{l_2 + 2d}{l}\right) \right| \cong \frac{rd}{l} \ \text{for} \ l >> d >> r. \tag{6.37}$$

The above processes define two contributions to the conductivity $\sigma_\perp^{(F)}$ of the $F$-layer (see Eq. (6.23a):

$$\sigma_\perp^{(F)} \cong (1 - Q)\sigma_0 \frac{r}{l} + \sigma_0 \frac{dr}{l^2}, \quad Q >> r/l, d/l. \tag{6.38}$$

In the case under consideration, the $N$-layer conductivity also contains two terms, one of which coincides with the first term in (6.38) and has the same origin. The second term is associated with the energy acquired by an electron directly in a nonmagnetic metal over a length $l_{eff} \cong dM$, where $M$ is the number of collisions with the boundary over a period $T$. For transmission probabilities close to unity, $M \cong l/(d + r)$. Hence the total conductivity $\sigma_\perp^{(N)}$ (see Eq. (6.24)) can be presented in the form

$$\sigma_\perp^{(N)} \cong \sigma_0 \frac{d}{r+d} + (1-Q)\sigma_0 \frac{r}{l}, \quad Q >> r/l, d/l. \tag{6.39}$$

If the MML has an antiferromagnetic structure, and for a high transparency of the boundaries $Q >> r/l$, $d/l$ electrons move along an open trajectory, being displaced along the electric field in the same direction. Their effective mean free path in the $F$-layer is $l_{eff} \cong rM$ and the conductivity $\sigma_\perp^{(F)}$ (see (6.25)) is defined as

$$\sigma_\perp^{(F)} \cong \sigma_0 \frac{r}{r+d}, \quad Q >> r/l, d/l. \tag{6.40}$$

The conductivity of the $N$-layer in the case under consideration is weakly sensitive to the relation between $d$ and $r$. Even for $r >> d$, when charge carriers spend most of the mean free time in $F$-layers, moving in the same direction in these layers, they acquire energy in an electric field. Hence $l_{eff}$ is always of the order of $l$, and the conductivity is comparable with the conductivity $\sigma_0$ of the bulk metal (see Eq.(6.25)). It follows from Eqs. (6.38)-(6.40) that the change in the magnetic order in MML (under the action of the external field) leads to a considerable increase in the conductivity of both $F$- and $N$-layers in a direction perpendicular to the magnetic field.

Thus, if the radius $r$ of the electron carrier trajectory in the magnetic field with spontaneous induction $\mathbf{B}$ is smaller than the ferromagnetic layer thickness, the conductivity of MML is sensitive to the direction of the current flowing parallel to the boundaries. The MML conductivity perpendicular to the vector $\mathbf{B}$ depends significantly on the probability of tunneling of electrons through the boundary, and also on the mutual orientation of the magnetic moments for quite large values of $Q$. The anisotropy of the MML resistance defined by these trajectory effects may well exceed the contribution emerging from a consideration of the anisotropy of matrix elements of spin-orbit interaction in a ferromagnetic, which is usually of the order of a few percent [67]. Thus, the experimental investigation of the longitudinal and transverse resistance (relative to the intrinsic magnetic field) for different types of magnetic order in a multilayer can provide information about the nature of interaction of charge carriers with the boundaries between layers.

## 7. BALLISTIC PHENOMENA IN THIN LAYERS OF A NORMAL METAL ADJACENT TO A SUPERCONDUCTOR.

As it was shown by Andreev [70], the reflection of charge carriers by the boundary between the normal ($n$) and superconducting ($s$) phases is accompanied by a reversal of the charge and the velocity of excitations if its energy $\xi(\mathbf{p}) = \varepsilon(\mathbf{p}) - \varepsilon_F$, is less than the gap in the superconductor. A lot of experimental results that was obtained to date confirm the existence of Andreev reflection. The most detailed information on the type of the interaction of quasiparticles with the $n-s$ interface can be obtained using effects, in which the principal role is played by a select group of charge carriers. These carriers incident on the boundary between the phases at a definite angle. Andreev reflection was directly observed, for example, using the radio frequency (RF) size effect [71], the transverse electron focusing [72] and the geometric resonance of the ultrasonic attenuation [73].

In this chapter, we consider the high-frequency (HF) impedance of a thin normal metal layer, which is bounded on one side by vacuum and on the other by the superconductor. The layer thickness $d$ is much less than the electron mean free path $l$ but substantially larger than the skin-layer depth $\delta$. This corresponds to the geometry used in the experiment by Krylov and Sharvin [73] (Fig.12). They have noted that at $d = r$ ($r$ is the characteristic Larmor radius) the carriers returning to the skin layer after one Andreev reflection (trajectory $b$ in Fig.12) contribute to a screening current of opposite sign. This is manifest by an additional RF size-effect line in a magnetic field

$$H = H_0 = \frac{cp_F}{ed}.$$

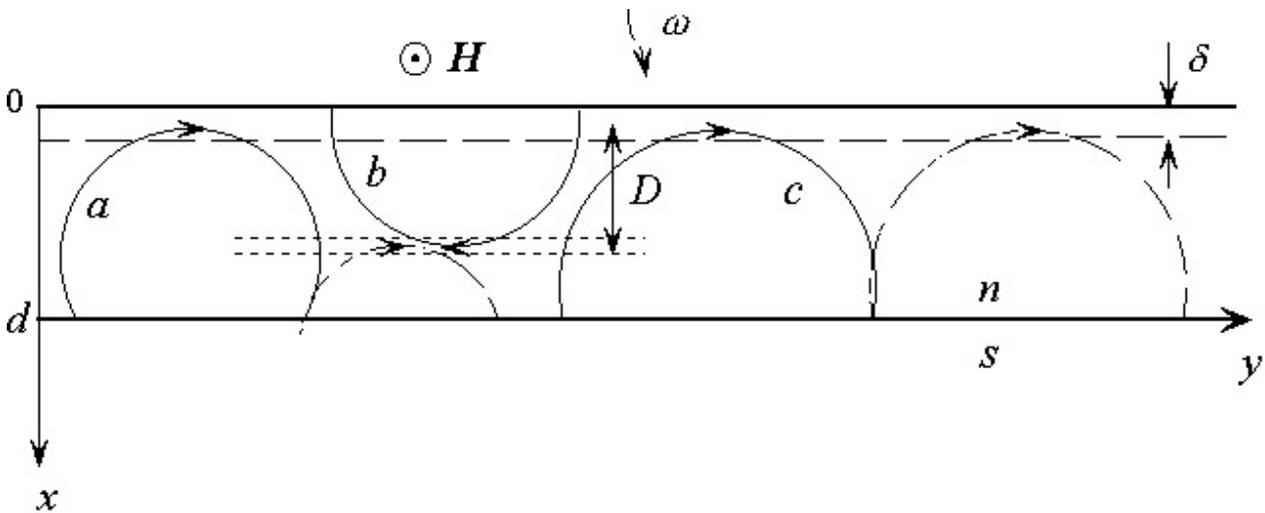

**Figure 12.** Onset of a HF field spike in a normal-metal layer at a depth $D$ in the case of Andreev reflection of carriers by the $n-s$ interface $x = d$. The electron and hole trajectories are designated by solid and dashed curves, respectively.

We have analyzed in details the effect of a change of the surface current on the layer impedance [74]. It was found that when the carriers are diffusely reflected by the sample boundary, the effect of the cutoff of the electron orbits with radius $r > d$ decrease the number of charges that efficiently interact with the RF wave.

A number of effects due to the Andreev reflection are possible in the microwave region. If the thickness of the normal layer satisfies the condition $r < d < 2r$, the period of the motion of the electrons that interact with the $n-s$ interface (trajectory $a$ in Fig.12) coincides with the Larmor period $T_H$ of the electron motion. Hence, the resonance frequencies of the external electromagnetic field $\omega = n\Omega$, $(\Omega = 2\pi/T_H)$ are the same as in the bulk conductors [51]. It is easily seen that the same carriers produce in the layer a narrow HF field spike, similar to the spikes produced in the bulk of the normal metal in the magnetic field parallel to the surface [16,45,47, 48].

If the electron reflection by the external surface $x = 0$ is close to specular, the electrons moving along this surface can return periodically, at a frequency $\Omega_0 = 2\pi/T_0$, to the spike (trajectory $b$ in Fig.12) and their interaction with the electromagnetic field in the spike is resonant. In the magnetic field $H < H_0$, when the Larmor radius exceeds the layer thickness, $r > d$, the situation depends essentially on the type of the electron interaction with the external boundary. At the diffusive reflection, the lines of the cyclotron resonance are cut off in fields $H \approx H_0$. At the specular reflection, the positions of the resonance lines, relative to the magnetic field scale, change because of the dependence of the period of the motion of the effective charges on the layer thickness $d$. Since the resonance line is formed by carriers with extremal periods of the motion, an investigation of the impedance in the microwave region yields detailed information on the Andreev reflection of quasiparticles belonging to select cross-sections of the Fermi surface.

We consider below a situation, in which a HF field spike does not come close to the $n-s$ boundary. Consequently, the amplitude of the electromagnetic wave at the superconductor boundary is small. In this case, the influence of the superconductor on the total impedance of the sample is connected mainly with the change of the dynamics of the electrons in the normal-metal layer, which collide with the boundary $x = d$ and interact with the HF field in the skin layer. Therefore at $d - D > \delta$ ($D$ is the distance from the HF field spike to the surface $x = 0$) there is no need to solve the microscopic problem of the penetration of the electromagnetic field into the superconductor. It suffices to take into account the presence of Andreev reflection from the plane $x = d$.

*7.1 Solution of Kinetic Equation*

Although the Andreev reflection is essentially a quantum effect, the motion of an excitation in an interval between two collisions with $n-s$ interface is quasiclassical [70]. The kinetic characteristics of the normal phase can be calculated by using the Boltzmann equation (2.1) for the increment

$$-\nu\left(\frac{\partial f_0^{\nu}}{\partial \xi}\right)\Psi(\mathbf{p},\mathbf{r},t_0)\,, \tag{7.1}$$

to the equilibrium distribution function

$$f_0^{\nu}(\xi)=\frac{1}{2}\left(1+\nu\quad th\left(\frac{\xi}{2T}\right)\right), \tag{7.2}$$

of the electrons $(\nu=-1)$ and of the holes $(\nu=1)$. The boundary condition that connects the distribution functions $\Psi$ of the incident and reflected carriers on the external surface of the sample can be written in the form (2.15).

The condition for the reflection of excitations at the $n-s$ boundary $x=d$ has the form [75]

$$\Psi^{ref}(d,\mathbf{p})=-Q_A\Psi^{inc}(d,-\mathbf{p})\,, \tag{7.3}$$

where $Q_A$ has the significance of the probability of Andreev reflection of carriers. At $Q_A=1$ the condition for Andreev scattering from the $n-s$ interface $x=d$ corresponds to the free flow of the current through the interface.

We consider the absorption of the monochromatic waves with the frequency $\omega$. In these cases the function $\Psi_{\mathbf{p}}(\mathbf{r},t_0)$ can be written in the form:

$$\Psi_{\mathbf{p}}(\mathbf{r},t_0)=\Psi_{\mathbf{p}}(\mathbf{r})\exp(-i\omega t_0)\,. \tag{7.4}$$

The solution (2.9) of the kinetic equation (2.1) contains an arbitrary function $F(\mathbf{r}-\mathbf{r}(t))$ of the characteristics. The boundary conditions (2.15) and (7.3) enable us to obtain an explicit expression for the function $F$.

In a magnetic field parallel to the external boundary and to the interface, the carrier motion in a plane perpendicular to the vector $H$ is periodic. The conditions (2.15) and (7.3) lead to a system of linear equations for the functions $F$, corresponding to motion along one of the segments of the trajectories (Fig.13) between two successive reflections. For carriers interacting only with the interface (Fig. 13a) we have

$$F_i=A_0Q_A\left[Q_A\alpha_k\varphi_i-\varphi_k\right],\qquad i,k=1,2;\qquad i\neq k\,, \tag{7.5}$$

where

$$A_0^{-1}=1-Q_A^2\alpha_i\alpha_k\,; \tag{7.6}$$

$$\alpha_i=\exp\left\{i\omega^*(\lambda_i'-\lambda_i)\right\}\,; \tag{7.7}$$

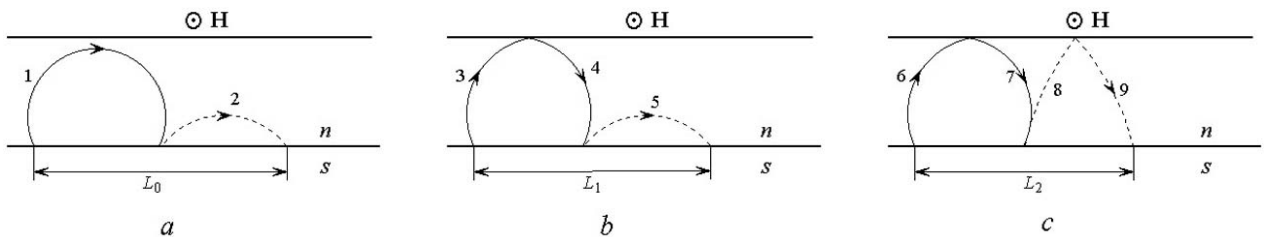

a                                    b                                    c

**Figure 13**  Possible types of periodic trajectories of carriers interacting with *n-s* boundary. The numbers 1-9 indicate, which of the function $F_i$ (Eqs. (7.6), (7.10), (7.11) corresponds to a given segment of the trajectory between two successive collisions of the carrier with layer boundaries.

$$\varphi_i = \int_{\lambda_i}^{\lambda_i'} dt' g(x + x(t') - x(t, \mathbf{p})) \alpha(\lambda_i', t') \ , \tag{7.8}$$

$$g(x, \mathbf{p}) = e\mathbf{E}(x)\mathbf{v}(\mathbf{p}). \tag{7.9}$$

Here $\omega^* = \omega + i/\tau$; $t$ is the "time" of the electron motion along a trajectory in the magnetic field; $\lambda_i$, and $\lambda_i'$ are the instants of two successive collisions of the quasiparticle with the boundaries $\lambda_i < \lambda_i'$.

For the carriers that interact with both boundaries of the normal layer and collide twice with the $n - s$ boundary after specular reflection from the surface $x = 0$ (Fig.13b), the function $\mathrm{E}_2(k)$ assumes three values:

$$F_3 = A_1 Q_A \left[ \alpha_5 Q_A \left( \varphi_4 + q_1 \alpha_4 \varphi_3 \right) - \varphi_5 \right];$$

$$F_4 = A_1 \left[ Q_A \alpha_3 \left( Q_A \alpha_5 \varphi_4 - \varphi_5 \right) + \varphi_3 \right]; \tag{7.10}$$

$$F_5 = A_1 Q_A \left[ q_1 \alpha_4 \left( Q_A \alpha_3 \varphi_5 - \varphi_3 \right) - \varphi_4 \right];$$

$$A_1^{-1} = 1 - q_1 Q_A^2 \alpha_3 \alpha_4 \alpha_5 \ .$$

If the carriers do not collide twice in succession with the same surface of the metallic layer (Fig.13c), $F_i$ takes on four different values:

$$F_i = A_2 Q_A \left\{ q_1' \alpha_m \left[ Q_A \alpha_l \left( \varphi_k + q_1 \alpha_k \varphi_i \right) - \varphi_l \right] - \varphi_m \right\};$$

$$F_k = A_2 q_1 \left\{ \varphi_i - Q_A \alpha_i \left[ \varphi_m - q_1' \alpha_m \left( Q_A \alpha_l \varphi_k - \varphi_l \right) \right] \right\}; \tag{7.11}$$

$$A_2^{-1} = 1 - q_1 q_1' Q_A^2 \alpha_6 \alpha_7 \alpha_8 \alpha_9; \qquad q_1' \equiv q_1 (\varepsilon_0 - p_z, \lambda_m) \ ;$$

$$i, l = 6, 8; \quad k, m = 7, 9; \quad i \neq l; \quad k \neq m; \quad k - i = 1; \quad m - l = 1,$$

where $q_1$ is a specular parameter of the external surface; $p_z$ is the projection of the momentum on the direction of the magnetic field. We shall not present here for the function $F$ the well known value corresponding to quasiparticles that move periodically along the metal surface (see Eq. (5.4)). The distribution function of the electrons in the bulk of the layer can be obtained from (2.9), in which we must put $\lambda = -\infty$. Below for simplicity we assume $Q_A = 1$.

*7.2 "Spike" of HF field in the normal layer*

Let us consider the monochromatic electromagnetic wave incidents on the metal surface. Maxwell equation in the Fourier representation

$$k^2 \mathrm{E}_\mu(k) + 2E'_\mu(0) - 2kE_\mu(d)\sin kd - 2E'_{\mu(}(d)\cos kd = \frac{4\pi i\omega}{c^2} j_\mu(k),\qquad(7.12)$$

where

$$\begin{Bmatrix} \mathrm{E}_\mu(k) \\ j_\mu(k) \end{Bmatrix} = 2\int_0^d dx \cos kx \begin{Bmatrix} E_\mu(x) \\ j_\mu(x) \end{Bmatrix},\qquad(7.13)$$

is an integral equation, since the relation between the Fourier components of the current $j_\mu(k)$ and of the electric field $E_\mu(k)$ is non-local:

$$j_\mu(k) = \int_0^\infty K_{\mu\nu}(k,k')\mathrm{E}_\nu(k')dk',\qquad \mu,\nu = y,z.\qquad(7.14)$$

The relation between the field $E_\mu(d)$ and its derivative $E'_\mu(d)$ is determined from the solution of the boundary-value problem at $x = d$. However, in the approximation in the small parameter $\delta/d \ll 1$, which will be considered below, the impedance terms containing $E_\mu(d)$ and $E'_\mu(d)$ are small and can be omitted. Solution of the kinetic equation yields the high-frequency conductivity tensor $K_{\mu\nu}(k,k')$, which is the kernel of the integral operator that relates $j_\mu(k)$ and $E_\mu(k)$. In the anomalous skin effect, when $\delta \ll (r,d) \ll l$, the significant values are $k \approx \delta^{-1}$ and to determine the surface-impedance tensor $Z_{\mu\nu}$ it suffices to know the asymptotic expression for $K_{\mu\nu}(k,k')$ at large $k$ and $k'$. In addition, we shall assume that the time of flight of the carriers through the narrow skin layer is substantially shorter than its effective free path time $\left|1/\omega^*\right|$, i.e., that the inequality (5.9) holds.

*1) The carriers reflection by the surface is close to specular,* i.e. the momenta of effective electrons satisfy to relation:

$$1 - q_1 \ll \left|\omega^*/\Omega\right|\left(\delta/r\right)^{1/2}.\qquad(7.15)$$

At $\delta \ll r$ this condition is not too stringent, since for electrons that do not leave the skin layer the angles of approach to the surface is of the order of $\left(\delta/r\right)^{1/2} \ll 1$. Inasmuch as at small $\alpha$ the diffuseness parameter $1 - q_1(\alpha) \approx q'_1(0)\alpha$, the forgoing inequality is equivalent to

$$q'_1(0) \ll \left|\omega^*/\Omega\right|.\qquad(7.16)$$

In this case the HF conductivity tensor can be represented by a sum of four terms:

$$K_{\mu\nu}(k,k') = K_n(k,k') + K_{ns}^{(1)}(k,k') + K_{ns}^{(2)}(k,k') + \tilde{K}_n(k,k'). \tag{7.17}$$

The kernel $K_n(k,k')$ is connected with the carriers that glide over the surface of the metal and make the main contribution to the formation of the screening HF current:

$$K_n(k,k') = \frac{k_0^{5/2}}{\beta(kk')^{1/2}} \left\{ \left\langle \alpha\hat{S}\left(\frac{T_H}{2} - \tau_0, \frac{T_H}{2}; x\left(\frac{T_H}{2}\right)\right)f_n - \frac{1}{(k+k')^{1/2}} \right\rangle \right\}. \tag{7.18}$$

Here

$$\hat{S}(a,b;x)f = \frac{1}{\pi}\int_a^b d\lambda f'(\lambda)\frac{\sin\left[(k-k')(x-x(\lambda))\right]}{k-k'}; \tag{7.19}$$

$$\alpha^{\pm} = \beta\rho_{\mu\nu}^{(\pm)}\left(\frac{T_H}{2}\right)k_0^{-5/2}; \quad \beta = \frac{4\pi\omega}{c^2}; \tag{7.20}$$

$$\rho_{\mu\nu}^{(\pm)}(t,p_z) = \frac{4\pi e^3 H}{ch^3}\frac{v_\mu(t,p_z)\ v_\nu(t,\pm p_z)}{\left|v_x'(t,p_z)\ v_x'(t,\pm p_z)\right|^{1/2}}; \tag{7.21}$$

$$f_n(\lambda) = \left\{\exp\left[-i\omega^*(T_H - 2\lambda)\right] - q_1(\lambda)\right\}^{-1}, \tag{7.22}$$

and $\tau_0$ satisfies the condition

$$d - x(\tau_0) + x(0) = 0. \tag{7.23}$$

The value $k_0$ is determined by Eq. (5.12). The angle brackets $\langle ...\rangle$ in (7.18) denote integration along that strip on the Fermi surface on which $v_x = 0$. We assume that on the trajectory of the carriers that do not interact with the layer boundary there are only two stationary phase points $t = 0$ and $T_H/2$, where $v_x(0) = v_x(T_H/2) = 0$, while $v_x'(0) > 0$, and $v_x'(T_H/2) < 0$.

The terms $K_{ns}^{(1)}(k,k')$ and $K_{ns}^{(2)}(k,k')$ in the HF electric conductivity are due to the carriers that interact with the $n-s$ boundary. For quasiparticles whose radius of Larmor trajectory $r(p_z) < d < 2r(p_z)$ $(\tau_0 < T_H/4)$ we have

$$K_{ns}^{(1)}(k,k') = \left\langle \theta(d - r(p_z))\theta(2r(p_z) - d)\ \rho_{\mu\nu}^{(+)}(0).\right.$$

$$\left.\left[\frac{1}{(kk')^{1/2}}\hat{S}\left(0,\frac{T_H}{2} - \tau_0; x(\tau_0)\right)f_{ns} - \frac{2}{\pi^2}f_{ns}(\tau_0)\frac{\ln(k/k')}{k^2 - k'^2}\right]\right\rangle; \tag{7.24}$$

$$K_{ns}^{(2)}(k,k') = -\frac{2}{\pi}\left\langle \theta(d - r(p_z))\ \theta(2r(p_z) - d)\rho_{\mu\nu}^{(-)}(0)e^{i\omega^*T/2}f_{ns}(\tau_0)\frac{\cos kD - \cos k'D}{k^2 - k'^2}\right\rangle; \tag{7.25}$$

where

$$f_{ns}(\lambda) = \left[\exp\left(-i\omega * (T_H - 2\lambda')\right) - 1\right]^{-1}, \tag{7.26}$$

and $\lambda'(\lambda)$ satisfies the equation

$$d - x(T_H/2 - \lambda) + x(\lambda') = 0 \quad \left(\lambda'(T_H/2 - \tau_0) = 0\right), \tag{7.27}$$

where $\theta(x)$ is the Heaviside function and

$$D(\tau_0; p_z) = x(\tau_0; p_z) - x(T_H/2 - \tau_0; -p_z). \tag{7.28}$$

The electric conductivity of the carriers for which $r(p_z) > d \quad (\tau_0 < T_H/4)$ can be written in the form

$$K_{ns}^{(1)}(k, k') = \left\langle \theta\left(r(p_z) - d\right) \rho_{\mu\nu}^{(+)}(0) \left[\frac{1}{(kk')^{1/2}} \hat{S}(0, \tau_0; x(\tau_0)) f_{ns}^{(1)} - \frac{2}{\pi^2} f_{ns}^{(1)}(\tau_0) \frac{\ln k/k'}{k^2 - k'^2}\right]\right\rangle; \tag{7.29}$$

$$K_{ns}^{(2)}(k, k') = \frac{4}{\pi^2}\left\langle \theta\left(r(p_z) - d\right) \rho_{\mu\nu}^{(-)}(0) \ e^{i\omega^* T_H/2} f_{ns}^{(1)}(\tau_0) \frac{C(kD) - C(k'D)}{k^2 - k'^2}\right\rangle; |kd|, |k'D| < 1; \tag{7.30}$$

$$K_{ns}^{(2)}(k, k') = 0; \ |kD|, \ |kD'| \gg 1, \tag{7.31}$$

where

$$f_{ns}^{(1)}(\lambda) = \left[\exp\left(-i\omega * \left(T_H - 2\lambda'(\lambda) - 2\lambda'(T_H/2 - \lambda)\right)\right) - 1\right]^{-1}; \tag{7.32}$$

$$C(kD) = \int_1^\infty \frac{dt}{t^2 - 1} \cos\left(kD \ (t^2 - 1)\right). \tag{7.33}$$

The term $\tilde{K}_n(k, k')$ takes into account the contribution made to $K_{\mu\nu}(k, k')$ by the electrons which do not collide with the $n-s$ boundary and whose orbit diameter is $2r(p_z) < d$.

*2) The reflection of the carriers by the metal surface is substantially different from specular,* i.e. a considerable part of the charges that make the main contribution to the high-frequency current satisfy the inequality

$$1 - q_1 \gg \left|\omega^*/\Omega\right|(\delta/r)^{1/2}. \tag{7.34}$$

In this case the contribution made to the HF electric conductivity by the subsurface electrons is small, and the kernel $K_{\mu\nu}(k, k')$ of integral operator (7.14) is determined mainly by the charges that interact with the $n-s$ boundary:

$$K_{\mu\nu}(k, k') = \left\langle \theta\left(d - r(p_z)\right) \theta\left(2r(p_z) - d\right) \rho_{\mu\nu}^{(+)}(0) \tilde{f}_{ns} \frac{1}{\pi(kk')^{1/2}} \left[\frac{\sin(k - k')D}{k - k'} - \frac{1}{k + k'}\right]\right\rangle -$$

$$\frac{1}{\pi}\left\langle \theta\big(d-r\big(p_z\big)\big)\theta\big(2r\big(p_z\big)-d\big)\rho^{(-)}_{\mu\nu}\big(0\big)\tilde{f}_{ns}\frac{1}{\big(kk'\big)^{1/2}}\left[\frac{\cos kD-\cos k'D}{k-k'}+\right.\right.$$

$$\left.\left.\frac{\sin kD+\sin k'D-\cos\big(k-k'\big)D}{k+k'}\right]\right\rangle-\frac{1}{2}\left\langle\theta\big(r\big(p_z\big)-d\big)\rho^{(+)}_{\mu\nu}\big(0\big)\left[\frac{1}{\big(kk'\big)^{1/2}}\hat{S}\big(0,\tau_0;x\big(\tau_0\big)\big)e^{2i\omega^*\lambda}\right.\right.$$

$$\left.\left.-\frac{2}{\pi^2}e^{2i\omega^*\lambda}\frac{\ln k/k'}{k^2-k'^2}\right]\right\rangle+\tilde{K}\big(k,k'\big),\qquad(7.35)$$

where

$$\tilde{f}_{ns}=\big(1/2\big)\cot\big(\omega^*T/2\big).\qquad(7.36)$$

To calculate the surface impedance

$$Z=\frac{4i\omega}{c^2}\frac{1}{E'(0)}\int\limits_0^\infty dk\mathrm{E}(k),\qquad(7.37)$$

of the normal metal layer bordering on a superconductor it is necessary to find the solution of Maxwell equation for the Fourier components of the electric field $E(k)$. To avoid lengthy equations we assume hereafter that the electric vector of the linearly polarized wave is directed along one of the axes for which the tensor $K_{\mu\nu}(k,k')$ is diagonal, and for simplicity, we shall omit the tensor indices.

If the electron reflection by the surface is close to specular, i.e. the momenta of effective electrons satisfy to Eq.(2.15), mainly electrons that do not leave the skin layer during the mean free time produce a large surface current. This permits a perturbation-theory solution of the Maxwell equation. Starting from the structure of the kernel $K(k,k')$ (7.17), it is convenient to seek the solution of Eq.(7.13) in the form of the sum

$$\mathrm{E}(k)=\mathrm{E}^*_0(k)+\Delta\mathrm{E}_0(k)+\mathrm{E}_1(k)+\mathrm{E}_2(k),\qquad(7.38)$$

where $\mathrm{E}^*_0(k)$ is the Fourier component that describes the field of the main skin layer (see, Eq.(5.35)), $\Delta\mathrm{E}_0(k)$ is a small increment to $\mathrm{E}^*_0(k)$ due to the carriers that undergo Andreev reflections and can resonantly interact with the HF wave, the function $\mathrm{E}_1(k)$ is responsible for the formation of the HF field spike at the depth $D$, and $\mathrm{E}_2(k)$ is a small addition that takes into account the influence of the spike on the field in the skin layer.

We consider first cylindrical Fermi surface with the axis coinciding with the magnetic field direction. In this case, which is apparently close to the conditions of the experiment [71], we have

$$E_1(x) = \frac{2i}{\pi} \frac{E'(0)}{k_0^2} \left(\frac{\eta}{k_1}\right)^3 \int_0^\infty \frac{dk}{\left(k/k_1\right)^3 - i} \left\{ F_0(k)\sin(k(D-x)) + \hat{G}F_0(k)\cos(k(D-x)) \right\}; \quad (7.39)$$

where

$$\eta^3 = \beta\rho(0)\exp\left(i\omega^* T_H/2\right) f_{ns}(\tau_0) b_z; \quad (7.40)$$

$$k_1^3 = \beta \quad \rho(0)\left[f_n(\tau_0') + f_{ns}(\tau_0)\right] b_z; \quad (7.41)$$

Here $\rho(t) = \rho_{yy}^+(t, p_z = 0)$ (see Eq.(7.21)); the functions $F_0(k)$ and $\hat{G}F_0(k)$ in Eq.(7.39) is

defined by Eqs.(5.34) and (5.38); $b_z/h$ is the period of the reciprocal lattice in the direction of the

$z$ axis, and $\tau_0'$ satisfies the equation

$$x\left(T_H/2\right) - x(\tau_0') - D(\tau_0) = 0, \quad (7.42)$$

and determines the period $T_0 = T_H - 2\tau_0'$ of the motion of the electrons that return to the spike after specular reflection from the surface $x = 0$.

The amplitude of the spike has the maximum near the center $(|D - x| \approx |k_0|^{-1})$, where it is equal

$$E_1(D) = -\frac{E'(0)}{k_0} \frac{1}{3\sqrt{2}} \left\{ \frac{\eta}{\sqrt{k_0 k_1}} \right\}^3 \exp\left(i\pi/4\right). \quad (7.43)$$

If the spike (7.39) is far from the surface $(|k_0 D| \gg 1)$, the influence of the carriers that interact with the $n - s$ boundary is described by the additions $\Delta\mathrm{E}_0(k)$, and $\mathrm{E}_2(k)$ to the function $\mathrm{E}_0^*(k)$. The corresponding corrections to the impedance $Z_0$ (5.48) can be found by perturbation theory. Thus,

$$\Delta Z_0 = \frac{4i\omega}{c^2} \frac{1}{E'(0)} \int_0^\infty dk \Delta\mathrm{E}_0(k) = C_1 Z^* \left(\frac{\eta_1}{k_0}\right)^3, \quad (7.44)$$

where

$$C_1 = 1.99 \cdot 10^{-2} e^{4\pi i/5}; \quad Z^* = \frac{8\omega}{c^2 k_0}; \quad \eta_1^3 = \beta\rho(0) f_{ns}(\tau_0) b_z. \quad (7.45)$$

The impedance correction because of the influence of the spike on the main skin layer appears in

second-order perturbation theory:

$$\Delta Z_2 = \frac{4i\omega}{c^2} \frac{1}{E'(0)} \int_0^\infty dk \mathrm{E}_2(k) = C_2 Z^* \left(\frac{\eta^3}{k_1 k_0^2}\right)^2; \quad (7.46)$$

$$C_2 = \frac{\pi}{48\sqrt{3}} \exp\left(i\frac{5\pi}{6}\right).$$

It is easily noted that $\Delta Z_2$ in $|k_0 r|^{1/6} \gg 1$ times smaller than $\Delta Z_0$ in the RF region $(\omega \ll \Omega)$.

In a narrow range of magnetic fields, in which the spike is near to the metal surface, its effect on the impedance is described by the terms $E_1(k)$. In this case, we obtain

$$\Delta Z_1 = A(k_0 D) Z^* \left( \frac{\eta}{k_0} \right)^3, \qquad (7.47)$$

where

$$A(k_0 D) \approx \begin{cases} -\dfrac{2}{\pi} \displaystyle\int_0^\infty d\xi \int_0^\infty d\xi' \, \dfrac{\cos(k_0 D\xi) - \cos(k_0 D\xi')}{\xi^2 - \xi'^2} F_0(\xi) F_0(\xi'), & |k_0 D| \leq 1, \ \tau_0 \geq T/4; \\[3mm] \dfrac{4}{\pi^2} \displaystyle\int_0^\infty d\xi \int_0^\infty d\xi' \, \dfrac{C(k_0 D\xi) - C(k_0 D\xi')}{\xi^2 - \xi'^2} F_0(\xi) F_0(\xi'), & |k_0 D| \leq 1, \ \tau_0 \leq T/4; \\[3mm] 0, & |k_0 D| \gg 1. \end{cases} \qquad (7.48)$$

Analysis of Eq.(7.48) shows $\arg A = \dfrac{4\pi}{5} + \dfrac{\pi}{2} \left[ 1 + sign\left( \tau_0 - T/4 \right) \right]$ changes by $\pi$ at $H = H_0$. The absolute value $|A|$ is of the same order as the absolute value of the constant $C_1$ if $|k_0 D| \approx 1$, and $|A| \ll C_1$ at $|k_0 D| \ll 1$ or $|k_0 D| \gg 1$. It follows from (7.48) that $A(0) = 0$. Consequently, for a cylindrical Fermi surface in magnetic fields $H \approx H_0$, satisfying the condition $r \cong d$, the addition $\Delta Z_1$ due to the spike approaching to the surface changes by an amount of the same order as $\Delta Z_0$. We note that in the case of specular reflection of the electrons from the metal surface the change of the impedance at $H = H_0$ is small compared with the main term $Z_0$.

If the surface scattering is diffuse, the current in the skin layer is made up of charges that collide with the *n-s* boundary and produce at $r < d < 2r$ $(d - r > \delta, \quad 2r - d \gg \delta)$ the HF field spike at a depth $D$. As the spike approaches the surface $(D \approx \delta)$, the surface current decreases just as it does in the case of the specular reflection from the surface $x = 0$. At $q_1 = 0$, however, what is more important is that in fields $H \leq H_0$ there are no carriers that return to the skin layer after Andreev reflections (the cutoff effect). At $kD$, $k'D \gg 1$ Maxwell equation (7.13) reduces to the integral equation solved in Ref. 76. By a standard calculation procedure, we find that the expression for the surface impedance in the case of the diffuse scattering by the surface is of the form

$$Z = \dfrac{8\pi\omega}{c^2} \begin{cases} \dfrac{1}{\sqrt{3}} e^{-i\pi/3} \dfrac{1}{k^*}, & H > H_0, \quad (d - r > |k^*|^{-1}), \\[3mm] \dfrac{\sqrt{3}}{4} \dfrac{1}{k_1^*}, & H \leq H_0, \end{cases} \qquad (7.49)$$

$$\left(k^*\right)^3 = \beta\rho(0)f_{ns}^* b_z; \qquad \left(k_1^*\right)^3 = \frac{1}{2}\beta\rho(0)e^{2i\omega^*\tau}b_z. \qquad (7.50)$$

In Eqs.(7.49) the cubic-root branch $k^*$ and $k_1^*$, must be chosen to satisfy the condition $\mathrm{Re}\,Z > 0$. When the Larmor radius becomes equal to the layer thickness $r(H_0) = d$, the cutoff of the periodic trajectories of the carriers takes place, and in the RF region ($\omega \ll \Omega$) the impedance of a conductor with a cylindrical dispersion law decreases by a factor $\frac{l}{r}$.

If the Fermi surface has no cylindrical sections, the HF field spike in the normal-metal layer is formed only by the small group of carriers located near the sections $p_z = p_{ze}$ the Fermi surface $\varepsilon(\mathbf{p}) = \varepsilon_F$. These sections correspond to the extrema $D(\tau, p_z)$, as functions of $p_z$. In this case, in magnetic fields, in which $r(p_{ze}) - d \leq \delta$, the contribution to the surface current is compensated for a small fraction $\left(\sim \left(\delta/r\right)^{1/2}\right)$ of all the charges entering the skin layer at small angles. Consequently, the change of the impedance is proportional to the small parameter $\left(\delta/r\right)^{1/2} \ll 1$. For the same reason the cutoff of the extremal radius $r(p_{ze}) = d$ at diffuse scattering by the metal surface does not lead to an abrupt decrease of the impedance, which receives contributions from carriers with all possible values of $p_z$.

*7.3 Resonance Phenomena*

More information can be obtained in investigations of the Andreev reflection from the high-frequency characteristics of the layer under resonance conditions. The reason is that the resonant singularities in the impedance are produced by select groups of carriers with extremal period of motion in a magnetic field. As follows from (7.24) and (7.27), in magnetic fields such that the radius $r(p_{z1})$ of the effective orbit is less than the thickness of the metal layer in the normal state but such that $d < 2r(p_{z1})$, a number of resonance lines is produced, which coincide with the cyclotron-resonance line in a bulk conductor [51]:

$$\omega = n\Omega(p_{z1}), \qquad \frac{\partial T_H}{\partial p_z}\Big|_{p_z = p_{z1}} = 0, \qquad n = 1, 2, 3\ldots \quad . \qquad (7.51)$$

At almost-specular reflection of the charges by the metal surface and at small detuning from the resonance $\omega T_H(p_{z1}) = 2\pi n(1-\Delta)(|\Delta| \ll 1)$, the resonant increment $\Delta Z_{0res}$ to the surface impedance takes the form

$$\Delta Z_{res} = \Delta Z_0^{res} + \Delta Z_1^{res}; \qquad (7.52)$$

$$\Delta Z_0^{res} = C_1 Z^* \rho^{(+)}(0) \beta \Psi(\Delta, T_H, \gamma)\big|_{p_z = p_{z1}} \; ; \tag{7.53}$$

$$\Delta Z_1^{res} = Z^* \rho^{(-)}(0) \begin{cases} (-1)^n A(k_0 D) \beta \Psi(\Delta, T_H, \gamma)\big|_{p_z = p_{z1}}, & 0 < |k_0 D| \leq 1; \\ \dfrac{C_2 \rho^{(-)}(0)}{k_0^2} \dfrac{(\beta \Psi(\Delta, T_H, \gamma))^{4/3}}{\left[\rho^{(+)}(0)\right]^{2/3}}\big|_{p_z = p_{z1}}, & |k_0| D \gg 1. \end{cases} \tag{7.54}$$

where

$$\Psi(\Delta, T_H, \gamma) = \frac{1}{2n\sqrt{2\chi}} \frac{\sqrt{\xi + s\Delta} + is\sqrt{\xi - s\Delta}}{\xi} \; ; \tag{7.55}$$

$$\chi = \frac{1}{2T_H} \frac{\partial^2 T_H}{\partial p_z^2}; \qquad s = sign\,\chi; \qquad \xi = \sqrt{\Delta^2 + \gamma^2}; \qquad \gamma = \frac{1}{\omega\tau}. \tag{7.56}$$

Besides the "bulk" cyclotron resonance, in a thin layer is possible to have a unique resonance due to the motion the electrons that land periodically in the field of spike (7.39) after specular reflections from the surface $x = 0$. The resonant frequencies are given by the relation

$$\omega T_0(p_{z2}) = 2\pi n; \qquad T_0 = T - 2\tau_0'; \qquad \frac{\partial T_0}{\partial p_z}\big|_{p_z = p_{z2}} = 0, \tag{7.57}$$

and the increment to the impedance (7.46), which describes the resonance (7.57) at detuning $|\Delta| \ll 1$ ($\omega T_0 = 2\pi n(1 - \Delta)$), can be represented in the form:

$$\Delta Z_2^{res} = C_2 Z^* \left(\frac{\chi^3}{k_0^2}\right)^2 (\beta \rho^{(+)}(0) \Psi(\Delta, T_0, \gamma))^{-2/3}\big|_{p_z = p_{z2}}. \tag{7.58}$$

Resonant abortion of the energy of the HF field in magnetic fields that satisfy Eq.(7.57) recalls cyclotron resonance in a thin plate [52], first theoretically investigated by one of us. In magnetic fields such that $r(p_{z1}) \geq d$ the period of motion of the carriers colliding with two boundaries depends on the layer thickness $d$, and the positions of the resonance lines on the magnetic-field scale differ from the values given by the condition (7.57):

$$\omega T_1(p_{z3}) = 2\pi n; \qquad T_1 = T - 2\tau'; \qquad \frac{\partial T_1}{\partial p_z}\big|_{p_z = p_{z3}} = 0, \tag{7.59}$$

where $\tau' = \lambda'(\tau)$, and the dependence of $\lambda' = \lambda'(\lambda)$ should be obtained by using Eq.(7.27). The impedance increment that describes the resonance (7.59) has the form

$$\Delta Z_{res} = C_1 Z^* \beta \rho^{(+)}(0) \Psi(\Delta, T_1, \gamma)\big|_{p_z = p_{z3}}. \tag{7.60}$$

If the reflection by the surface of the sample is close to diffuse, the resonant dependence of the surface impedance on the magnetic field is preserved at $r(p_{z1}) < d < 2r(p_{z1})$ and is connected as before with the carriers undergoing Andreev reflection. Near the resonant frequencies (7.51), expression (7.50) for the impedance $Z$ takes the form

$$Z = \frac{2}{\sqrt{3}}\beta e^{-i\pi/3}\left\{\beta\rho^{(+)}(0)\Psi(\Delta,T_H,\gamma)\right\}^{-1/2}\Big|_{p_z=p_{z1}}. \tag{7.61}$$

At $d = r(p_z)$, cutoff of the lines of the cyclotron resonance (7.51) takes place in fields satisfying the condition $d = 2r(p_{z1})$.

In Eqs. (7.53), (7.54), and (7.61), which describe the behavior of the impedance near the resonance (7.51), we have assumed that the probability $Q_A$ of the Andreev reflection is equal to unity. If, however $Q_A < 1$, it can be taken into account by introducing an additional broadening of the resonance lines $(1-Q_A^2)\big/2\pi n$, i.e., $\gamma$ in Eqs. (7.53), (7.54), and (7.61) must be replaced by $\gamma' = \gamma + (1-Q_A^2)\big/2\pi n$.

Andreev reflection of carriers from $n-s$ boundary leads thus to an entirely different dependence of the surface impedance of a thin normal-metal layer on the magnetic field compared with the impedance of a thin metallic plate. At $r < d < 2r$, a narrow HF-field spike is produced inside the layer at a distance $D(H)$ from its surface. If the electrons are specularly reflected from the surface, the carriers gliding over the boundary and landing periodically in the spike, produce the resonance that is not observed in either bulk or thin conductors in the normal state. In the same magnetic-field range, at any electron scattering from the layer surface, resonance should be observed at frequencies corresponding to the cyclotron resonance [51]. In weak field $H$, at which $r \geq d$, the behavior of the impedance as a function of the magnetic field depends essentially on the state of the sample boundary. Thus, in the case of specular reflection the resonant dependence of $Z$ on $H$ is preserved, where as for diffuse scattering the cyclotron resonance vanishes for $r \geq d$. At radio frequencies in the magnetic field interval $|r-d| < \delta$ an abrupt change takes place in the contribution made to the HF current by the carriers that interact with the $n-s$ interface. This manifests itself in the onset of an RF size-effect line at $r = d$. Such a line is most intense when it is due to motion of excitations belonging to cylindrical parts of the Fermi surface.

An experimental investigation of the high-frequency properties of thin normal-metal layers bordering on superconductors makes it thus possible not only to observe directly Andreev reflection of carriers, but also to gauge its probability and the temperature dependence from the amplitude and width of the resonance lines.

### 7.4 Ultrasound attenuation

In this chapter, we consider the absorption of the ultrasound in a thin normal-metal layer on a superconducting substrate layer [77]. The thickness of the layer is much smaller than the carrier mean free path $l$, but considerably larger than the sound wavelength $\lambda_s$. A magnetic field $\mathbf{H}$ is parallel to the outer surface of the sample $x = 0$ and the $n - s$ boundary $x = d$ (Fig.14). We analyze the influence of the interaction of electrons with $n - s$ boundary on the dependence of the longitudinal-wave absorption coefficient on the magnitude of $\mathbf{H}$ in the cases when the wave vector $\mathbf{k} \perp \mathbf{H}$ and is directed along the normal to the metal layer $\mathbf{k} \parallel x$, or parallel to its surface ($\mathbf{k} \parallel y$). If the characteristic Larmor radius $r$ of trajectory of electrons satisfies the condition $r < d < 2r$ and if the vector $\mathbf{k} \parallel y$, geometrical resonance set in, with the period of its oscillation determined by the condition $kD_e = 2\pi n$ $(kD_e \gg 1)$, where $D_e$ is the extremal distance between points of the carrier trajectory tangent to the surface of the sample (Fig.14). In these points the velocity vector lies in a plane of the constant phase of the sound. At higher ultrasonic frequencies $\omega$, such $\omega\tau > 1$, the acoustic cyclotron resonance set in. In the range of magnetic fields $r < d < 2r$ the resonance set in at frequencies $\omega = n\Omega$, since the period of the motion of excitations undergoing the Andreev reflection coincides with Larmor period $T_H$.

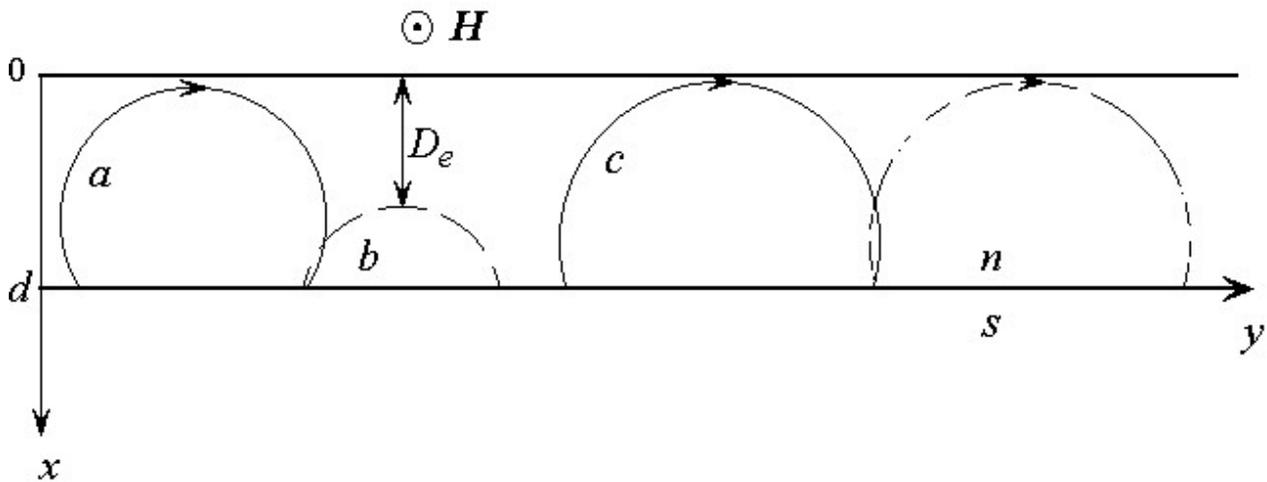

**Figure 14** Change carrier trajectory (electrons are represented by solid curves, and holes by dashed curves) generating geometrical oscillations of ultrasound absorption coefficient.

In the case of the perturbation of the electron system by the ultrasonic wave the nonequilibrium distribution function $\Psi_{\mathbf{p}}(\mathbf{r}, t_0)$ and the function $g(\mathbf{r}, \mathbf{p}, t_0)$ in the kinetic equation (2.1) takes the form

$$\Psi_{\mathbf{p}}(\mathbf{r}, t_0) = \Psi_{\mathbf{p}}(\mathbf{r}) \exp(i\mathbf{k}\mathbf{r} - i\omega t_0);$$ 
(7.62)

$$g(\mathbf{r}, \mathbf{p}, t_0) = \Lambda_{ik} \dot{u}_{ik} + e\mathbf{v}\left(\mathbf{E} + \frac{1}{c}[\mathbf{uH}]\right); \qquad (7.63)$$

$$u_{ik} = u_{ik}^{(0)} \exp(i\mathbf{kr} - i\omega t_0).$$

Here $\Lambda_{ik}$ is the deviation of the deformation potential from its mean value on the Fermi surface, and $u_{ik}$ is the amplitude of the strain tensor, $\mathbf{u}$ is the ion displacement. The induction electric field $\mathbf{E}$ produced by the transmission of the sound through the sample must be determined from the Maxwell equations with the inclusion of the nonequilibrium electric current. In the given case of a longitudinal sound wave, however, it has been shown [78] that allowance for the field $\mathbf{E}$ is of little consequence, and we shall restrict the right-hand side (7.63) of the kinetic equation (2.1) to the deformation term $\Lambda_{ik} \dot{u}_{ik}$ only.

Substituting the functions (7.62), (7.63) into Eq.(2.1) we can present its solution in the form:

$$\Psi_{\mathbf{p}}(\mathbf{r}) = F(x - x(t)) \exp(i\omega^*(t - \lambda) - i\mathbf{k}(\mathbf{r}(t) - \mathbf{r}(\lambda))) +$$
$$\int_{\lambda}^{t} dt' g(\mathbf{p}, x + x(t') - x(t)) \exp(i\omega^*(t - t') - i\mathbf{k}(\mathbf{r}(t) - \mathbf{r}(t'))). \qquad (7.64)$$

Function $F$ must be determined by means of boundary conditions (2.15) and (7.3) and takes the finite number of values $F_i$ corresponding to motion along one of arcs of the trajectories (Fig.13) between two successive reflections. Values $F_i$ have the same form as Eqs.(7.5), (7.10) and (7.11), if to replace the functions $\alpha_i$ (7.7) and $\varphi_i$ (7.8) by functions

$$\alpha_i = \exp\left(i\omega^*(\lambda_i' - \lambda_i) - i\mathbf{k}(\mathbf{r}(\lambda_i') - \mathbf{r}(\lambda_i))\right); \qquad (7.65)$$

$$\varphi_i = \int_{\lambda_i}^{\lambda_i'} dt' g(\mathbf{p}, x + x(t') - x(t)) \exp(i\omega^*(\lambda_i' - t') - i\mathbf{k}(\mathbf{r}(\lambda_i') - \mathbf{r}(t'))). \qquad (7.66)$$

Knowing the nonequilibrium increment to the distribution function of the excitations, we can calculate the sound attenuation coefficient in a thin layer of a normal metal on a superconducting substrate:

$$\Gamma = \mathrm{Re}(2\rho s \left|\dot{\mathbf{u}}\right|^2 d)^{-1} \int_0^d dx \langle g^* \Psi \rangle \equiv \mathrm{Re} \frac{eH}{ch^3 \rho s \left|\dot{\mathbf{u}}\right|^2 d} \iint_{\varepsilon(\mathbf{p}) = \varepsilon_0} dt dp_z \int_0^d dx g^* \Psi. \qquad (7.67)$$

Here $\rho$ is the density of the crystal, $\mathbf{u}$ is the velocity of the atoms under the action of the sound wave. To avoid cumbersome expressions in the ensuing discussion, we shall assume that the Fermi surface represents a body of revolution with its axis parallel to the $p_z$ axis.

*a) The sound propagating perpendicular to the normal-metal layer.*

Making use of the fact that an electron interacts effectively with ultrasound in the vicinity of points with $\mathbf{k v} = \omega$, we calculate the integral with respect to $t$ in Eq.(7.67) by the stationary-phase method. As a result, we obtain the following for the sound absorption coefficient $\Gamma_{ns}$ associated with Andreev-reflected electrons in the principal approximation with respect to the parameter $\lambda_s \big/ r \ll 1$,

$$\Gamma_{ns} = \Gamma_0 \operatorname{Re} \frac{1}{d} \int \frac{dp_z}{m\, v_\perp} \int_0^{T_H} d\lambda\, v_x(\lambda) \left\{ \frac{S_1(\lambda, p_z)}{1 - Q_A^2 e^{i\omega^* T_H}} + \frac{S_2(\lambda, p_z)}{1 - q_1 Q_A^2 e^{i\omega^*(T_H - 2\lambda)}} \right\}, \qquad (7.68)$$

where $\Gamma_0$ is of the same order of magnitude as the value of $\Gamma$ for a bulk sample in the absence of a magnetic field:

$$\Gamma_0 \cong \frac{4\pi}{h^3 \rho s k \left| \overset{\cdot}{\mathbf{u}} \right|^2} \frac{\left( \langle m | g | \rangle \right)^2}{\langle 1 \rangle} \approx \frac{n_0 \varepsilon_F \omega}{\rho s v_F}, \qquad (7.69)$$

$n_0$ and $m$ are the density of electrons and their effective mass, $v_\perp$ is the projection of the electron velocity onto the plane $p_z = const$; $\lambda$ is the "instant" of a last collision of a carrier with the boundary of the layer. The values of $\lambda$ are measured relative to the stationary-phase point $\lambda = 0$ ($k v_x(0) = \omega$). The unit function $S_\gamma(\lambda, p_z)$ differs from zero for the values of $\lambda$, and $p_z$ for which the charge moves along trajectories of the type $a$ ($\gamma = 1$), $b$ ($\gamma = 2$) or $c$ ($\gamma = 3$) in Fig. 13.

In the case of low-frequency sound, $\omega \tau \ll 1$, the sound field is practically constant during the mean free transit time, and $i\omega^*$ can be set equal to $-\frac{1}{\tau}$ in Eq. (7.68). Making use of the fact that the specular reflection parameter $q_1(\varphi) \approx 1 - |q_1'(0)| \varphi$ for small angles $\varphi$ of incidence of electrons on the sample surface [79], and integrating with respect to $\lambda$ and $p_z$ in Eq.(7.68), we obtain the following asymptotic forms for $\Gamma_{ns}$:

$$\Gamma_{ns} \cong \begin{cases} \Gamma_0, & r \gg l^2 \big/ d, \\ \Gamma_0 \left( r \big/ d \right)^{1/2} \left\{ |q_1'(0)| + r \big/ l \right\}^{-1}, & d \ll r \ll l^2 \big/ d, \\ \Gamma_0\, l \big/ d, & r \leq d, \end{cases} \qquad (7.70)$$

where $r = c p_{\perp extr} \big/ eH$ and numerical factors of the order of unity are omitted. A comparison of Eqs.(7.70) with the results of an investigation [80] of the absorption of ultrasound in a thin normal-metal plate shows that the presence of the $n-s$ boundary does not alter the dependence of the monotonic part of the absorption coefficient $\Gamma$ on the magnetic field.

In fields **H** for which the thickness of the normal layer satisfies the inequality $r < d < 2r$, the quantity $\Gamma_{ns}(H)$ experiences geometrical oscillations, which are generated by electrons colliding with the $n-s$ boundary. The carriers interact most effectively with the sound field if the distance $D(\lambda, p_z) = x(\lambda) - x\left(\dfrac{T_H}{2} - \lambda\right)$ along the $x$ axis between stationary-phase points is an integer multiple of the wavelength $\lambda_s$. Equation (7.68) can be used to show that electron orbits tangent to the boundary $x = 0$ (Fig. 14) for electrons belonging to the Fermi surface cross section $p_z = p_{ze}$ for which $\dfrac{\partial D}{\partial p_{ze}} = 0$, are distinguished orbits. The small oscillating increment $\Gamma_{ns}^*$ to the quantity $\Gamma_{ns}$ has the form

$$\Gamma_{ns}^* \cong a_0 Q \Gamma_0 \left[1 - Q_A^2 \exp\left(-T_H(p_{ze})/\tau\right)\right]^{-1} (kD_e)^{-3/2} \sin\left(kD_e + \gamma_{1e} + \gamma_{2e}\right) , \qquad (7.71)$$

where $D_e \equiv D(\tau_0, p_{ze}) \gg \lambda_s$ and $\tau_0$ satisfies the equation

$$d - x(\tau_0) + x(0) = 0. \qquad (7.72)$$

Here

$$a_0 = \frac{1}{m\, v_\perp d} \left(\frac{2\pi D^3}{\left|\dfrac{\partial^2 D}{\partial p_z^2}\right|}\right)^{1/2}_{\lambda = \tau_0;\, p_z = p_{ze}} ; \quad \gamma_{1e} = \frac{\pi}{2} sign(v_x(p_{ze})); \quad \gamma_{2e} = \frac{\pi}{4} sign\left(\frac{\partial^2 D}{\partial p_{ze}^2}\right).$$

Equation (7.71) is valid for $\sqrt{\lambda_s r} \ll l$, i.e., when the arc of the trajectory on which the charge moves almost parallel to the wave front is much smaller than the mean free path $l$.

At sound frequencies $\omega$ greater than the Larmor frequency $\Omega$ $(\Omega\tau \gg 1)$, the sound field can no longer be considered steady-state, and the carriers can enter into resonance interaction with the sound wave. It follows from Eq.(7.68) that a series of resonance lines coinciding with the acoustic cyclotron resonance lines in the bulk conductor occur. In fields H for which the characteristic orbit radius is smaller than the thickness of the layer in the normal state [81] (even though $d < 2r$ ):

$$\omega T_H(p_{z0}) = 2\pi n; \quad \frac{\partial T_H}{\partial p_z}\Big|_{p_z = p_{z0}} = 0; \qquad n = 1, 2, 3, \dots \quad . \qquad (7.73)$$

Resonance oscillations of the coefficient $\Gamma_{ns}$ are generated by a distinguished group of carriers with an extremal period of motion in the magnetic field. Integrating the first term in the braces in Eq. (7.68), we obtain

$$\Gamma_{ns} = \Gamma_{mon}(H) + a_1 \Gamma_0 \sum_{n=1}^{\infty} \frac{Q_A^{2n} \cos(n\omega T_H(p_{z0}) + \pi s(T_H)/4)}{(n\omega T_H(p_{z0}))^{1/2}} \exp\left(-\frac{nT_H(p_{z0})}{\tau}\right), \quad (7.74)$$

where

$$s(T_H) = sign T''_{H\,p_z^2}\Big|_{p_z = p_{z0}}; \quad a_1 = \frac{D}{dm v_\perp} \left(\frac{2\pi T_H}{\left|T''_{H\,p_z^2}\right|}\right)^{1/2}\Big|_{p_z = p_{z0};\lambda = \tau_0};$$

$$\Gamma_{mon}(H) \approx \Gamma_0 \frac{2r}{d}.$$

The factor $a_1 \approx D/d$ characterizes the relative fraction of charges colliding only with the $n - s$ boundary. For $D \approx d$ the amplitude of the resonance oscillations is of the same order of magnitude as the amplitude of the acoustic cyclotron resonance in the bulk sample [82]. For small deviations from resonance $\Delta = (H - H_{res})/H_{res}$, the equation for $\Gamma_{ns}$ takes the form

$$\Gamma_{ns} \cong a_1 \Gamma_0 \frac{[\varsigma - s(T_H)\Delta]^{1/2}}{n\varsigma_1}. \quad (7.75)$$

Here

$$\varsigma_1 = \left\{\Delta^2 + \left(1/\omega\tau + \frac{(1 - Q_A^2)}{2\pi n}\right)^2\right\}^{1/2}. \quad (7.76)$$

In the investigated interval of magnetic fields, where $r < d < 2r$, the acoustic cyclotron resonance is modulated by geometrical resonance oscillations (7.71). As in the case of resonance in the bulk metal, [82] the modulating term $\Gamma_{ns}^*$ is smaller than $\Gamma_{ns}$ (7.74) with respect to the parameter $(kD_e)^{-3/2} \ll 1$ if $\omega \ll v_F/s\tau$. Only in the range of very high frequencies $\omega \gg v_F/s\tau$ does geometrical resonance induce an appreciable variation of the absorption coefficient.

Cutoff of the "bulk" acoustic cyclotron resonance lines takes place in a magnetic field $H_0 = cp_{\perp,extr}/ed$. If the scattering of electrons by the sample surface is almost specular $1 - q_1 \ll 1$, then for $r > d$ the dominant factor will be resonance oscillations of $\Gamma_{ns}$ generated by carriers moving along periodic trajectories (with a period $T_1$) such as those shown in Fig. 13b. The resonance values of the magnetic field are determined from the condition

$$\omega T_1(\tau_0, p_z') = 2\pi n; \quad T_1(\tau_0) \equiv T_H - 2\tau_0; \quad \partial T_1/\partial p_z' = 0; \quad (7.77)$$

and $\tau_0$ once again satisfies Eq.(7.72). The calculation of $\Gamma_{ns}^{spec}$ according to Eq. (7.68) in the case $\omega T_1(\tau_0, p_z') \gg 1$ yields the following result

$$\Gamma_{ns}^{spec} = \Gamma_{mon}^{spec} + a_2 \Gamma_0 \sum_{n=1}^{\infty} (n\omega T_1)^{-3/2} \exp\left(-\frac{nT_1}{\tau}\right) q_1^n Q_A^{2n} \sin(n\omega T_1 + s(T_1)\pi/4). \qquad (7.78)$$

Here

$$a_2 = \left\{ \frac{2\pi T_1}{\left| T_{1p_z''}'' \right|} \right\}^{1/2} \frac{\left| v_x(\lambda) \right| T_1}{m\, v_\perp d}; \qquad \Gamma_{mon}^{spec} \cong \Gamma_0 .$$

All the quantities in Eq.(7.78) are evaluated at the point $p_z = p_z'$, $\lambda = \tau_0$. Additional selection with respect to the instant $\lambda$ of collisions of electrons with the surface leads to an appreciable reduction in the amplitude of the resonance oscillations in comparison with the case $r < d$, and the derivative $\partial\,\Gamma_{ns}^{spec}/\partial H$ contains the root singularity for $\tau \to \infty$ and $q_1 = Q_A = 1$:

$$\frac{\partial \Gamma_{ns}^{spec}}{\partial \Delta} \approx -s a_2 \frac{\Gamma_0}{n} \frac{\left| \varsigma_2 - s(T_1)\Delta \right|^{1/2}}{\varsigma_2}, \qquad (7.79)$$

where

$$\varsigma_2 = \left[ \Delta^2 + \left( 1/\omega\tau + (1 - q_1 Q_A^2)/2\pi n \right)^2 \right]^{1/2}; \qquad \Delta << 1.$$

If the scattering of electrons by the surface $x_s = 0$ is almost diffuse (i.e., if the inequality $q_1(\lambda) << 1$ holds for the majority of the carriers interacting effectively with the sound wave), acoustic cyclotron resonance is associated with a nonextremal radius $r(p_z)$ of the trajectory. In contrast with the case of a thin normal-metal plate, [83] two series of resonance lines must now be observed:

$$\omega T_H(p_{z1}) = 2\pi n; \qquad 2r(p_{z1}) = d; \qquad (7.80)$$

$$\omega T_H(p_{z2}) = 2\pi n; \qquad r(p_{z2}) = d; \qquad n = 1, 2, 3 \dots \ . \qquad (7.81)$$

The amplitude of the oscillations of $\Gamma_{ns}^{dif}$ for $q_1 << 1$ is much smaller than in the specular reflection ($q_1 \approx 1$) of carriers:

$$\Gamma_{ns}^{dif} = \Gamma_{mon}^{dif} + \Gamma_0 \sum_{k=1}^{2} \sum_{n=1}^{\infty} a_3^{(k)} Q_A^{2n} (n\omega T_H(p_{zk}))^{-2} \exp\left(-\frac{nT_H(p_{zk})}{\tau}\right) \cos(n\omega T_H(p_{zk})); \ k = 1, 2, \qquad (7.82)$$

where

$$a_3^k = (-1)^k \left( T_H / T_{H\,p_z''}'' \right)^2 \frac{1}{dm\, v_\perp} \frac{d}{dp_z} \left( D - (2-k)2r \right) \bigg|_{\lambda = \frac{T}{2k}; p_z = p_{zk}}; \qquad (7.83)$$

$\Gamma_{mon}^{dif}$ is the monotonic part of $\Gamma_{ns}^{dif}$. Near the resonance frequencies (7.80), (7.81), the term $\delta\Gamma_{ns}$ in $\Gamma_{mon}^{dif}$ containing the singularity of the derivative with respect to $H$ for $\tau \to \infty$ and $Q_A = 1$ has the form

$$\delta\Gamma_{ns} \cong \frac{1}{4\pi n}\Gamma_0\sum_{k=1}^{2}a_3^{(k)}\gamma_1\ln(\Delta_k^2 + \gamma_1^2).\tag{7.84}$$

Here $\Delta_k = {(H - H_{res}^{(k)})}\big/{H_{res}^{(k)}}$ is the deviation from the $k$-th resonance (7.80), (7.81)

$\gamma_1 = {1}\big/{\omega\tau} + {(1 - Q_A^2)}\big/{2\pi n}$.

We note that resonance at frequencies satisfying the relations (7.80), (7.81) exists for any type of surface scattering.

*a ) Sound propagation along the layer.*

When sound propagates along the boundaries of the normal-metal layer ($k = k_y$), all three types of electron trajectories shown in Fig.13 that contain stationary-phase points $kv_y(t_\beta) = \omega$ contribute to the absorption. Carrying out the integration with respect to $t$ in Eq.(7.67) on the assumption that $kd^2 >> r$, we obtain

$$\Gamma_{ns} \cong \Gamma_0 \operatorname{Re}\frac{1}{d}\int\frac{dp_z}{m}\frac{1}{v_\perp}\int_0^{T_H}d\lambda v_x(\lambda)\left\{\frac{S_1(\lambda, p_z)\alpha_0(\lambda, p_z)}{1 - Q_A^2\exp\left(i\omega^*T_H - ikL_0\right)} + \frac{S_2(\lambda, p_z)}{1 - q_1Q_A^2\exp\left(i\omega^*T_1 - ikL_1\right)}\right.$$

$$\left. + \frac{S_3(\lambda, p_z)\alpha_2(\lambda, p_z)}{1 - q_1^2Q_A^2\exp\left(i\omega^*T_2 - ikL_2\right)}\right\}.\tag{7.85}$$

Here

$L_0(\lambda) = 2\rho(\lambda) = 2\left[y(\lambda) - y(-\lambda)\right];\qquad L_1(\lambda) = L_0(\lambda) - \rho(\lambda_1);\qquad L_2(\lambda) = L_1(\lambda) - \rho(\lambda_2)$ are the displacements of the electrons along the $n - s$ boundary $x = d$, (see Fig.13), and $T_H$, $T_1 = T_H - 2\lambda_1$, $T_2(\lambda) = T_1 - 2\lambda_2$ are the periods of the motion along trajectories of the types *a, b,* and *c* in Fig.13. The dependence of the instant $\lambda_1$ and $\lambda_2$ of collision of electrons with the surface $x = 0$ on the instant $\lambda$ of collision with the $n - s$ boundary $x = d$ is given by the relations

$$x(\lambda_1) = x\left({T_H}\big/{2} - \lambda\right) - d;\quad x(\lambda_2) = x(\lambda) - d;\tag{7.86}$$

$$\alpha_n = \begin{cases} 1, & \Omega\left|t_\beta - \lambda\right| >> (kr)^{-{1}/{2}}; \\ \frac{1}{2}\left(1 - q^n\exp\left({1}/{2}\left[i\omega^*T_n - ikL_n\right]\right)\right), & \Omega\left|t_\beta - \lambda\right| << (kr)^{-{1}/{2}}, \end{cases}\tag{7.87}$$

where $n = 0, 2$; $T_0 = T_H$. We take the turning point $t = 0$, $(v_x(0) = 0)$ as the initial reference point of the carrier phase on each arc of the orbit. Expanding the denominators of the terms in the braces in Eq.(7.85) in power series in $\exp\left(-\dfrac{T_n}{\tau}\right) \ll 1$ and carrying out the integration with respect to $\lambda$ by the stationary-phase method, we can show that the monotonic part of $\Gamma_{ns}$ is of the order of $\Gamma_0$ for $d < 2r$ and depends weakly on the magnetic field. The term $\Gamma_{ns}^{osc}$ of $\Gamma_{ns}$ that oscillates with the variation of $H$ has the following form for $\omega\tau \ll 1$

$$\Gamma_{ns}^{osc} = \Gamma_0 \frac{b_1}{(kL_2)^{3/2}} \sum_{n=1}^{\infty} \frac{(-1)^n}{n^{3/2}} (q_1 Q_A)^n e^{-nT_2/2\tau} \sin\left(nkL_2/2 + \pi/4\right); \quad kL_2 \gg 1; \quad d \ll r \ll kd^2 \quad (7.88)$$

$$\Gamma_{ns}^{osc} = -\Gamma_0 \frac{b_2}{2kr} \sum_{n=1}^{\infty} \frac{(-1)^n}{n} Q_A{}^n e^{-nT_H/\tau} \cos 2nkr, \quad d > r; \quad (7.89)$$

where

$$b_1 = \frac{4v_x(\lambda)}{dm \, v_\perp L'_{2\lambda}} \left(\frac{L_2^3}{\left|L''_{2p_z^2}\right|}\right)^{1/2}; \qquad b_2 = \frac{4\pi v_x(\lambda)}{dm v_\perp} \left(\frac{L_0^2}{\left|L''_{0p_z^2} L''_{0\lambda^2}\right|}\right)^{1/2}.$$

All the quantities in Eq.(7.88) are evaluated at the point $\lambda = t_\beta$, $p_z = p_z^*$ $\left(\dfrac{\partial L_2}{\partial p_z^*} = 0; \dfrac{\partial L_2}{\partial t_\beta} \neq 0\right)$ and those in Eq.(7.89) are evaluated at the point $\lambda = t_\beta$, $p_z = p_{ze}$ $\left(\dfrac{\partial L_0}{\partial p_{ze}} = 0; \dfrac{\partial L_0}{\partial t_\beta} = 0\right)$. The results (7.88) and (7.89) show that the periods of the oscillations of the sound absorption coefficient are the same as in the case of a normal-metal plate [84, 85]. However, the odd harmonics in the nonmonotonic dependence of $\Gamma_{ns}$ on $H$ are shifted by $\pi$.

For the propagation of transverse sound in the layer in magnetic fields for which $kr \gg 1$, an important role is played by solenoidal electromagnetic fields, which induce a significant variation in the amplitudes of the considered resonance and oscillation effects. Nonetheless, the resonance frequencies and periods of the geometrical oscillations, which are determined by the geometry of the Fermi surface and the thickness of the normal layer, remain the same as in the case of longitudinal ultrasound discussed above.

Thus, a number of resonance and oscillation acoustic effects nonexistent in the thin plates are found to occur in thin layer of a normal metal adjacent to a superconductor. For $r < d < 2r$, the absorption coefficient of ultrasound propagating perpendicular to the surface of the sample and to the $n - s$ boundary undergoes geometrical oscillations, and, at sound frequencies $\omega$ larger than the Larmor frequency $\Omega$, acoustic cyclotron resonance with the "bulk" period sets in. In weaker magnetic fields, such that orbits with the extremal radius are cut off, the behavior of the absorption coefficient $\Gamma$ as a function of $\mathbf{H}$ depends strongly on the state of the surface of the sample. For example, in specular reflection the resonance dependence of $\Gamma$ on $\mathbf{H}$ is formed mainly by electrons interacting both with the $n-s$ boundary and with the outer boundary of the conductor. For surface scattering of a diffuse nature, the resonance oscillations of $\Gamma$ are associated with carriers having a trajectory of nonextremal radius. If ultrasound propagates along the layer, an oscillatory dependence of the coefficient $\Gamma$ on a weak magnetic field $(r > d)$ is possible only for the near-specular reflection of charge carriers by the surface of the metal. In pure conductors, the amplitudes of the geometrical oscillations and resonance lines and the width of the resonance lines are determined by the probability of Andreev reflection. Consequently, the experimental investigation of the effects discussed in the present study, which take place in connection with the absorption of ultrasound by normal-metal layers adjacent to a superconductor, will make it possible to acquire information on Andreev reflection.

## 8. Effect of Interdiffusion on Kinetic Phenomena in Double-Layer Films

### 8.1 Electrical conductivity

Extensive applications of metallic multilayer films in electronics gave rise to a problem of stability of its properties. Therefore, diffusion processes in these films are of interest. A possible way to obtain plausible information on the diffusion coefficients is to investigate the annealing-time dependence of kinetic coefficients of thin double-layer films (with the thickness being smaller than the electron mean free path) [86-90]. The possibility is due to the formation of a region with a high concentration of impurities diffused into the metal near the interface between the layers. These impurities cause diffuse electron scattering. Therefore, the positions of lines corresponding to radio frequency size effect [86-88], Sondheimer magnetoresistance oscillations [89, 90], etc. are changed on the magnetic-field scale. They are determined not by the layer thickness $d_i$, but by the width of the impurity-free region $d_i - x_{0i}$, where $x_{0i}$ is the characteristic penetration depth of impurity atoms. The size-dependent phenomena mentioned above arise in strong magnetic fields such that the Larmor radius of electron trajectories is of the order of the sample thickness. Along with these phenomena, more simple experiments on classical size effect in conductivity of DLF's can be used to determine the bulk and grain - boundary diffusion coefficients $D_b$ and $D_g$. However, in this case, obtaining information on diffusion processes requires comparing the experimental data and theoretically calculated dependence of conductivity on the diffusion-annealing time.

In this chapter the dependence of conductivity of metal DLF's on the annealing time is calculated at arbitrary ratios between layer thicknesses, grain size, and the mean three paths of electrons [91].

The conductivity of a DLF is given by the expression (3.5), where the function $\Psi_i(x, \mathbf{p})$ ($i = 1, 2$) satisfies the Boltzmann kinetic equation (2.1). The characteristic frequency of collisions in the bulk of the sample $\tau_i^{-1}(x, \mathbf{p})$ in Eq.(2.1) can be represented in the form [87,92,93]

$$\tau_i^{-1}(x, \mathbf{p}) = \tau_{0i}^{-1} + \tau_{1i}^{-1}(x) + \tau_{2i}^{-1}(x, \mathbf{p}), \qquad (8.1)$$

where $\tau_{0i}^{-1}$ is $x$ independent and accounts for collisions with phonons and point defects, whereas $\tau_{1i}^{-1}(x)$ describes the electron scattering by impurities diffused into the bulk of the metal. The presence of the term $\tau_{2i}^{-1}(x, \mathbf{p})$ in equation (8.1) corresponds to the electron scattering by grain boundaries (grain-boundary diffusion is taken into account).

The solution of equation (2.4) has the form

$$\Psi_i(x, \mathbf{p}) = F_i \exp\left\{ -\frac{1}{v_{xi}} \int_{x_s}^{x} \frac{dx'}{\tau_i(x', \mathbf{p})} \right\} + \frac{1}{v_{xi}} \int_{x_s}^{x} dx' e v_{yi} E \exp\left\{ -\frac{1}{v_{xi}} \int_{x'}^{x} \frac{dx''}{\tau_i(x'', \mathbf{p})} \right\}, \qquad (8.2)$$

where $x_s$, is the coordinate of the point where an electron scatters at the outer surface ($x_s = -d_1, d_2$) or at the interface between the layers ($x_s = 0$). The solution (8.2) involves arbitrary functions $F_i$ that should be determined through boundary conditions. To describe the interaction of electrons with the interface and the external surface of a DLF, we use the boundary conditions (2.14) and (2.15). By substituting functions $F_i$ in the form (3.2) into solution (8.2) of kinetic equation, it is possible to calculate the electrical current in the DLF, and hence the conductivity in the presence of interdiffusion. The calculation gives:

$$\sigma = \frac{2e^2}{h^3 dE} \sum_{i \neq j} \left\langle \left( v_y^2 \middle/ \Delta v_x \right) \left\{ 2J_i + q_i I_i^2 (1 - q_j P W_j^2(0)) + I_{0i}(P + q_j(Q^2 - P^2) W_j^2(0)) \right. \right.$$

$$\left. \left. \cdot (I_{0i} + 2q_i W_i(0) I_i) + Q(I_{0j} + q_j W_j(0) I_j)(I_{0i} + q_i W_i(0) I_i) \right\} \right\rangle_+, \qquad (8.3)$$

where the subscript "+" at the angular brackets signifies that integration is carried out over the part of the Fermi surface where $v_x > 0$,

$$\Delta = 1 - P(q_i W_i^2(0) + q_j W_j^2(0)) - q_i q_j (Q^2 - P^2) W_i^2(0) W_j^2(0); \qquad (8.4)$$

$$J_i = \int\limits_0^{d_i} dx W_i(x) \int\limits_x^0 dx' W_i^{-1}(|x'|); \qquad I_i = \int\limits_0^{d_i} dx W_i(x); \qquad (8.5)$$

$$I_{0i} = \int\limits_0^{d_i} dx W_i(x) W_i^{-1}(0); \quad W_i(x) = \exp\left\{-\frac{1}{|v_x|}\int\limits_{|x|}^{d_i}\frac{dx'}{\tau_i(x',\mathbf{p})}\right\}. \qquad (8.6)$$

The last formula gives the probability that an electron starting from a point in the $i$-th layer with coordinate $x$ will come to the sample surface $x_s = -d_1,\ d_2$, without collisions with diffused impurities.

The results of further calculation depend essentially on whether the layers are single-crystal or polycrystalline.

### 8.1.1 Bulk diffusion in single-crystal film

Let us consider the case of the grain size being much larger than the layers thickness and the mean free path of electrons (Fig. 15). In such a situation, the electron scattering by grain boundaries can be neglected and $\tau_{2i}^{-1}$ can be set equal to zero in formula (8.1).

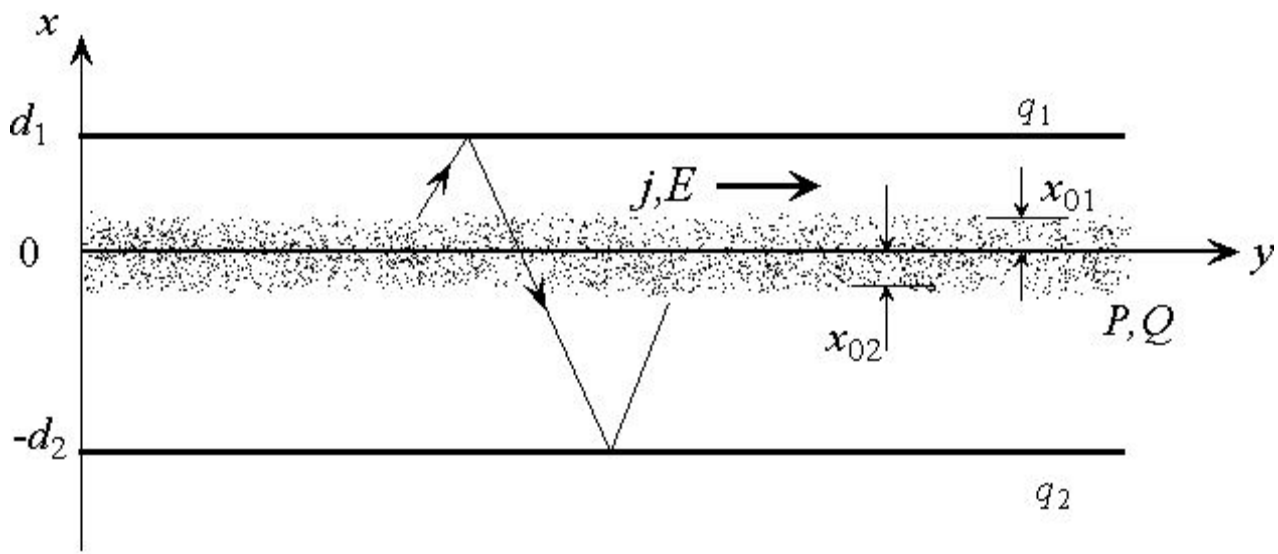

**Figure 15.** Model of a double layer single-crystal film in the presence of metal interdiffusion. The kinked arrowed line schematically shows one of the possible trajectory of an electron being scattered in the impurity layer as well as at the interface between the layers and at outer surfaces.

In layers, whose thickness $k_i^*(t_D)$ is much less than the mean-free path $l_{0i} = v_F \tau_{0i}$ of the charge carriers in the pure region of the single crystal, the presence of a scattering layer of atoms near the interface gives rise to the fact that the effective path of most electrons between two scattering acts becomes of the order of $d_i - x_{0i}$, where $x_{0i}$ is the effective decrease in the thickness of the layer owing to the diffusion of impurities (Fig. 15). For this reason, independent of the nature of the interface scattering, just as in a plate with diffuse boundaries, the electrical conductivity depends on the quantity $d_i - x_{0i}$, and its change after diffusion annealing enables us to determine the coefficient of diffusion.

We assume that the bulk diffusion coefficient is constant, the impurity concentration changes gradually at the layer interface, and metal solubility is limited. Then, $\tau_{1i}^{-1}(x)$ has the form [87 -89]

$$\tau_{1i}^{-1}(x) = v_{bi} C_{bi}(x, t_D),\qquad(8.7)$$

where $v_{bi} \cong v_F \sigma_{eff} n_{0i}$; $\sigma_{eff}$ is the effective cross section for scattering of electrons by impurity atoms, $n_{0i}$ is the density of atoms of the pure sample. The distribution $C_{bi}(x, t_D)$ of impurity atoms in each layer may be represented in the following form [94]:

$$C_{b1}(x, t_D) = C_0 \left\{ \gamma - (1 - \gamma) erf \left( \frac{x}{2\sqrt{D_{b1} t_D}} \right) \right\}, \quad x < 0;$$

$$C_{b2}(x, t_D) = C_0 \gamma erfc \left( \frac{x}{2\sqrt{D_{b2} t_D}} \right), \quad x > 0,\qquad(8.8)$$

where

$$\gamma = \frac{\sqrt{D_{b1}}}{\sqrt{D_{b1}} + \sqrt{D_{b2}}}$$

and $D_{bi}$ is the bulk – diffusion coefficient in the *i-th* layer of the DLF.

At small diffusion times $\sqrt{D_{bi} t_D} \ll d_i$, the derivative of the function $W_i(x)$ is the sharply changing function of coordinate $x$ compared to $\exp(-x/v_x \tau_{0i})$. Therefore, integrals in (8.3) can be calculated asymptotically at $d_i / \sqrt{D_{bi} t_D} \gg 1$. By integrating over the Fermi surface, we obtain the following expression for the conductivity of double-layer single-crystal film:

$$\sigma = \sum_i \sigma_{0i} \begin{cases} 1 - \dfrac{3(2 - q_i)}{16 k_i(t_D)}, & k_i \gg 1; \\ \dfrac{3}{4}(1 + q_i) \dfrac{l_{0i}}{d_i} k_i^2(t_D) \ln \dfrac{1}{k_i(t_D)}, & k_i \ll 1; \end{cases}\qquad(8.9)$$

where

$$k_i(t_D) = \frac{d_i - x_{0i}(t_D)}{l_{0i}}; \quad x_{0i}(t_D) \cong a_i \sqrt{D_{bi} t_D}; \quad a_i \cong 2 \ln^{1/2} \left( \frac{2\sqrt{D_{bi} t_D}}{v_F \tau_{1i}(0)} \right). \tag{8.10}$$

The quantity $x_{0i}$ is the effective decrease in the thickness of the *i-th* layer caused by bulk interdiffusion of metals.

At large annealing times $d_i / \sqrt{D_{bi} t_D} \approx 1$, the distribution of impurities across the sample becomes practically uniform. Therefore, one may consider the impurity concentration in each layer to be coordinate independent and put it equal to its average value:

$$\overline{C_{bi}}(t_D) = \frac{1}{d_i} \int_0^{d_i} dx \, C_{bi}(x, t_D);$$

$$\overline{C_{b1}} = \frac{C_0}{d_1} \left\{ d_1 \gamma - (1-\gamma) \left[ d_1 erf \frac{d_1}{2\sqrt{D_{b1} t_D}} - 2\sqrt{\frac{D_{b1} t_D}{\pi}} \left( 1 - \exp\left( -\frac{d_1^2}{4D_{b1} t_D} \right) \right) \right] \right\};$$

$$\overline{C_{b2}} = \frac{C_0 \gamma}{d_2} \left\{ d_2 erf \left( \frac{d_2}{2\sqrt{D_{b2} t_D}} \right) + 2\sqrt{\frac{D_{b2} t_D}{\pi}} \left( 1 - \exp\left( -\frac{d_2^2}{4D_{b2} t_D} \right) \right) \right\}. \tag{8.11}$$

This facilitation allows for calculating the integrals in formula (8.3). Calculations give the following expression for the conductivity of the DLF $\sigma$:

$$\sigma = \frac{1}{d} \sum_{i=1}^{2} d_i \sigma_{0i} \Phi_i, \tag{8.12}$$

where

$$\Phi_i = 1 - \frac{3}{4 k_i^*(t_D)} \int_0^1 d\xi (\xi - \xi^3)(1 - \varepsilon_i) G_i; \tag{8.13}$$

$$G_i = \frac{1}{\Delta} \left\{ (2 - q_i - P + \varepsilon_i(q_i + P - 2q_i P))(1 - q_j P \varepsilon_j^2) - \right.$$

$$\left. q_j Q^2 \varepsilon_j^2 (1 - \varepsilon_i + 2q_i \varepsilon_i) - Q \tau_{0j,i} (1 - \varepsilon_j)(1 + q_i \varepsilon_i)(1 + q_j \varepsilon_j) \right\}; \tag{8.14}$$

$$\beta_i^{-1} << d_i$$

$$\varepsilon_i = \exp\left\{ -\frac{k_i^*(t_D)}{\xi} \right\}; \quad \tau_{0j,i} = \frac{\tau_{0j}}{\tau_{0i}}; \quad k_i^*(t_D) = \frac{d_i}{\bar{l}_{bi}(t_D)};$$

$$\sigma_i(t_D) = \sigma_{0i} \frac{\bar{l}_{bi}(t_D)}{l_{0i}}; \quad \bar{l}_{bi}(t_D) = \frac{l_{0i}}{1 + \tau_{0i} v_{bi} \overline{C_{bi}}(t_D)}, \tag{8.15}$$

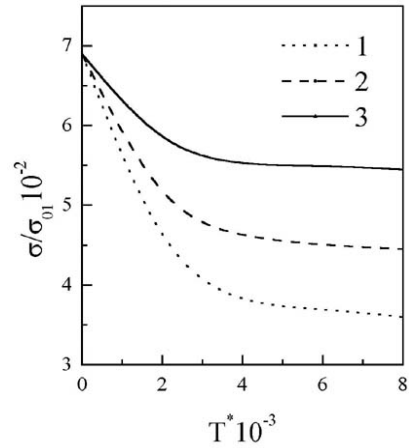

**Figure 16** Conductivity of a double-layer single-crystal film versus dimensionless annealing time $T^* = 4D_{b1}t_D d_1^{-2}$ at the parameter values $k_1 = 10^{-2}$, $k_2 = 10^{-4}$ $q_i = 0.6$, $P = 0.2$, and different values of $Q$: $1 - Q = 0.2$, $2 - Q = 0.5$, $3 - Q = 0.8$.

is the effective mean free path of electrons in the impure sample.

Formula (8.13) can be simplified in the limiting case of a DLF with thick layers ($k_i^* \gg 1$) or thin layers ($k_i^* \ll 1$). The asymptotic expressions for the function $\Phi$ have the same form as Eqs. (3.11)-(3.14) after replacement $k_i$ by $k_i^*(t_D)$.

At large annealing times $\sqrt{D_{bi}t_D} > d_i$, the relationship between $\sigma$ and the bulk-diffusion coefficient $D_{bi}$ essentially more complicated. Therefore, to describe experimental results the precise formula (8.3) should be used. The impurity concentration entering into this equation should be determined with an account of the sample boundaries (see [94]). The curves shown in Fig. 16 are obtained by numerical calculation via (8.12). They illustrate the annealing-time dependence of conductivity of a double layer single-crystal film.

### 8.1.2 Grain-boundary diffusion in polycrystalline films

A theoretical analysis of the effect of grain-boundary diffusion on the conductivity of double layer polycrystalline films (Fig. 17) may be carried out using modified Mayadas-Shatzkes model [29]. This model takes into account a change in the grain-boundary reflection factor $R_i$ of electrons caused by the migration of impurity atoms along them in the course of a grain-boundary interdiffusion. This method of solving the problem on annealing-time dependence of conductivity in double-layered polycrystalline films was suggested in [92, 93].

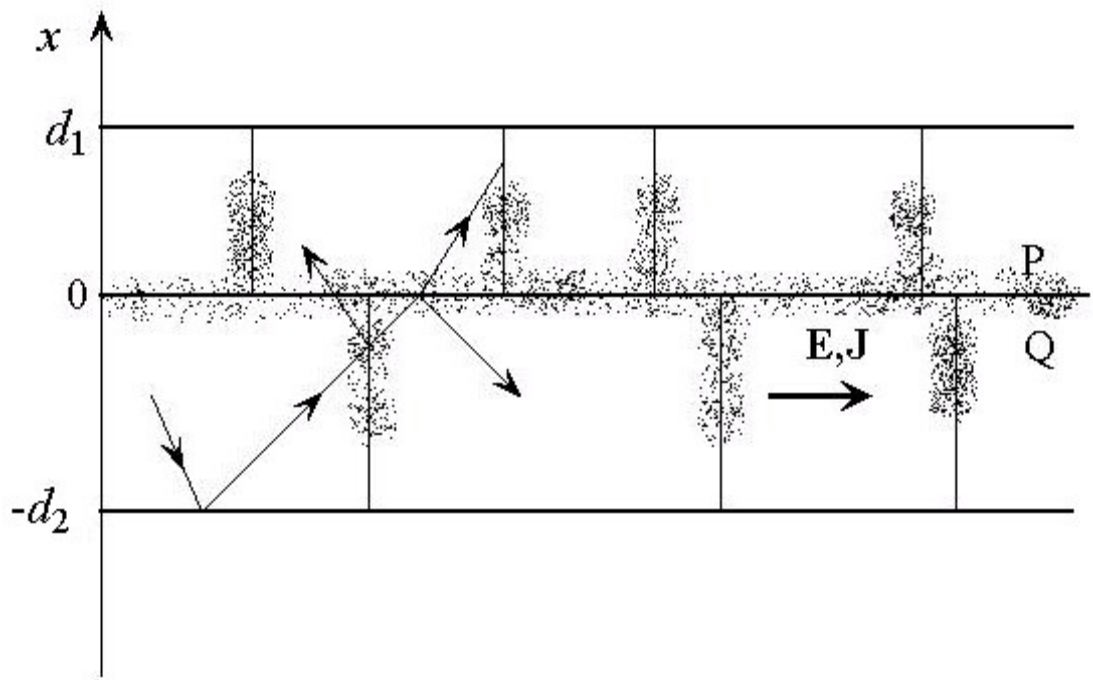

**Figure 17.** Model of a double-layer polycrystalline film in the presence of grain-boundary interdiffusion of metals. The kinked arrowed line schematically shows one of the possible electron trajectories.

At sufficiently low temperatures $T$ of annealing ($T < 0.38 \; T_m$, where $T_m$ is the melting temperature), mass transfer in polycrystalline films occurs mainly along grain boundaries [95]. Therefore, during the course of interdiffusion the electrical resistance of the film, caused by electron scattering at external surfaces and in the bulk of the sample, remains practically unchanged, although the resistance of grain boundaries is essentially changed by segregation of impurity atoms at them.

At low concentration of diffusing impurity atoms at grain boundaries $C_{gi}(x, t_D) << 1,$, the grain-boundary-reflection factor of electrons can be represented in the following form [92, 93]

$$R_i(x, t_D) = R_{0i} + \gamma_{gi} C_{gi}(x, t_D),$$ (8.16)

where the value $R_{0i}$ describes the grain-boundary reflection of electrons in the absence of impurity atoms. The coefficient $\gamma_{gi}$ is of an order of unity and has an arbitrary sign because the penetration of impurity atoms into grain boundaries can both decrease and increase the electron reflection factor $R_i$. If the grain-boundary-diffusion process is accompanied by forming solid solutions [96], the conductivity of a DLF decreases with time, i.e., $\gamma_{gi} > 0$. Along with scattering immediately at grain boundaries, electrons may be scattered by elastic-deformation fields near the grain boundaries. The impurity atoms give rise to the relaxation of these fields, and this leads to negative values of $\gamma_{gi}$, and, therefore, to an increase in the conductivity of the film [92,93].

If the inequality $\sqrt{D_{bi}t_D} << \delta_i$, holds ($\delta_i$, is the width of diffusion grain boundary), the diffusion of impurity atoms out of grain boundaries into the bulk of the sample can be neglected [97] and the diffusion flux can be considered as one-dimensional [98,99]. Therefore,

$$C_{gi}(x,t_D) = C_{0i}\exp\left\{-\eta_i x\right\};$$

$$\eta_i = \left\{\frac{2}{\delta_i D_{gi}}\left(\frac{D_{bi}}{\pi t_D}\right)^{1/2}\right\}^{1/2}, \tag{8.17}$$

where $D_{gi}$, is the grain-boundary diffusion coefficient in the *i-th* layer. The $\eta_i^{-1}$, has the meaning of characteristic penetration depth of impurities into the bulk of metal near a grain boundary.

The conductivity of a double-layer polycrystalline film with account of metal interdiffusion is determined by expression (8.3) with characteristic mean free time $\tau_i(x, p_y)$ given by

$$\tau_i^{-1}(x,p_y) = \tau_{0i}^{-1} + \frac{l_{0i}p_F}{L_i|p_y|}\frac{R_i(x,t_D)}{1-R_i(x,t_D)}, \tag{8.18}$$

where $L_i$, is the mean grain size in the plane of the i-th layer.

At small diffusion times ($\eta_i^{-1} << d_i$), the integrals in formulas (8.3) and (8.4) are readily calculated asymptotically and the following expressions for the conductivity can be obtained:

$$\sigma = \sum_i \sigma_{0i}\begin{cases} 1 - \dfrac{3}{2}\beta_{0i} - \dfrac{3\beta_{0i}}{2R_{0i}\chi_i(t_D)}\ln\left(1 + \gamma_{gi}C_{0i}\dfrac{1-e^{-\chi_i}}{1-R_{0i}-\gamma_{gi}C_{0i}}\right), & \beta_{0i} << 1; \\[4mm] \dfrac{3}{4\beta_{0i}}\left[1 - \dfrac{1}{(1-R_{0i})\chi_i(t_D)}\ln\left(1 + \gamma_{gi}C_{0i}\dfrac{1-e^{-\chi_i}}{R_{0i}+\gamma_{gi}C_{0i}e^{-\chi_i}}\right)\right], & \beta_{0i} >> 1, \end{cases} \tag{8.19}$$

for a thick film ($d_i >> l_{0i}$), and

$$\sigma = \sum_i \frac{3}{4}\sigma_{0i}(1+q_i)\frac{l_{0i}}{d_i}k_{gi}^2\left\{\ln\frac{1}{k_{gi}} - \frac{4}{\pi}\beta_{0i}\right\}; \tag{8.20}$$

$$\beta_{0i} = \frac{l_{0i}}{L_i}\frac{R_{0i}}{1-R_{0i}}; \quad k_{gi}(t_D) = \frac{d_i - x_{0i}^*(t_D)}{l_{0i}}, \quad \chi_i(t_D) = \eta_i(t_D)\left(d_i - x_{0i}^*(t_D)\right),$$

for thin layers ($d_i \ll l_{0i}$, $\beta_{0i} \ll 1$), where

$$x_{0i}^{*}(t_D) \approx \frac{1}{\eta_i(t_D)} \ln \left| \frac{\gamma_{gi} C_{0i}}{\eta_i(t_D) L_i} \right|. \tag{8.21}$$

Notice that $x_{0i}^{*}$ is the effective decrease in the thickness of the *i-th* layer caused by grain-boundary interdiffusion of the metals.

At large annealing times ($\eta_i^{-1} \approx d_i$), the above-mentioned average – concentration approximation can be used, i.e., the impurity distribution across grain boundaries can be considered as uniform. Therefore, the impurity concentration is given by the expression

$$\overline{C_{gi}}(t_D) = \frac{C_{0i}}{\eta_i d_i} \left\{ 1 - \exp\left( -\eta_i d_i \right) \right\}. \tag{8.22}$$

Calculating the integrals in (8.3) shows that the DLF conductivity is given by equation (8.12) with $\Phi_i^{*}$ substituted for $\Phi_i$:

$$\Phi_i^{*} = M(\overline{\alpha_i}) - \frac{3}{\pi k_{gi}^{*}} \int\limits_{0}^{\pi/2} d\varphi \cos^2\varphi \int\limits_{0}^{1} \frac{d\xi(\xi - \xi^3)(1 - \varepsilon_i^{*})}{H_i^2(\xi, \varphi, t_D)} G_i^{*}; \tag{8.23}$$

$$M(\overline{\beta_i}) = 1 - \frac{3}{2}\overline{\beta_i} + 3\overline{\beta_i}^2 - 3\overline{\beta_i}^3 \ln\left(1 + \frac{1}{\overline{\beta_i}}\right). \tag{8.24}$$

Here

$$H_i(\xi, \varphi, t_D) = 1 + \frac{\overline{\beta_i}(t_D)}{\cos^2\varphi\sqrt{1 - \xi^2}}; \qquad \varepsilon_i^{*} = \exp\left( -\frac{k_{gi}^{*}H_i}{\xi} \right); \qquad k_{gi}^{*} = \frac{d_i}{l_{0i}};$$

$$\overline{\beta_i}(t_D) = \frac{l_{0i}}{L_i} \frac{\overline{R_i(t_D)}}{1 - \overline{R_i(t_D)}}. \tag{8.25}$$

The functions $G_i^{*}$ can readily be obtained from (8.14) by substituting $\varepsilon_i^{*}$ for $\varepsilon_i$, and $\tau_{0j}H_i/(\tau_{0i}H_j)$ for $\tau_{0j}/\tau_{0i}$.

If the thickness of the film layers is much larger than the mean free path of electrons, i.e., the inequality $d_i \gg l_{0i}$ holds, the exponents in equation (8.23) are small and can be neglected. The angular integration gives the expressions (3.23)-(3.26) for the function $\Phi_i^{*}$ (8.23), in which $\tau_{j,i}$ and $\beta_i$ must be replaced by $\tau_{0j,i}$ and $\overline{\beta_i}$.

At arbitrary ratio between the values $\eta_i^{-1}$ and $d_i$, the experimental data must be analyzed using numerical calculation via equation (8.3). The effective relaxation frequency of electrons at grain boundaries entering into this expression is given by equation (8.1). Figure 18 shows the computer - calculated curves that describe the variation of a DLF conductivity with annealing time with account of grain-boundary interdiffusion at different values of parameter $\gamma_g$.

In this way, metal interdiffusion essentially influences the conductivity of DLF's. At small diffusion-annealing times $t_D$, the characteristic penetration depth of impurities ($\sqrt{D_{bi}t_D}$ for a single-crystal film and $\left\{\left(\dfrac{2}{D_g\delta_i}\right)\left(\dfrac{D_b}{\pi t_D}\right)^{1/2}\right\}^{-1/2}$ for a polycrystalline film) is smaller than the thickness of the layer, and the size effect in DLF's is determined by the width of the impurity-free region. The

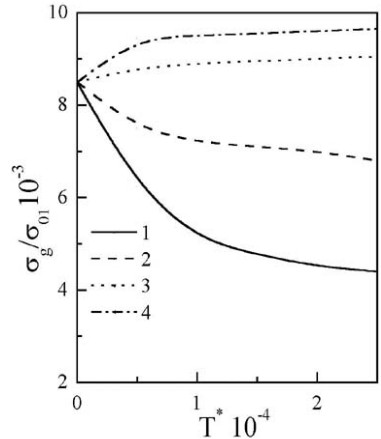

**Figure 18.** Conductivity of a double-layer polycrystalline film versus dimensionless annealing time $T^* = \delta_1^2 D_{g1}^2 t_D \pi / 4 d_1^4 D_{b1}$ at the parameter values

$$k_{g1} = 10^{-2}, \quad k_{g2} = 10^{-4}, \, q_i = 0.5, \, P = Q = 0.3, 1 - \gamma_{gi} = 0.5; \quad 2 - \gamma_{gi} = 0.2;$$

$$3 - \gamma_{gi} = -0.2; \quad 4 - \gamma_{gi} = -0.8.$$

diffusion coefficients $D_b$ and $D_g$ may be obtained by measuring $t_D$ dependencies of electrical resistance and using equations (8.10) and (8.21) for impurity penetration depth $x_{0i}$, in the case of bulk interdiffusion and grain-boundary interdiffusion, respectively. If the characteristic penetration depth of impurities is of the order of the layer thickness, and yet the effective mean free path of electrons (with account of grain-boundary scattering) is still larger than the value $d_i$ or comparable to it, the size effect also takes place. In this case, the relationships between the DLF conductivity and the parameters $\bar{l}_{bi}(t_D)$ and $\overline{\beta}_i(t_D)$ also allow for determining the bulk and grain-boundary diffusion coefficients.

## 8.2 Galvanomagnetic phenomena

For simplicity we shall analyze the effect of a diffusing impurities on the magnetoconductivity of DLF's in the case $d_1 \gg d_2$ $\left(d_1 \approx d\right)$, and below omit the indexes, which numerate the layers.

The presence of a strong magnetic field **H** (the radius of curvature of the electron trajectory $r$ in the field **H** is much less than $d$ and $l$) greatly expands the possibility of studying impurity diffusion in metals. Thus, if the vector **H** is tilted away from the surface of the sample by an angle $\theta > r/d$, then the resistance is an oscillatory function of the magnitude of the field **H**, as first predicted by Sondheimer [33]. In the case under study, however, electron scattering at one of the boundaries occurs not at the surface, but rather at a distance $x_0$ from it, equal to the characteristic thickness of the layer containing the impurity (Fig. 19). The period of Sondheimer oscillations is determined by the difference $d - x_0$.

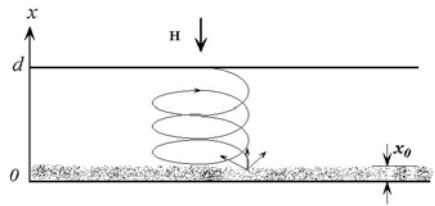

**Figure 19.** Trajectory of an electron, scattered in the impurity layer in perpendicular magnetic field.

In thin films of compensated metals in a magnetic field oriented parallel to the surfaces of the conductor and perpendicular to the electric current density vector **j,** there arises a static skin effect [40, 41], in which almost the entire current flows at the boundary of the sample. The presence of the impurity layer, generally speaking, does not cause the static skin effect to vanish, since the effective mean-free path length of the electrons near the boundaries is of the order of $r$ (Fig. 20), and their rela tive number $r/d$ and the conductivity of a layer with thickness $2r$ at a distance $x_0$ from the boundary is $l/d$ times greater than in the core of the sample. If the radius $r$ is much larger than the mean-free path of the charge carriers in the "dirty" region $l_1$ then its contribution to the total transverse conductivity of the film may be very substantial. The change in the electrical conductivity with the diffusion of the impurity atoms enables determining the value of $D_b$.

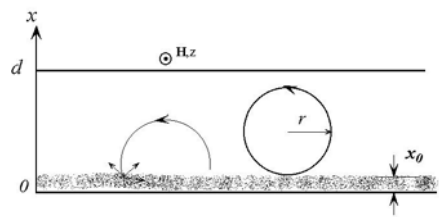

**Figure 20** Trajectory of an electron, scattered in the impurity layer in parallel magnetic field.

We shall analyze in detail below the magnetoconductivity in a thin film ($d << l$), near whose surface $x = 0$ the layer of impurities diffusing into the sample is located [89]. We shall assume that the mean-free path of electrons in this layer $l_1$ is less than not only the mean-free path length in the "pure" region, but also the characteristic thickness of the layer $x_0$. We note that in the case under study, when $l_1 << x_0$, the charge carriers after interacting with the boundary $x = 0$ undergo scattering in the impurity layer, and the electrical conductivity of the film is virtually independent of the state of the surface $x = 0$.

We shall present below the results of the calculation of the magnetoconductivity under different conditions, for simplicity assuming that the dispersion law for the charge carriers is quadratic and isotropic.

If a strong magnetic field ($r < d < l$) is oriented perpendicular to the surface of a film, the components of the tensor of manetoconductivity $\sigma_{\alpha\beta}$    ($\alpha, \beta = y, z$) can be presented in the form

$$\sigma_{\alpha\beta} = \begin{pmatrix} \operatorname{Re}\sigma^{(+)} & \operatorname{Im}\sigma^{(+)} \\ -\operatorname{Im}\sigma^{(+)} & \operatorname{Re}\sigma^{(+)} \end{pmatrix}, \tag{8.26}$$

where

$$\sigma^{(+)} = \frac{2e}{dh^3} \left\langle \frac{v_\perp^2}{v_x} \cos\Omega t \; e^{i\Omega t} \; \left( q_2 \; I_1^2 + 2I \right) \right\rangle_+ ; \tag{8.27}$$

$$I_1 = \int_0^d dx \; W(x) \; \exp\left\{ -\left( \frac{1}{\tau_0} + i\Omega \right) \frac{d-x}{v_x} \right\}; \tag{8.28}$$

$$I = \int_0^d dx \; W(x) \int_x^d dx' \; W^{-1}(x') \; \exp\left\{ -\left( \frac{1}{\tau_0} + i\Omega \right) \frac{x'-x}{v_x} \right\}; \tag{8.29}$$

$v_\perp = (v^2 - v_x^2)^{1/2}$; $\Omega = eH/mc$ and $x_0\left(v_x/v_F\right) = a\left(v_x\tau_1(0)/2\sqrt{D_b t_D}\right)\sqrt{D_b t_D}$ . The probability $W(x)$ is determined, as before, by the expression (8.6). For $\Omega\tau_0 >> 1$ the effective truncation of the electron trajectories at the boundary of the layer of impurity atoms occurs only if $\sqrt{D_b t_D} << r$ . Calculating the integrals $I_1$ (8.28) and $I$ (8.29) asymptotically for $r/\sqrt{D_b t_D} >> 1$, it can be shown that in this case $\sigma^{(+)}$ has the form

$$\sigma^{(+)} \cong \sigma_0 \frac{d-\overline{x_0}}{d} \left\{ \frac{1}{i\Omega\tau_0 + 1} - \frac{3}{16} \frac{l}{d-\overline{x_0}} \frac{1}{\left(1 + i\Omega\tau_0\right)^2} (2-\overline{q_2}) \right\} +$$

$$3\sigma_0 \frac{l^3}{d(d-x_0(1))^2} \frac{1}{(1+i\Omega\tau_0)^4} \left\{ (1-q_2(1)) \exp\left[-\left(\frac{1}{\tau_0}+i\Omega\right)\frac{d-x_0(1)}{v_F}\right] + \right.$$

$$\left. \frac{1}{8}q_2(1) \exp\left[-2\left(\frac{1}{\tau_0}+i\Omega\right)\frac{d-x_0(1)}{v_F}\right]\right\}. \tag{8.30}$$

Here $q_2(1)$ is the specularity parameter of the surface $x=d$ for electrons at the reference point on $\left(v_x = v_F\right)$ the Fermi surface;

$$\overline{x_0} = \frac{3}{2}\int_0^1 d\xi(1-\xi^2) \ x_0(\xi); \qquad \overline{q_2} = 4\int_0^1 \xi(1-\xi^2)q_2(\xi)d\xi; \qquad \xi = \frac{v_x}{v_F}.$$

Knowing the components of the conductivity tensor $\sigma_{\alpha\beta}$ (8.26), it is not difficult to calculate the transverse magnetoconductivity of the film $\sigma_\perp$. Its nonmonotonic part $\sigma_\perp^{osc}$ is most informative:

$$\sigma_\perp^{osc} \cong \sigma_\perp^{mon} \frac{r^2}{d^2}\left\{(1-q_2(1)) \exp\left(-\frac{d-x_0(1)}{l}\right)\cos\frac{d-x_0(1)}{r}\right.$$

$$\left. +\frac{1}{8}q_2(1)\exp\left[-2\frac{d-x_0(1)}{l}\right]\cos 2\frac{d-x_0(1)}{r}\right\}, \tag{8.31}$$

where $\sigma_\perp^{mon} \cong \sigma_0(d-\overline{x_0})l^{-1}$ is the monotonic part of $\sigma_\perp$; and $r = v_F/\Omega$, and $l = v_F\tau_0$. The result (8.31) shows that by studying the oscillating part of the magnetoconductivity of the film it is possible to determine the magnitude of the effective decrease in the thickness of the sample $x_0(1)$ and, therefore, the coefficient of diffusion $D_b$.

When the magnetic field tilts away from the normal to the surface of the sample by an angle $\varphi \sim 1$ the nature of the oscillatory dependence of $\sigma_\perp$ on $H$ does not change. In this case $d-x_0(1)$ in the formulas (8.30) and (8.31) must be replaced by $\left(d-x_0(1)\right)/\cos\varphi$.

Let us consider a magnetic field parallel to the surface. We shall study a film made of a compensated metal (the number of electrons is equal to the number of holes: $n_1 = n_2$), placed in a strong magnetic field $(r < d < l)$ oriented along the $z$ axis $(H = H_z)$ and perpendicular to the electric current $j = j_y$ (Fig.20). As it was shown in Refs. 5 and 6, the contribution of the bulk electrons to the transverse electrical conductivity is small, in this case, and the electric current is concentrated primarily over the surface of the conductor (static skin effect). We analyzed the contribution of charge carriers moving near the surface $x = 0$ and scattered in the layer of diffused impurities to $\sigma_\perp$. The term in the current density linked with the surface electrons can be written in the form

$$\Delta j_y = \frac{2e^2}{dh^3} \sum_{i=1}^{2} \int_{0}^{d} dx \left\langle v_y^{(i)}(t) \ W_i^{-1}(x,t) \int_{}^{t} dt' \ W_i(x,t') \ \mathbf{v}^{(i)}(t') \ \mathbf{E}(x+x^{(i)}(t')-x^{(i)}(t)) \right\rangle. \qquad (8.32)$$

Here the summation extends over groups of charge carriers;

$$W_i(x,t) = \exp\left\{ -\int_{t}^{\pi/\Omega_i} dt' \ \tau_1^{-1}\left( x+x^{(i)}(t')-x^{(i)}(t) \right) \right\}, \qquad (8.33)$$

is the probability that electrons, having a phase $\Omega_i t$ at the point with the coordinate $x$, reach a point on the trajectory farthest away from the surface $x=0$, at which we take the phase to be equal to $\pi$, without scattering ($\Omega_i = eH/m_i c$; $m_{1,2}$ is the effective mass of electrons (holes)). Analysis of the equation of electrical neutrality (2.13) shows that the nonuniform component of the electric field $E_x(x) \sim E_y$ differs substantially from zero only if the distance from the layer of diffused atoms is less than or of the order of $2r = \max(2r_1, 2r_2)$. Substituting the solution of Eq. (2.13) for $E_x(x)$ into the formula (8.32), we obtain an expression for the surface correction to the transverse electrical conductivity $\Delta\sigma_\perp = \Delta j_y / E_y$. After simple, though very cumbersome, calculations we arrive at the result

$$\Delta\sigma_\perp = \alpha_1 \sigma_0 \sum_i \frac{r_i^2}{l_i \ d} + \alpha_2 \ \sigma_0 \sum_i \frac{x_{0i} l_{1i}}{l_i d}, \qquad (8.34)$$

where $r_i = v_{Fi}/\Omega_i$; $l_i = v_{Fi}\tau_0$; $l_{1i} = v_{Fi}\tau_1(0)$; $v_{Fi}$ is the Fermi velocity of the electrons (holes); and, the constants $\alpha_1$ and $\alpha_2$ are of order unity. The effective decrease in the thickness of the film $x_{0i}$ is determined from the condition

$$(W_i)''_{t^2}(x_{0i}, 0)\big|_{p_z=0} = 0. \qquad (8.35)$$

The first term in the formula (8.34) is the contribution of charges, whose characteristic free flight time is of the order of $1/\Omega_i$ while the distance between the center of their orbit and the boundary of the layer of impurity atoms $x = x_{0i}$ does not exceed $r_i$, to the conductivity. The second term, depending on the coefficient of diffusion ($x_{0i} = a_i\sqrt{D_b t_D}$), is described by the conductivity of the "dirty" layer with thickness $x_0 = \max(x_{01}, x_{02})$, in which the mean-free path of the charge carriers $l_{1i} << x_{0i} << r_i$.

We have shown that the presence of a diffusing layer of impurities substantially affects the magnrtoconductivity of thin metal films. For short diffusion times, when the characteristic distance $\sqrt{D_b t_D}$ at which the impurity concentration decreases is less than the distance over which the electron distribution function changes, size effects determined by the thickness of the "pure" region $d - x_0$, where $x_0 = A\sqrt{D_b t_D}$ is the effective decrease in the quantity $d$, appear in the film. The coefficient $A$, differing for each of the effects studied, generally speaking, also depends on $D_b$, since the increase in the diffusion time changes the mean-free path of the charge carriers in the "dirty" region.

Thus, measurements of the magnetoconductivity of thin DLF's enable determining efficiently the rate of diffusion of impurity atoms in metals. In particular, Fourier analysis of the dependence of the conductivity on the magnitude of the magnetic field oriented perpendicular (or inclined) to the surface enables finding with high accuracy the period of Sondheimer oscillations and, therefore, calculating the depth to which the impurity atoms diffused.

The method of the coefficient $D_b$ by using Sondheimer oscillations magnetoconductivity was realized in experiment [90]. The effect of the concentration of electrical current near a surface region with a high density of defects was experimentally found in Ref.100 in compensated metals placed in the strong magnetic field parallel to the surface [100].

### 8.3 High frequency effects

Gudenko and Krylov [86,87] showed that the observation of a radio frequency (RF) size effect [46] in thin single crystal plates when a diffused impurity layer exists near one of the surface makes it possible to determine the diffusion coefficient $D_b$. The essence of the method they propose [88,89] is that in a plate with a film of another metal deposited on its surface the RF size effect lines shift after diffusion annealing because of the cutoff of electron trajectories in the surface layer of impurity atoms (Fig. 21). To calculate the shape and position of the RF size effect lines from the initial lines (up to diffusion annealing) Gudenko and Krylov [86] proposed a method that would allow information to be obtained about the coefficient $D_b$ from the experimental data.

In this chapter we construct a consistent theory of high-frequency phenomena in a double layer single-crystal metal film when near the interface is an impurity layer that arises because of interdiffusion [88].

In the considered situation of a nonunlform impurity distribution, the probability $W$ of electron motion over a periodic trajectory without scattering depends essentially on the magnetic field $\mathbf{H}$ parallel to the surface. If the diffusion coefficient $D_b = const$ and if a solubility limit exists, i.e.. the concentration of impurity atoms at the boundary of the sample $C(0, t_D) = C_0 < 1$ ($t_D$ is the diffusion-annealing time), the function $W$ for carriers that touch the surface of the sample has the form

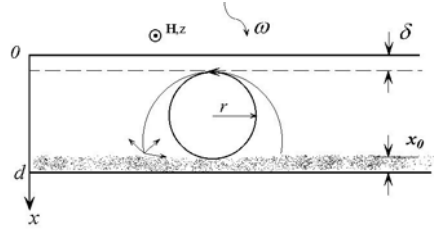

**Figure 21.** The cut-off of electron trajectories in the surface layer of impurity atoms in parallel magnetic field.

$$W(r,\Omega) = \exp\left\{-\frac{1}{\Omega\tau_1}\int\limits_0^{2\pi} d\varphi\, Erfc\left(\frac{d - r(1 - \cos\varphi)}{2\sqrt{D_b t_D}}\right)\right\}. \tag{8.36}$$

Here $\tau_1$ denotes the characteristic mean free time of electrons near the boundary $x = d$, $\tau_1^{-1} \cong v_F \sigma_{eff}\, n_0 C_0$. For simplicity, we assume that the Fermi surface is a body of revolution with its axis coinciding with the direction of the vector **H**. The probability $W$ is an almost step function of the Larmor radius and decreases sharply as $r$ increases over a narrow interval $\delta_W = 2a^{-1}(D_b t_D)^{1/2}$ near the value $r = r_1 = 1/2\left(d - 2a\left(D_b t_D\right)^{1/2}\right)$, where $r_1$ is the root of the equation $W''_{r^2} = 0$. When $l_1 = v_F \tau_1 << \left(D_b t_D\right)^{1/2}$ coefficient

$$a \cong \ln\left\{\frac{1}{\Omega\tau_1}\left(\frac{2\sqrt{D_b t_D}}{d}\right)^{1/2}\right\} >> 1, \tag{8.37}$$

and the quantity $\delta_W$ is substantially smaller than the thickness $x_0$ of the, layer with a high impurity concentration $x_0 = 2a\left(D_b t_D\right)^{1/2}$.

Under the conditions of an anomalous skin effect, when the depth of the skin layer $\delta$ is substantially smaller than the radius $r$ and the characteristic carrier mean free path $l$ $(l >> d,\ r)$ in the bulk of the sample, in order to calculate the surface impedance Z the RF electric field must be determined from Maxwell equations (5.7). The relation between the corresponding components of the field and current (5.8) is found by solving the Boltzmann kinetic equation (2.1) in which the frequency $\tau^{-1}(x)$ of bulk collisions is a function of the $x$ coordinate,

$$\tau^{-1}(x) = \tau_0^{-1} + \tau_1^{-1} Erfc\left(\frac{d - x}{2\sqrt{D_b t_D}}\right),\ d >> x_0, \tag{8.38}$$

where $\tau_0^{-1}$ describes the "background" electron scattering in the bulk of the sample. The second term on the right-hand side of formula (8.38) is due to the carrier scattering in the layer of diffused atoms and is considerably different from zero when the distances from the surface $x = d$ are smaller than $x_0$ or of the same order of magnitude.

If the scattering of carriers that glance along the metal surface $x = 0$ and do not leave the skin layer during their mean free time $\tau_0$ is nearly specular, then the formula for the impedance $Z$ can be written as

$$Z = Z_0 + \Delta Z, \tag{8.39}$$

where $Z_0$ is the impedance of the metal in the main approximation with respect to the anomaly parameter $\delta / r$ which is a smooth function of the magnetic field (see Eq.(5.48)) and $\Delta Z$ is a small correction to because of carriers that execute periodic motion in the bulk of the metal and enter the skin layer.

In the geometry under discussion, when there is a layer of impurities at the surface $x = d$ that scatters electrons diffusely, the singularity discontinuity in the derivative of the impedance with respect to the magnetic field is due mainly to the cutoff of electron orbits with extremal diameter $2r$. Therefore, even when analyzing the surface impedance of the film we assume that the electron time of flight through the skin layer is much shorter than the effective electron mean free time $\left| \omega^* \right|^{-1}$, i.e., the inequality (5.9) is satisfied.

First, we consider the radio-frequency range $\left( \left| \omega^* / \Omega \right| << 1, \; \omega^* = \omega + \tau_0^{-1} \right)$. If the characteristic distance $\delta_W$, at which the probability $W(r)$ changes, is considerably smaller than the skin-layer depth $\delta$, then when $t_D << l_1^2 / D_b$ ($l_1$ is the carrier mean free path near the interface $x = d$; $l_1 << l$) a situation arises similar to that in a thin slab, where one face is specular and the other is diffuse.

The appearance of RF size effect line is attributable to an abrupt change in the number of bulk electrons (noninteracting with the boundaries of the sample) that return to the skin layer in a field $H = H_1$, in which

$$2r_e(H_1) = 2r_1 = d - x_0,$$

$$r_e = r(p_{ze}); \; \partial r / \partial p_z = 0 \; \text{for} \; p_z = p_{ze}. \tag{8.40}$$

The amplitude and width of the RF size effect line are determined by the distribution of the RF field in the skin layer and the position of the line as a function of the magnetic field (and the hence quantity $D_b$) is determined from formula (8.40). When $r_e - r_1 << \delta$ the derivative $\partial Z / \partial H$ is

$$\frac{\partial Z}{\partial H} \cong C_1 \frac{8\omega}{ce} \frac{1}{k_0 H^2} \left( \frac{k_{10}}{k_0} \right)^3 \left( \frac{\partial r}{\partial p_z} \right)^{-1} cth\pi\gamma, \tag{8.41}$$

where

$$k_0^{5/2} = \frac{4\pi^3}{(ch)^3} \frac{\omega}{\omega^*} \frac{e^3 H}{} \int dp_z \Omega(p_z) \quad r^{3/2}; \; k_{10}^3 = \frac{8\pi^2 e^4 \omega}{c^4 h^3} \frac{H^2 r^2}{};$$

$$\gamma = \frac{1}{\Omega \tau_0}; \; C_1 = 0.95 \cdot \quad 10^{-2} \exp\left\{\frac{4\pi i}{5}\right\}; \; |k_0|^{-1} = \delta.$$

Formula (8.41) for $\partial Z / \partial H$ has a singularity at $r(H) = r_e$ because the derivative $\partial r / \partial p_z$ equals to zero for electrons with an extremal radius of trajectory.

In the case of long diffusion annealing times, when $\delta_W$ becomes greater than $\delta$ but as before the inequality $\delta_W \ll x_0 \ll d$ is satisfied, the intensity and width of the RF effect line depend on the nature of the impurity distribution in the boundary region,

$$\frac{\partial Z}{\partial H} \cong C_2 \frac{8\omega}{c^2 k_0} \left(\frac{k_{10}}{k_0}\right)^3 r \quad F'(r) \quad \frac{1}{H} \begin{cases} \sqrt{2\pi} \dfrac{c}{eH} \left(\dfrac{\partial r}{\partial p_z}\right)^{-1} \Phi(r), \; \delta_W \ll r_e - r_1 \ll r_e; \\[3mm] \dfrac{1}{2^{1/4}} \quad \Gamma\left(\dfrac{1}{4}\right) \Phi^{1/2}(r), \; |r_e - r_1| \ll \delta_W, \end{cases} \quad (8.42)$$

where

$$F(r) = \frac{1 + W(r) \quad \exp(-2\pi\gamma)}{1 - W(r) \quad \exp(-2\pi\gamma)}; \; \Phi(r) = \frac{1}{F}\left(\frac{F'(r)}{F''(r)}\right)^{1/2};$$

$$F''(r_1) = 0; \; C_2 = 1.99 \cdot 10^{-2} \exp\left(\frac{4\pi i}{5}\right).$$

In the microwave region $(\omega > \Omega)$, the absorption of energy by an electromagnetic wave is a resonant process. If the magnetic field is so weak that none of the orbits with an extremal period $T_H$ of electron motion fits within the pure region of thickness $d - x_0$, then besides electrons near the reference point on the Fermi surface, carriers whose orbit diameter $2r$ and Larmor frequency $\Omega$ of motion satisfy the equalities

$$2r(p_1) = d - x_0; \; \omega = n\Omega(p_1), \; n = 1, 2, 3, \dots , \quad (8.43)$$

can also participate in the resonance. The effective cutoff of the electron trajectories by the impurity layer occurs only when $\delta_W / r < 1 / \omega\tau_0$ i.e., when the width $\delta p_z = p_F / \omega\tau_0$ of the strip on the Fermi surface that corresponds to the resonance electrons is considerably greater than its smearing $p_F \delta_W / r$ as a consequence of the indeterminacy of the position of the center of the cutoff orbit. When the inequality $r / \delta \gg \omega\tau_0$ is satisfied, as in a slab with a pure surface [101], the RF impedance has a logarithmic singularity near the resonance (8.43). The resonant correction $\Delta Z$ to the part of the impedance that is a monotonic function of the magnetic field for $\gamma \ll 1$ is determined by the parameter $w = \delta_W \omega\tau_0 / r$, which characterizes the decrease in the amplitude and width of the resonance line caused by the smearing of the boundary of the layer of impurities. If the resonance mismatch is small, $\Delta << 1$ $\left[\omega = n\Omega(1-\Delta)\right]$, then the expression for $\Delta Z$ has the form

$$\Delta Z \cong C_2 \frac{8\omega}{c^2 k_0} \left(\frac{\tilde{k}_{10}}{k_0}\right)^3 \ln\left(i\tilde{\gamma} + \left(\frac{\chi}{\chi_1} - 1\right)\Delta\right), \qquad (8.44)$$

where

$$\tilde{k}_{10}^3 = k_{10}^3 \frac{2c}{ieHr\chi}; \quad \chi = -\frac{1}{\Omega}\frac{\partial \Omega}{\partial p_z}; \quad \chi_1 = \frac{1}{r}\frac{\partial r}{\partial p_z}; \quad \tilde{\gamma} = \frac{1}{n}(\gamma + w).$$

All the functions of the momentum in formula (8.44) have been taken for the value $p_z = p_1$, as determined by Eq. (8.43). In deriving Eq. (8.44), we assumed that $\left(r/\delta\right)^{1/2} >> \omega\tau_0$, but this result remains qualitatively valid when $r/\delta >> \omega\tau_0$, as well. If the opposite inequality, i.e., $r/\delta << \omega\tau_0$ is satisfied the derivative of the impedance has a logarithmic singularity.

Thus, the observation of a size-effect cyclotron resonance besides RF size effect in a DLF, which has undergone diffusion annealing, allows the diffusion coefficient $D$ to be calculated from the shift of the resonance frequencies. Formulas (8.41), (8.42), and (8.44) give the solution of the formulated problem, since they determine the amplitude of the RF size effect lines and resonance lines, whose position is found from the formulas (8.40) and (8.43), respectively.

## CONCLUSION

The main feature of electron transport in multilayers distinguishing them from bulk conductors, is the interaction of charge carriers with the internal boundaries, which affects significantly the dependence of the kinetic coefficients on the layer thickness and external fields.

The dependences of the specific conductivity $\sigma$ of a double-layer conducting film (DLF) on its size differ considerably from the corresponding dependences for a monocrystalline sample. For a small thickness of the deposited layer $\left(d_2 << d_1\right)$, its contribution to the conductivity of a DLF is insignificant, but its absolute magnitude $\sigma$ differs from the corresponding value for a single-layer plate in view of the possibility of electron scattering the interface. With increasing layer thickness $\left(d_2 \le d_1\right)$, the value of $\sigma$ decreases since the increase of $d_2$ leads to a simultaneous growth of in the relative number of charge carriers scattered at the interface (their mean free path is of the order of $d_2$). In the range $d_2 > d_1$, the variation of the conductivity $\sigma$ depends on the purity of the conductor layer being deposited. If the bulk value $\sigma_0$ of the deposited layer is smaller than the value $\sigma$ for a DLF with a thin coating $\left(d_2 \to 0\right)$, the quantity $\sigma$ decreases monotonically ($\sigma \to \sigma_0$ for $d_2 \to \infty$). If the inverse inequality is satisfied, the conductivity $\sigma$ attains a minimum for layer thicknesses of the same order of magnitude, and then increase monotonically, approaching asymptotically the bulk value $\sigma_0$ for the deposited layer.

The analysis of the oscillatory dependence $\sigma_{osc}(H)$ of the conductivity of a DLF on the perpendicular to the surface magnetic field $H$ (Zondheimer oscillations) makes it possible to determine the degree of diffuseness of the interface and the probability of tunneling of electrons. The amplitude of the oscillations associated with the dimensions of the individual layer does not depend on the ratio between the probabilities $P$ and $Q$ of reflection and transmission, respectively. Therefore, the investigation of $\sigma_{osc}(H)$ may prove to be the method of obtaining the information on the diffuseness parameter for electron scattering at the interface. In the strong magnetic field applied parallel to the surfaces, the presence of interface improves the conducting properties of the DLF due to the concentration of electrical current near the interface (static skin-effect).

If the radius $r$ of the electron trajectory in the magnetic field of spontaneous induction **B** is comparable with the ferromagnetic layer thickness, the conductivity of a magnetic multilayer (MML) is sensitive to the direction of the current flowing parallel to the boundaries. The MML conductivity perpendicular to the vector **B** depends significantly on the probability of tunneling of electrons through the boundary, and on the mutual orientation of the magnetic moments for quite large values of tunneling probability $Q$.

Andreev reflection of carriers at $n-s$ boundary leads to an entirely different dependence of the high-frequency (HF) surface impedance of a thin normal-metal layer on the magnetic field compared with the impedance $Z$ of a thin metallic plate. When the layer thickness $d$ and Larmor radius satisfy the inequality $r < d < 2r$, a narrow HF-field spike is produced inside the layer. If the electrons are specularly reflected from the surface, the carriers, gliding over the boundary and landing periodically in the spike, produce the resonance that is not observed in either bulk or thin conductors in the normal state. In the same magnetic-field range, at any electron scattering from the layer surface, resonance should be observed at frequencies corresponding to the cyclotron resonance frequencies. In a weak field $H$, at which $r > d$, the behavior of the impedance as a function of the magnetic field depends essentially on the state of the sample boundary. In the case of the specular reflection the resonant **H** dependence of $Z$ is preserved, while in the case of diffuse scattering the cyclotron resonance vanishes for $r > d$. At radio frequencies (RF) in the magnetic field interval $|r - d| < \delta$ an abrupt change of impedance $Z$ takes place due to the contribution made to the HF current by the carriers that interact with the $n-s$ interface ($\delta$ is a skin depth). This manifests itself in the onset of an RF size-effect line at $r = d$.

A number of resonance and oscillation acoustic effects nonexistent in the thin plates are found to occur in thin a layer of a normal metal adjacent to a superconductor. For $r < d < 2r$, the absorption coefficient of ultrasound propagating perpendicular to the surface of the sample and to the $n - s$ boundary undergoes geometrical oscillations, and at sound frequencies $\omega$ being greater than the Larmor frequency $\Omega$ acoustic cyclotron resonance with the "bulk" period sets in. In weaker magnetic fields, such that orbits with the extremal radius are cut off, the behavior of the absorption coefficient $\Gamma$ as a function of $\mathbf{H}$ depends strongly on the state of the external surface of the layer. For example, for the specular reflection the resonance dependence of $\Gamma$ on $\mathbf{H}$ is formed mainly by electrons interacting both with the $n - s$ boundary and with the outer boundary. For the diffuse surface scattering, the resonance oscillations of $\Gamma$ are associated with carriers having a trajectory of nonextremal radius. If ultrasound propagates along the layer, an oscillatory dependence of the coefficient $\Gamma$ on a weak magnetic field ($r > d$) may take place only for the near-specular reflection of charge carriers by the surface of the metal.

A metal interdiffusion influences essentially on the kinetic properties of DLF's. At small diffusion-annealing times $t_D$, the characteristic penetration depth of impurities $x_0$ is smaller than the layer thickness $d_i$ and the size effects in DLF's are determined by the width of the impurity-free region $d_i - x_0$. The coefficients $D_b$ and $D_g$ of bulk and grain-boundary interdiffusion, respectively may be obtained by measuring $t_D$ dependencies of electrical conductivity. The Fourier analysis of the dependence of conductivity on the magnitude of the strong magnetic field oriented perpendicular (or inclined) to the surface makes possible to find the period of Sondheimer oscillations with high accuracy and, therefore, the diffusion depth for impurity atoms to be calculated. The study of a size - dependent cyclotron resonance and RF size effect in a DLF, which has undergone diffusion annealing, allows the diffusion coefficient $D_g$ to be found from the shift of the resonance frequencies.

The experimental investigation of kinetic phenomena in metal multilayers and double-layer films can provide the most convenient method for obtaining information on the nature of scattering of conduction electrons at the interface and on the probability of penetration through it. The relatively simple measurements of the electrical conductivity and high-frequency impedance of DLF's enable determining efficiently the rate of diffusion of impurity atoms in metals. The study of the high-frequency properties of thin normal-metal layers on superconducting substrate makes it possible to observe directly Andreev reflection of carriers, and to gauge its probability and the temperature dependence from the amplitude and width of the resonance lines.

REFERENCES

1. Camley, R.E. and Stamps, J. (1993) *J. Phys. Cond. Mat.* **5**, 3727.

2. Baibich, M.N., Broto, J.M., Fert, A. *et. al*. (1988) *Phys. Rev. Lett.* **61**, 2472.

3. Fert, A. and Campbell, J. (1976) *J. Phys. Metal Phys.* **6**, 849.

4. Vonsovskii, S.V. (1981) *Magnetism,* John Wiley, New York.

5. Radovic, Z. *et al.* (1991) *Phys. Rev. B* **44**, 759.

6. Buzdin, A.I., Vuinchich, B., and Kupriyanov, M.Yu, (1992) *Zh. Eksp. Teor. Fiz*. **74**, 124.

7. Muhge, Th., Westerholt, K., Zapel, H., Garis, Yanov, N.N., Goryunov, Yu.V., Garisfullin, I.A., and Khalinullin, G.G. (1997) *Phys. Rev. B* **55**, 8945.

8. Izyumov, Yu.A., Proshin, Yu.N., and Khusainov, M.G. (2000) *Pis'ma Zh. Eksp. Teor. Fiz.* **71**, 202.

9. Ando, T., Fowler, A., and Stern, F. (1982) *Rev. Mod. Phis.* **54**, 1.

10. Bass, F.G., Bulgakov, A.A., and Teterev, A.P. (l989) *High-Frequency Properties of Semiconducting Superlattice,* Moscow: Nauka (in Russian).

11. Menashe, D. and Laikhman, B. (1996) *Phys. Rev.* **54**, 11561.

12. Lukas, M.S.P. (1964) *Appl. Phys. Lett.* **4**, 73.

13. Kaganov, M.I. and Fiks, V.B. (1977) *Zh. Eksp. Teor. Fiz. ***73***,* 753 (*Sov. Phys. JETP* **46**, 393 (1977)).

14. Ustinov, V.V. (1980) *Fiz. Met. Metalloved.* **49**, 31 (in Russian).

15. Fuchs, K. (1938) *Proc. Camb. Phil. Soc.* **34**, 100.

16. Lifshits, I.M., Azbel', M.Ya., and Kaganov, M.I. (1973) *Electron Theory of Metals,* New York: Consultants Bureau.

17. Chopra, K.L. and Randlett, M.R. (1967) *J. Appl. Phys.* **38**, 3144.

18. Pariset, C. and Chauvineau, P. (1978) *Surf. Sci.* **78**, 478.

19. Fisher, B., Moske, M., and Minnigerode, G.V.Z. (1983) *Z. Phys.* **B51**, 327.

20. Schumacher, D. and Stark, D. (1982) *Surf. Sci.* **123**, 384.

21. Lukas, M.S.P. (1965) *J. Appl. Phys.* **36**, 1632.

22. Bezak, V., Kedro, M., and Pevala, A. (1974) *Thin Solid Films* **23***,* 305.

23. Bergman, G. (1978) *Phys. Rev. Lett.* **41**, 1619.

24. Bergman, G. (1979) *Phys. Rev. B* **19**, 3933.

25. Dimmich, R. and Warkusz, F. (1983) *Thin Solid Films* **109**, 103.

26. Khater, F. (1983) *Acta Phys. Slov.* **33**, 43.

27. Verchenko, V.I., Grishayev, V.I., Dekhtyaruk, L.V., Kolesnichenko Yu.A., and Shermergor, T.D. (1990) *Fiz.* Met. Metalloved. **69**, 102 (*Sov. Phys. Met. Metall.* **69**, 100, (1990)).

28. Dekhtyaruk, L.V. and Kolesnichenko, Yu.A. (1993) *Fiz. Nizk. Temp.* **19**, 1013 (*Sov. J. Low Temp. Phys.* **19** (9), 720, (1993)).

29.  Dimmich, R. (1985) *J. Phys. F. Metal Phys.* **15**, 2477.

30.  Mayaadas, A.F. and Shatzkes, M. (1970) *Phys. Rev. B. Cond.  Matter.* **1**, 1382.

31.  Dekhtyaruk, L.V. and Kolesnichenko, Yu.A. (1997) *Fiz.  Met. Metallotled.* **84**, 37 (*Sov. Phys. Met. Metall.* **84** (2), 118, (1997)).

32.  Dekhtyaruk, L.V. and Kolesnichenko, Yu.A. (1997) *Ukr. Fiz. Zh.* **42**, 1094 (in Ukrainian).

33.  Sondheimer, E.H. (1950) *Phys. Rev.* **80**, 401.

34.  Gurevich, V.L. (1958) *Zh. Eksp. Teor. Fiz.* **35**, 668 (*Sov. Phys.  JETP* **8**, 464 (1959)).

35.  Kirichenko, O.V. (1974) In: *Condensed State Physics* **30***,* Khar'kov, 51 (in Ukrainian).

36.  Kirichenko, O.V., Peschanskii, V.G., and Savel'eva, S.N. (1974) *Zh. Eksp. Teor. Fiz.* **67***,* 1451 (*Sov. Phys. JETP* **8**, 722 (1974)).

37.  Kirichenko, O.V, and Kolesnichenko, Yu.A. (1982) *Fiz. Nizk. Temp.* **8**, 276 (*Sov. J. Low Temp. Phys.* **8**, 138 (1982)).

38.  Kolesnichenko, Yu.A. and Peschansky, V.G. (1984) *Fiz. Nizk. Temp.* **10**, 1141 (*Sov. J. Low Temp. Phys.* **10**, 595 (1984)).

39.  Kolesnichenko, Yu.A., Peschansky, V.G., and Sinolitskii, V.V. (1981) *Fiz. Nizk. Temp.* **7**, 66 (*Sov. J. Low Temp. Phys.* **7**, 328 (1981)).

40.  Azbel', M.Ya. and Peschansky, V.G. (1965) *Zh. Eksp. Teor. Fiz.* **49**, 572 (*Sov. Phys.  JETP* **22**, 399 (1966)).

41.  Peschansky, V.G. and Azbel', M.Ya. (1968) *Zh. Eksp. Teor. Fiz.* **55**, 1980 (*Sov. Phys.  JETP* **28**, 1045 (1969)).

42.  Kulesko, G.I. (1979) *Pis'ma Zh. Eks. Teor. Fiz.* **30**, 641 (*JETP Lett.* **30**, 607 (1979)).

43.  Kulesko, G.I. (1983) *Fiz. Tverd. Tela* **25**, 76 (In Russian).

44.  Gantmakher, V.F. (1962) *Zh. Eksp. Teor. Fiz.* **43**, 345 (*Sov. Phys.  JETP* **16**, 247 (1963)).

45.  Azbel', M.Ya. (1960) *Zh. Eksp. Teor. Fiz.* **39**, 400 (*Sov. Phys.  JETP* **12**, 283 (1961)).

46.  Kaner, E.A. and Gantmakher, V.F. (1968) *Usp, Fiz. Nauk* **43**, 345 (*Sov. Phys. Usp.* **11**, 81 (1968)).

47.  Peschansky, V.G. and Yasemidis, K. (1980) *Fiz. Nizk, Temp.* **6***,* 541 (*Sov. J. Low Temp. Phys.* **6**, 260, (1980)).

48.  Peschansky, V. G., Kardenas, V., Lur'e, M.A. and Yasemidis, K. (1981) *Zh. Eksp. Teor. Fiz.* **80**, 1645 (*Sov. Phys.  JETP* **53**, 849, (1981)).

49.  Sharvin, Yu.V. and Sharvin, D.Yu. (1979) *Zh. Eksp. Teor. Fiz*, **77**, 2153 (*Sov. Phys.  JETP* **50**, 1033 (1979)).

50.  Kolesnichenko, Yu.A. and Lur'e, M.A. (1981) *Fiz. Nizk. Temp.* **7**, 1267 (*Sov. J. Low Temp. Phys.* **7**, 614 (1981)).

51.  Azbel', M.Ya.. and Kaner, E.A. (1957) *Zh. Eksp. Teor. Fiz.* **32**, 896 (*Sov. Phys.  JETP* **5**, 730 (1957)).

52.  Peschansky, V.G. (1968) *Pis'ma Zh, Eks. Teor. Fiz.* **7**, 489 (*JETP Lett.* **7**, 375 (1968)).

53.  Kirichenko, O.V., Lur'e, M.A., and Peschansky, V.G. (1976) *Fiz. Nizk. Temp.* **2**, 858 (*Sov. J. Low Temp. Phys.* **2**, 421 (1976)).

54.  Meierovich, B.E. (1970) *Zh. Eksp. Teor. Fiz* **58**, 324 (*Sov. Phys. JETP* **31**, 175 (1970)).

55.  Fal'kovskii, L.A. (1970) *Zh. Eksp. Teor. Fiz* **58**, 1830 (*Sov. Phys. JETP* **31**, 981 (1970)).

56.  Zherebchevskii, D.E. and Kaner, E.A. (1972) *Zh. Eksp. Teor. Fiz.* **63**, 1858 (*Sov. Phys. JETP* **36**, 983 (1973)).

57.  Camley, R.E. and Barans, J. (1989) *Phys. Rev. Lett.* **63**, 664.

58.  Barans, J., Fuss, R.E., Camley, R.E. *et. al.* (1990) *Phys. Rev. B* **44**, 8110.

59.  Dieny, B. (1992) *J. Phys. Cond. Mat.* **4**, 8009.

60.  Ustinov, V.V. (1994) *Zh. Eksp. Teor. Fiz.* **106**, 207 (*Sov. Phys. JETP* **79**, 113 (1994)).

61.  Okulov, V.I. (1994) *Fiz. Nizk. Temp.* **20**, 100 (*Sov. J. Low Temp. Phys.* **20**, 81, (1994)).

62.  Kolesnichenko, Yu.A. and Dekhtyaruk, L.V. (1997) *Fiz. Nizk. Temp.* **23**, 936 (*Sov. J. Low Temp. Phys.* **23**, 702, (1997)).

63.  Azbel', M.Ya. (1963) *Zh. Eksp. Teor. Fiz.* **44**, 983 (*Sov. Phys. JETP* **17**, 667 (1963)).

64.  Peschansky, V.G. (1992) *Sov. Sci. Rev. A. Phys.* **16**, 1.

65.  Kirichenko, O.V., Peschansky, V.G., and Savel'eva, S.N. (1979) *Zh. Eksp. Teor. Fiz.* **77**, 2045 (*Sov. Phys. JETP* **50**, 976 (1979)).

66.  Zakharov, Yu.V. and Man'kov, "Yu.I., (1986) *Phys. Stat. Solidi, (b)* **125**, 197.

67.  Zakharov, Yu.V., Man'kov, Yu.I., and Titov, L.S. (1986) *Fiz. Nizk. Temp.* **12**, 408 (*Sov. J. Low Temp. Phys.* **12**, 232, (1986)).

68.  Kaganov, M.I. and Peschansky, V.G. (1958) *Zh. Eksp. Teor. Fiz.* **35**, 1052 (*Sov. Phys. JETP* **8**, 734 (1958)).

69.  Sondheimer, E.H. (1950) *Adv. Phys.* **1**, 1.

70.  Andreev, A.F. (1964) *Zh. Eksp. Teor. Fiz.* **46**, 1823 (*Sov. Phys. JETP* **19**, 1228 (1964)).

71.  Krylov, I.P. and Sharvin, Yu.V. (1970) *Pis'ma Zh. Eks. Teor. Fiz.* **12**, 102 (*JETP Lett.* **12**, 71 (1970)).

72.  Bozhko, S.I., Tsoi, V.S., and Yakovlev, C.E. (1982) *Pis'ma Zh. Eks. Teor. Fiz.* **36**, 123 (*JETP Lett.* **30**, 153 (1982)).

73.  Shepelev, A.G., Ledenev, O.P. and Filimonov, G.D. (1971) *Pis'ma Zh. Eks. Teor. Fiz.* **14**, 428 (*JETP Lett.* **14**, 290 (1971)).

74. Kolesnichenko, Yu.A. and Peschansky, V.G. (1983) *Zh. Eksp. Teor. Fiz.* **85**, 1409 (*Sov. Phys. JETP* **58** (4), 817 (1983)).

75. Galaiko, V.P., Bezuglyi, E.V., and Puppe, E. (1974) *Fiz. Met. Metalloved,* **37**, 478 (in Russian).

76. Fal'kovskii, L.A. (1982) *Poverkhnost* **1**, 13 (in Russian).

77. Kolesnichenko, Yu.A. (1985) *Fiz. Nizk. Temp.* **11**, 703 (*Sov. J. Low Temp. Phys.* **11**, 386 (1985)).

78. Gurevich, V.L., Lang, O.G., and Pavlov, S.T. (1970) *Zh. Eksp. Teor. Fiz.* **59**, 1679 (*Sov. Phys. JETP* **32**, 914 (1971)).

79. Fal'kovsky, L.A. (1983) *Adv. Phys.* **32**, 753.

80. Gokhfeld, V.M., Kirichenko, O.V., and Peschansky, V.G. (1980) *Zh. Eksp. Teor. Fiz.* **70**, 538 (*Sov. Phys. JETP* **52**, 271 (1980)).

81. Mikoshyba, N.J. (1958) *Phys. Soc. Jpn.* **13**, 759.

82. Kaner, E.A. (1962) *Zh. Eksp. Teor. Fiz.* **43**, 215 (*Sov. Phys. JETP* **16**, 154 (1963)).

83. Gokhfeld, V.M. (1975) *Fiz. Nizk. Temp.* **1**, 268 (*Sov. J. Low Temp. Phys.* **1**, 131 (1975)).

84. Peschansky, V.G. and Gokhfeld, V.M. (1969) *Ukr. Fiz. Zh.* **14**, 461 (in Ukrainian).

85. Gokhreld, V.M. and Peschansky, V.G. (1971) *Zh. Eksp. Teor. Fiz.* **61**, 762 (*Sov. Phys. JETP* **34**, 407 (1972)).

86. Gudenko, S.V. and Krylov, I.P. (1978) *Pis′ma Zh. Eks. Teor. Fiz.* **28**, 243 (*JETP Lett.* **28**, 224 (1978)).

87. Gudenko, S.V. and Krylov, I.P, (1984) *Zh. Eksp. Teor. Fiz.* **86**, 2304 (*Sov. J. Phys. JETP* **59**, 1343 (1984)).

88. Kolesnichenko, Yu.A. (1986) *Fiz. Nizk. Temp.* **12**, 632 (*Sov. J. Low Temp. Phys.* **12**, 358, (1986)).

89. Kolesnichenko, Yu.A. (1985) *Fiz. Nizk. Temp.* **11**, 1165 (*Sov. J. Low Temp. Phys.* **11**, 641 (1985)).

90. Koval', V.F., Vatamannyuk, V.I., Ostroukhov, Yu.S., and Panchenko, O.A. (1986) *Fiz. Nizk. Temp.* **12**, 880 (*Sov. J. Low Temp. Phys.* **12** (8), 500 (1986)).

91. Dekhtyaruk, L.V. and Kolesnichenko, Yu.A. (1993) *Fiz. Met. Metalloved.* **75**, 21 *(Sov. Phys. Met. Metall.* **75**, 474 (1993)).

92. Volkova, R.P., Palatnik, L.S., and Pugachev, A.T. (1982) *Fiz, Tverd. Tela* **24**, 1161 (in Russian).

93. Yolkova, Yu.A., Yolkova, R.P., and Pugachev, A.T. (1986) *Fiz,. Met. Metalloved.* **62**, 298 (in Russian).

94. Raichenko, A.I. (1981) *Mathematical Theory of Diffusion in Appllications* Kiev: Naukova Dumka (in Ukrainian).

95. Poate, J.M., Tu, K.N., and Mayer, J.W. (eds.) (1978) *Thin Films-Interdiffussion and Reactions,* New York: Wiley Interscience.

96. Cahn, J.W., Pan, J.D. and Balluffi, R.W. (1979) *Scr. Metall.* **13**, 503.

97. Kaur, J. and Gust, W. *Fundamentals of Grain and Interphase Boundary Diffusion,* Stuttgart: Ziegh, 1988.

98. Fisher, J.C. (1951) *J. Appl. Phys.* **22**, 74.

99. Klotsman, S.M. (1991) *Usp. Fiz. Nauk* **160**, 99 (in Russian).

100. Kolesnichenko, Yu.A. (1985) *Fiz. Niz. Temp.* **11**, 1165 (*Sov. J. Low Temp.* Phys. **11**, 641 (1985)).

101. Cherepanov, A.N., Marchenkov, V.V., Startsev, V.E., and Volkenshtein, N.V. (1986) *Fiz. Nizk. Temp.* **12**, 1181 (*Sov. J. Low Temp. Phys.* **12** (11), 666 (1986)).

102. Kirichenko, O.V., Lur'e, M.A., and Peschansky, V.G. (1976) *Zh. Eksp. Teor. Fiz.* **70**, 337 (*Sov. Phys. JETP* **43**, 175 (1976)).